\documentclass[ reprint,superscriptaddress,amsmath,amssymb,aps,pre,nofootinbib]{revtex4-2}

\usepackage{amsmath, mathrsfs}
\usepackage{amssymb}
\usepackage{mathtools}
\usepackage{graphicx}
\usepackage{hyperref}
\usepackage[arrow, matrix, curve]{xy}
\usepackage{bm}
\usepackage{yhmath}
\usepackage{subdepth}

\usepackage{textcomp}

\usepackage{amsthm}
\theoremstyle{definition}

\usepackage{mdframed}
\newmdenv[
    linecolor=blue,
    linewidth=1pt,
    roundcorner=5pt,
    backgroundcolor=blue!5,
    skipabove=10pt,
    skipbelow=10pt
]{definitionbox}

\newmdenv[
    linecolor=magenta,
    linewidth=1pt,
    roundcorner=5pt,
    backgroundcolor=magenta!5,
    skipabove=10pt,
    skipbelow=10pt
]{fundamentalbox}

\newmdenv[
    linecolor=red,
    linewidth=1pt,
    roundcorner=5pt,
    backgroundcolor=red!5,
    skipabove=10pt,
    skipbelow=10pt
]{propositionbox}

\newcommand\diff{\mathrm{d}}

\renewcommand\vec[1]{\boldsymbol{\mathrm{#1}}}

\DeclareMathOperator\Real{Re}
\DeclareMathOperator\Imag{Im}

\usepackage[usenames,dvipsnames]{xcolor}
\hypersetup{colorlinks=true, linkcolor=BrickRed, urlcolor=blue!50!black, citecolor=blue!50!black}

\usepackage{amsmath,amsthm,amsfonts,amssymb,times,bbm,graphicx,epsfig,bm}
\usepackage{url} 
\usepackage{hyperref}

\usepackage[normalem]{ulem}

\makeatletter
\newcommand\hide@visible[1]{%
  \bgroup\fboxsep=.3ex\colorbox{Gray}{begin hide}%
  #1\colorbox{Gray}{end hide}\egroup%
}
\newcommand\hide@hidden[1]{%
  \bgroup\fboxsep=.3ex\colorbox{Gray}{hidden text}%
}
\newcommand\hide@invisible[1]{}
\newcommand\makevisible{\let\hide\hide@visible}
\newcommand\makehidden{\let\hide\hide@hidden}
\newcommand\makeinvisible{\let\hide\hide@invisible}

\makeatother
\makehidden

\usepackage{pgfplots}
\usepackage{tikz,animate}
\usetikzlibrary{arrows,shapes}
\usepackage{pgfplots}
\usetikzlibrary{shapes.arrows}
\usepackage{bm}

\begin{document}

\title{Fundamental problems in Statistical Physics XIV: Lecture on Correlation and response functions in statistical physics }


\hfill {\tt Published in  J. Stat. Mech.(2026)104002}

\author{Thomas Franosch}
\affiliation{Institut f\"ur Theoretische Physik, Universit\"at Innsbruck, Technikerstra{\ss}e, 21A, A-6020 Innsbruck, Austria}
\email{thomas.franosch@uibk.ac.at}


\date{\today}

\begin{abstract}

In the first part of these short lecture notes, we will present an introduction to (auto-)correlation functions and linear-response functions in the language of a physicist. In particular,  the fluctuation-dissipation theorem in classical physics is presented underlining the central role of correlation functions.  The fundamental importance of (auto-)correlation functions raises the natural question on how they are characterized in general without referring to the concrete underlying dynamical laws. 
Perhaps unexpectedly -- despite being elegant and long established in the mathematical literature (Bochner's theorem for correlations; Herglotz-Nevanlinna representations for response) -- this answer is not widely appreciated in physics, partly because the requisite tools lie outside the standard curriculum.
 In the second part we adopt a more rigorous viewpoint: we state the key structural properties of correlation functions and provide selected derivations of these results. Finally, we return to linear response and discuss general characterization results for response functions.
\end{abstract}

\maketitle

\section{Introduction}\label{Sec:Introduction}

These notes bridge the gap between physical intuition and the underlying mathematics by summarizing the key properties of correlation and response functions without delving into the full formalism. Our objective is to make the insights from the relevant theorems accessible to physicists and practitioners so they can be applied effectively in practice. This approach both deepens understanding and helps ensure that models and data-driven reconstructions are physically and mathematically consistent.

We provide an accessible introduction to correlation functions in statistical physics and their connection to linear-response experiments. While the topic is well covered in standard textbooks (see, e.g., \cite{Kubo:Statistical_physics_II:2012, Hansen:Theory_of_simple_liquids:2013}), our focus is to go beyond the traditional treatment and offer a more rigorous characterization of correlation functions. Specifically, we ask: Can one determine whether a given function could arise as the correlation function of a stationary stochastic process?

This question is particularly relevant to theorists constructing simple models for experimental or simulation data, and to data-driven approaches more broadly. For such models to be meaningful, they must respect the constraints of probability theory. Perhaps unexpectedly, the answer is both elegant and long established in the mathematical literature; yet the necessary framework often lies outside a typical physics curriculum, drawing on advanced concepts from probability and stochastic processes. By collecting and explaining the key structural results in a self-contained way, we aim to make these tools usable in day-to-day modeling, thereby promoting models that are both faithful to the physics and mathematically sound.

\section{Correlation functions}

\subsection{Importance of correlation functions in statistical physics}

Statistical physics deals with systems that have an enormous number of degrees of 
freedom, making a complete microscopic description practically impossible, even when the underlying
 dynamics are deterministic. Instead, statistical characterizations are employed to provide 
insight into the relevant processes governing the system's behavior. Among the simplest and most fundamental of these tools are time-dependent correlation functions, which describe how observable quantities at different times are related in equilibrium systems or nonequilibrium steady states.

Correlation functions are central to statistical physics for several reasons. First, they often carry intrinsic physical meaning and are directly measurable in both experiments and simulations. For example, in molecular-dynamics simulations, observables such as particle velocities, positions, or forces can be monitored over time, and their correlation functions can be computed. These results can then be compared to predictions from analytical theories, providing a direct test of theoretical models. Similarly, in experimental set-ups, microscopic information is often accessible, allowing correlation functions to be evaluated directly from recorded data. A prominent example is found in soft-matter systems, where mesosized particles, such as colloids, can be tracked using video microscopy, enabling the computation of time-dependent correlations.

Second, in thermal equilibrium, correlation functions determine the linear (and, with appropriate generalizations, weakly nonlinear) response of many-body systems to small, time-dependent perturbations. Via the fluctuation-dissipation theorem, equilibrium fluctuations are linked to response functions, allowing key transport and material properties -- such as electrical and thermal conductivity, shear and bulk viscosity, and self- and inter-diffusion coefficients -- to be computed from equilibrium correlation functions. Beyond the linear regime, additional information (e.g., higher-order response functions or full nonequilibrium dynamics) is generally required.

Third, correlation functions play a central role in interpreting scattering experiments, such as light scattering, neutron scattering, and photon-correlation spectroscopy. In these experiments, the measurable quantity is the angle- and energy-resolved scattering cross section, which is directly related to the dynamic structure factor. The dynamic structure factor is the Fourier transform of a spatio-temporal correlation function, providing a bridge between microscopic dynamics and experimental observables. This connection allows experimentalists to probe the dynamics of systems at different length and time scales.

Fourth, correlation functions are invaluable in computer simulations. Instead of driving a system out of equilibrium to measure its response, one can monitor the system's spontaneous equilibrium fluctuations and compute the corresponding correlation functions. Similarly, rather than simulating an incident beam of particles and tracking their scattering, one can calculate the relevant spatio-temporal correlation functions and perform a Fourier transform to obtain the scattering cross section. This approach simplifies computational efforts while providing direct access to experimentally relevant quantities.

Finally, (auto-)correlation functions are essential for theoretical modeling and are subject to rigorous mathematical constraints. Since they arise from actual stochastic processes, correlation functions must satisfy symmetry properties, such as time-reversal symmetry in equilibrium systems, and positivity conditions, such as the non-negativity of the power spectrum (the Fourier transform of the correlation function). They must also decay appropriately over time to reflect the system's relaxation to equilibrium. These constraints are not merely mathematical curiosities but serve as essential guides for constructing physically meaningful models and theories. General theorems, such as the Bochner-Khinchin theorem, ensure that theoretical predictions remain consistent with the principles of probability and stochastic processes. For example, the Bochner-Khinchin theorem relates the Fourier transform of an autocorrelation function to a non-negative spectral density, providing a rigorous foundation for the analysis of correlation functions.

Correlation functions are a cornerstone of statistical physics, providing a bridge between microscopic dynamics and macroscopic observables. They are directly measurable in experiments and simulations, encode essential material properties through their connection to linear-response theory, and play a pivotal role in interpreting scattering experiments. For theorists, correlation functions serve as a testing ground for models and approximations, constrained by the rigorous mathematical framework of probability theory. By combining experimental, computational, and theoretical approaches, correlation functions offer a powerful tool for understanding the complex behavior of many-body systems in thermal equilibrium and non-equilibrium steady states.

\subsection{Formal definition of correlation functions}

Many physical systems can only be described statistically because they involve an overwhelming number of degrees of freedom, making it impossible to predict the precise value of an individual observable. This limitation applies not only to systems with intrinsically random time evolution but also to those governed by deterministic microscopic laws, where the dynamics exhibit high complexity or chaotic behavior. Examples include many-body systems with Newtonian or quantum time evolution, time series of weather data, or fluctuations in the stock market. In all these cases, the sheer complexity of the system necessitates a statistical approach.

In physics, the quantities of primary interest are observables -- measurable outcomes of experiments or computer simulations. In the language of probability theory, these correspond to random variables, that is, measurable maps from a sample space $\Omega$ (the set of all possible outcomes) to the real numbers:
$
X: \Omega \to \mathbb{R}.
$
Deterministic experiments (e.g., Newtonian dynamics with fixed initial data) fit into this framework as a degenerate case in which $X$ takes a single value with probability one.

For mathematical consistency, the sample space $\Omega$ is endowed with a $\sigma$-algebra of measurable subsets, which specifies the events to which probabilities can be assigned, and the random variable $X$ is required to be measurable, meaning that for any $a\in\mathbb{R}$ the preimage $X^{-1}((-\infty,a])$ is measurable in $\Omega$. These standard conditions ensure that the theory is well defined; we do not dwell on them here. For rigorous treatments, see, for example, \cite{Feller:Probability:1991,Shiryaev:Probability:2016}. The discussion extends straightforwardly to complex-valued observables modeled as random variables $X:\Omega\to\mathbb{C}$ or vector-valued random variables $X:\Omega\to \mathbb{R}^n$.

In practice, a random variable $X$ assigns a value $X(\omega)$ to each outcome $\omega\in \Omega$. When the experiment is repeated, the outcome $\omega\in \Omega$ changes, leading to random fluctuations in the observed value of $X(\omega)$. Examples of random variables in physics include the position of a particle in a fluid, the velocity of a particle,
 or the kinetic energy of a collection of particles. These quantities fluctuate due to the inherent randomness (e.g., in quantum physics), or complexity of the system (many-body systems in classical physics) by lack of control of all degrees of freedom.

 \begin{figure}[htp!]
\includegraphics[width=\linewidth]{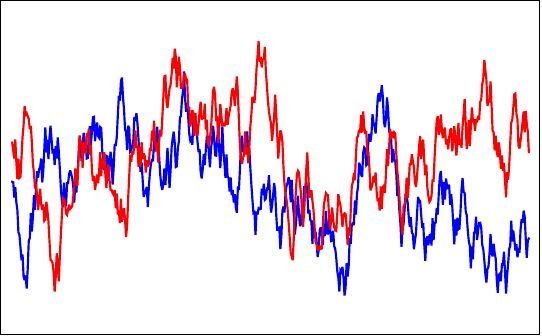}
\caption{Two sample paths $X(t,\omega)$ of a stochastic process.
  \label{fig:trajectories}}.
\end{figure}

Rather than focusing on a single random variable, we are often interested in stochastic processes, which are families of random variables indexed by a parameter set $T$. A stochastic process is denoted as $(X(t): t\in T )$, where $t$ is the index parameter. In most physical applications, $t$ represents time, and the process describes the evolution of the system over time. Depending on the context, the time parameter $t$ can take different forms: $T=\mathbb{N}$ or $T=\mathbb{Z}$ in discrete time, or $T=\mathbb{R}$ in continuous time.  
For simplicity, we will focus on continuous time ($T=\mathbb{R}$) in the following discussion, as it is most relevant for many physical systems.

For each outcome $\omega\in \Omega$, the stochastic process generates a function $X(t,\omega)$, which is called a realization, sample path, or trajectory, see Fig.~\ref{fig:trajectories} for illustration. This function represents the time evolution of the observable for a specific outcome of the experiment. For example, in a molecular-dynamics simulation, $X(t,\omega)$ could represent the position of a particle as a function of time for a given initial condition.
    In a weather model, $X(t,\omega)$ could represent the temperature at a specific location as a function of time for a particular realization of the atmospheric conditions. Each realization corresponds to a single "run" of the experiment or simulation, while the stochastic process as a whole describes the ensemble of all possible realizations.

To describe stochastic processes quantitatively, we assume that probabilities can be assigned to the outcomes of the process. In mathematical terms, the expectation value of a random variable is denoted by  $\mathbb{E}[ \ldots ]$. However, in the statistical physics community, it is customary to use angular brackets $\langle \ldots \rangle$ to denote the expectation value  and we shall follow the physicist's notation.
 The expectation value represents the \emph{average} or \emph{mean value} over the ensemble of initial conditions, realizations of the stochastic processes or disorder, etc. 
 Experimentally, this average can (at least in principle) be determined by repeating the random experiment many times and computing the empirical mean. For a stochastic process $( X(t): t\in \mathbb{R})$ , the one-time average is given by 
$\langle X(t)\rangle$ for times $t\in \mathbb{R}$. 
 Similarly, two-time averages describe correlations between the values of the process at two times $\langle X(t) X(s)\rangle$ with $t,s\in \mathbb{R}$. For two stochastic processes $( X(t): t\in \mathbb{R}), (Y(t): t\in \mathbb{R})$ defined on the same probability space, we can also compute the cross-correlation $\langle X(t) Y(s) \rangle$. The generalization to $n$-time averages is straightforward, but for simplicity, we will focus on one-time and two-time averages. 
Throughout this discussion, we also assume that all these averages exist and are finite. 

The particularly important class of  stochastic processes for physical applications is that of a  \emph{stationary process}. 
A stochastic process is stationary if its statistical properties are invariant under a shift in time. In other words, the process "looks the same" at any point in time, at least in a statistical sense. This reflects the idea that the results of an experiment should not depend on the specific time at which the experiment is performed.
Mathematically, stationarity  implies the following
\begin{subequations}
\begin{align}
\langle X(t) \rangle &= \langle X(0) \rangle, \\
\langle X(t) Y(s) \rangle &= \langle X(t-s) Y(0) \rangle.
\end{align}  
\end{subequations}
The first identity states that the average of a random quantity is independent of time. The  second implies 
that the two-time correlation depends only on the time difference (or lag) $t-s$, rather than the absolute times $t$ and $s$. This allows us to simplify the notation by shifting one of the times to zero.

From now on, we will restrict the discussion to  stationary processes. For such processes, if no time is explicitly specified, it is understood to refer to time zero. For example,  $X=X(0)$ and $\langle X \rangle = \langle X(t) \rangle$. Since the averages of the stationary stochastic processes are time-independent and finite, 
it is often convenient to focus on the fluctuations of the process around its mean value. The fluctuation of a process $X(t)$
is defined as $\delta X(t) \coloneq X(t) 
- \langle X \rangle$. 
This decomposition allows us to isolate the random, time-dependent part of the process from its mean behavior. For later applications, we generalize this framework to complex-valued stochastic processes, which do not require any fundamentally new concepts but are useful in many physical contexts.

With all the preliminaries introduced, we are now ready to define the fundamental quantity of interest in this lecture. 
\begin{definitionbox}
\textbf{Definition:}  
For two time-dependent complex-valued observables $( X(t): t\in \mathbb{R} )$, $(Y(t): t\in \mathbb{R})$, the 
\textbf{time-dependent correlation function} between the fluctuations of $X$ and $Y$ is 
$$C_{XY}(t) \coloneq \langle \delta X(t) \delta Y^* \rangle,$$
where $\delta X(t) \coloneq X(t) - \langle X\rangle$ and $\delta Y^*$ is the complex conjugate of the fluctuation of $Y$ at time zero.  
\end{definitionbox}
This definition captures the statistical relationship between the fluctuations of two observables at different times. The correlation function is a central object in statistical physics, as it encodes information about the dynamics, memory, and interactions within the system. The correlation function of the fluctuations coincides with the covariance function of the original observables.

Let us delve into the important example of a system evolving deterministically under Newtonian dynamics. The state of the system at time $t=0$ is fully specified as a point in phase space, denoted by $\omega = (q,p)\in \Omega$, where $q=(q^1,\ldots, q^f)$ represents the collection of generalized coordinates and $p=(p_1,\ldots, p_f)$ the associated generalized momenta, and $f$ is the number of degrees of freedom. Since Newtonian dynamics is deterministic, the randomness here enters solely by the initial condition, either because the exact $(q,p)$ are imperfeclty known or because they are sampled from a prescribed probability distribution on $\Omega$ (e.g., microcanonical, canonical, or another ensemble).
The time evolution of the phase space point $\omega_t = (q_t, p_t)$ is driven by a Hamilton function $\mathcal{H}=\mathcal{H}(q,p)$ via the canonical equations of motion
\begin{subequations}
\begin{align}
\dot{q}^i &= \frac{\partial \mathcal{H}}{\partial p_i}, \\
\dot{p}_i &= -\frac{\partial \mathcal{H}}{\partial q^i},  
\end{align}
\end{subequations}
 for $i=1,\ldots, f$. These equations describe how the generalized coordinates and momenta evolve over time.
 
Since we are aiming for a stationary stochastic process, the Hamilton function is assumed to be time-independent, ensuring that the dynamics are autonomous and do not explicitly depend on time. In this framework, observables 
are functions defined on phase space, $X: \Omega\to \mathbb{C}$, which maps a phase-space point $\omega=(q,p)$ to a complex number.  In short, an observable is written as $X=X(\omega) = X(q,p)$. The time-dependence of an observable 
is inherited from the time evolution of the  phase-space point $\omega_t$ under the canonical equations of motion. Specifically, the time-dependent observable is provided by   $X(t) \coloneq X(\omega_t) = X(q_t, p_t)$. Here, $X(t)$ represents the value of the observable at time $t$, evaluated along the trajectory of the system in phase space. 
To ensure stationarity, we consider only observables that do not have an explicit time dependence, meaning that any time dependence in $X(t)$ is inherited entirely from the phase space dynamics.

The randomness in this deterministic system arises from the choice of the initial conditions. In equilibrium statistical physics, the initial phase space point $\omega \in \Omega$ 
is often  drawn from the canonical equilibrium probability density, $\rho^{\text{eq}} : \Omega \to [0,\infty)$, which is given by:
\begin{subequations}
\begin{align}
\rho^{\text{eq}}(\omega) &= Z^{-1} \exp[ - \mathcal{H}(\omega)/k_B T] ,\\
Z &= \int \exp[ - \mathcal{H}(\omega)/k_B T] \diff \omega .
\end{align} 
\end{subequations}
Here $T$ denotes the temperature, $k_B$ is Boltzmann's constant. $Z$ 
 is the partition function, which normalizes the probability density.
 To properly connect this formalism  to thermodynamics, the  phase-space element $\diff \omega$ is made dimensionless and includes the Gibbs factor to account for  the indistinguishability of  particles. 

In equilibrium, the average of an observable $X=X(\omega)$ is defined as the expectation value with respect to the equilibrium distribution
\begin{align}
\langle X \rangle \coloneq 
\int X(\omega) \rho^{\text{eq}}(\omega) \diff \omega.
\end{align}
This average corresponds to the ensemble mean, where the contributions of all possible initial conditions are weighted by their equilibrium probabilities. Importantly, the normalization factor of the phase-space element $\diff \omega$ cancels out in the computation of averages, simplifying the evaluation. 

The time-dependent correlation function quantifies the statistical relationship between two observables $X$ and $Y$ at different times and is provided by
\begin{align}
C_{XY}(t) = \int \delta X(\omega_t) \delta Y(\omega)^* \rho^{\text{eq}}(\omega) \diff \omega. 
\end{align}
In words, an initial state $\omega\in \Omega$ in phase space is drawn according to the canonical distribution, then the fluctuation of $Y^*$ is evaluated for this point. The initial state $\omega$ evolves according to the canonical equations of motion to $\omega_t$. The fluctuation of $X$ is evaluated at the evolved phase space point $\omega_t$.
The product of these fluctuations is averaged over all possible initial conditions, weighted by the equilibrium distribution.

As a second example, we consider a generalization of deterministic motion in phase space to a more abstract dynamics in a general state space $S$. In this framework, the observables are restricted to functions that map state points to complex numbers, $X: S\to \mathbb{C}$.
 Each outcome of the experiment yields an entire trajectory in the state space, denoted by $\omega = (\omega_t: t\in \mathbb{R})$, where at each instant of time $t$, the system  is in state  $\omega_t \in S$. 
The time dependence of the observable arises solely from the time evolution of the system in the state space, and is given by $X(t) \coloneq X(\omega_t)$. 
In particular, the observables are restricted to depend only on the current state of the system, in contrast to  the general probabilistic framework where observables may include the entire history of the stochastic process.  The current set-up is particularly useful for the case of Markov processes~\cite{Shiryaev:Probability:2016}, where the probability to transition to a state $\omega_t \in S$ at time $t$ given that the system is in state $\omega_s \in S$ at some earlier time $s\leq t$ is independent of the history on how the system arrived in $\omega_s$ before.
For a stationary process, the probability density  to find  the system in a particular state point $\omega_t$ at time $t$  is time-independent, meaning $p^{\text{st}}(\omega_t) = p^{\text{st}}(\omega_0)$ where $p^{\text{st}}(.)$ is the stationary probability density.  This reflects the fact that the statistical properties of the system do not change over time.

The time-dependent correlation function for two observables $X$ and $Y$ in this generalized setting can be computed as
\begin{align}
C_{XY}(t) = \int \diff \omega_t \diff \omega_0 \delta X(\omega_t) \mathbb{P}(\omega_t, t | \omega_0, 0 ) \delta Y(\omega_0)^* p^{\text{st}}(\omega_0) ,
\end{align}
where $\mathbb{P}(\omega_t, t | \omega_0, 0)$ denotes the conditional probability density for the system of being in state point $\omega_t$ at time $t$ given it  started in $\omega_0$ at time $t=0$. The interpretation is analogous to the Newtonian case. First a random initial state point $\omega_0\in S$ is drawn from the stationary distribution $p^{\text{st}}(\omega_0)$. 
At the initial state $\omega_0$, the fluctuation of the observable $Y^*$ is evaluated as $\delta Y(\omega_0)^*$.
The system evolves from the initial state $\omega_0$ to a new state $\omega_t$ at time $t$, according to the dynamics of the system. The probability of reaching $\omega_t$ at time $t$ is given by the conditional probability density $\mathbb{P}(\omega_t, t | \omega_0, 0)$. At this new state point $\omega_t$  the fluctuation of $X$ is evaluated as $\delta X(\omega_t)$. The product of the fluctuations $\delta X(\omega_t)$ and $\delta Y(\omega_0)^*$  is averaged over all possible initial states $\omega_0$ and final states $\omega_t$ weighted by the stationary distribution $p^{\text{st}}(\omega_0)$ and the conditional probability $\mathbb{P}(\omega_t, t | \omega_0, 0)$. 

This formulation is highly general and applies to a wide range of systems, including those with stochastic dynamics, Markov processes, or deterministic dynamics in a probabilistic setting. It provides a powerful framework for analyzing time-dependent correlations in systems where the state space is not necessarily a phase space but a more abstract configuration space.

\section{Linear Response}\label{Sec:Linear_Response}

In this section, we discuss some elementary properties of response functions and establish their connection to correlation functions in equilibrium (classical) statistical physics via the celebrated \emph{fluctuation-dissipation theorem}. 
To illustrate these concepts, we  provide a series of instructive examples that demonstrate how correlation and response functions encode transport properties of macroscopic systems. 

A fundamental strategy for probing the  properties of a system is to apply a small time-dependent perturbation  and monitor the response of an observable. For definiteness, we denote  the perturbation a 'force' $f(t)$\footnote{In the following, we use sloppy notation and do not carefully distinguish between a function $f$ and its value $f(t)$ assumed at a certain instant of time $t$.} and the  macroscopic observable as $X(t)$. By 'macroscopic', we mean that fluctuations are negligible, and later we will identify $X(t)$ with the ensemble average in a  slightly perturbed system.

In general, the linear response $X(t)$ must be a linear functional of the force history $\{f(\bar{t}) : \bar{t} \leq t\}$. This is expressed as  
\begin{propositionbox}
\textbf{Linear response}
\begin{align}\label{eq:lin_resp}
X(t) = (\chi \ast f)(t) \coloneq \int_{-\infty}^\infty \chi(t-\bar{t}) f(\bar{t}) \diff \bar{t}. 
\end{align}
\end{propositionbox}
Equation~\eqref{eq:lin_resp} represents the most general form of a functional that satisfies the principles of linearity and  time-translational invariance. Specifically, if the entire force history is shifted in time, the repsonse must shift accordingly. This requirement implies that the response function $\chi$ depends only on the lag time $t-\bar{t}$, which is the time difference between the moment $t$ when the response is evaluated and the time $\bar{t}$ where the external force acts. Furthermore, the principle of  causality imposes an additional constraint:  the response function must vanish for negative times, $\chi(t)= 0$ for $t<0$. This ensures that the response at time $t$ depends only on the force applied at earlier times $\bar{t}$, and not on any future forces. In particular, the integration domain in Eq.~\eqref{eq:lin_resp} reduces to the interval $(-\infty, t]$. 

A cornerstone of non-equilibrium statistical physics is the fluctuation-dissipation theorem (FDT)~\cite{Kubo:Statistical_physics_II:2012, Hansen:Theory_of_simple_liquids:2013}, which  establishes a profound connection between the macroscopic response  and the equilibrium correlations  of microscopic fluctuations. Consider a  classical system with  time-independent Hamiltonian $\mathcal{H}_0$ that is perturbed by a generalized force $f(t)$ conjugate to an observable $Y$,
\begin{align}
\mathcal{H}(t) = \mathcal{H}_0 - f(t) Y.
\end{align} 
The expectation value of another observable $X$ becomes time-dependent in the driven ensemble; denote it by 
$\langle X(t) \rangle_f$. We focus on the deviation from equilibrium and, for simplicity assume $\langle X \rangle =0$ (otherwise replace $X$ by $\delta X = X -\langle X\rangle$. 
 A microscopic calculation (see  Appendix \ref{Sec:Appendix_FDT}) shows that, to linear order in the perturbing force, the deviation is given by
\begin{align}\label{eq:micro_lin_resp}
 \langle X(t) \rangle_f = \int_{-\infty}^\infty \chi_{{XY}}(t-\bar{t}) f(\bar{t}) \diff \bar{t},
\end{align}
where the response function is 
\begin{align}\label{eq:FDT}
\chi_{XY}(t) = -\frac{1}{k_B T} \Theta(t) \frac{\diff }{\diff t} C_{XY}(t),
\end{align}
with the equilibrium correlation function $C_{XY}(t) = \langle \delta X(t) \delta Y^* \rangle$. 
Here, $\Theta(t)$ is the Heaviside function, defined as  $\Theta(t) = 1$ for $t \geq 0$ and $0$ otherwise. 
 Equation~\eqref{eq:micro_lin_resp} is the microscopic analog of Eq.~\eqref{eq:lin_resp}. The connection between the response function $\chi(t)$ and the equilibrium correlation function $\langle \delta X(t) \delta Y^* \rangle$ is the essence of   the fluctuation-dissipation theorem. 

The fluctuation-dissipation theorem is remarkable in several respects. First it provides a microscopic expression for the 
linear response in terms of an equilibrium correlation function. 
This is particularly valuable for computer simulations. Instead of simulating the system under external perturbations, one can extract material properties -- such as how the system responds to small external forces -- by analyzing equilibrium fluctuations. This approach is far more efficient, as the equilibrium system is in a stationary state, allowing for time averaging in addition to ensemble averaging.

Second, the two sides of the fluctuation-dissipation theorem, Eq.~\eqref{eq:FDT},  initially appear  to describe very different phenomena. The left-hand side represents the deterministic macroscopic response of a system to an external perturbation, while the right-hand side encodes microscopic fluctuations of the system in equilibrium. The thermal energy $k_BT $ serves as the bridge connecting these two realms.
The fluctuation-dissipation theorem is reminiscent of other unifying principles in physics. For example, 
de Broglie's relation $\mathbf{p} = \hbar \mathbf{k}$ or Einstein's relation $E =\hbar \omega$ links the world of mechanics (momentum $\mathbf{p}$, energy $E$) to the world of waves (wave vector $\mathbf{k}$, angular frequency $\omega)$ via Planck's constant $\hbar$. Einstein's  relation $D= k_B T \mu$  connects the diffusion coefficient $D$ of a particle to its mobility $\mu$ and the thermal energy $k_BT$. This relation can be viewed as a special case of the fluctuation-dissipation theorem.

Einstein's genius lay in unifying seemingly unrelated concepts, such as fluctuations and response in statistical physics, mechanics and waves in quantum theory, and geometry and gravity in general relativity. The fluctuation-dissipation theorem continues this tradition by uniting macroscopic response and microscopic fluctuations into a single, elegant framework.

The fluctuation-dissipation theorem is not limited to the realm of classical mechanics. It also applies to systems governed by Brownian motion as underlying microscopic dynamics~\cite{Doi:Polymer_Dynamics:1988}, with minor modification for general Markov processes~\cite{Kampen:Stochastic_Processes:1992} and quantum dynamics~\cite{Kubo:Statistical_physics_II:2012}. In all these cases correlation functions in the unperturbed reference system encapsulate the relevant information about the system's response to external perturbations.

The connection between linear response and correlation functions becomes even more evident when considering a pure step perturbation of the form $f(t) = F \Theta(t)$. Using the fluctuation-dissipation theorem, Eq.~\eqref{eq:FDT}, the integral in Eq.~\eqref{eq:micro_lin_resp} can be explicitly evaluated, yielding  the 'onset behavior'
 \begin{align}
  \langle X(t) \rangle_f = \frac{F}{k_B T} [ C_{XY}(0) - C_{XY}(t) ].
 \end{align}
 This result demonstrates that correlation functions can be measured directly in a step-force experiment. 
 
If the correlation function decays to zero at long times, $\lim_{t\to\infty} C_{XY}(t) =0$, the response at large times simplifies to 
\begin{align}
 \langle X(t\to\infty) \rangle_f = (F/k_B T) \langle \delta X  \delta Y^*\rangle.
\end{align}
This relation is the static analogue of the fluctuation-dissipation theorem and is commonly referred to as the fluctuation-response theorem. It provides a direct link between the static response of a system to a constant perturbation and the equilibrium fluctuations of the corresponding observables.

We illustrate the linear-response formalism with a concrete example. Consider a many-body system described by the Hamiltonian $\mathcal{H}_0$. A single particle is selected, and a time-dependent force $\vec{f}(t) = f(t) \vec{e}_x$  is applied to it. The Hamiltonian is then modified to $\mathcal{H}(t) = \mathcal{H}_0 - f(t) X$, where $X$ is the observable corresponding to the $x$-coordinate of the selected particle's position. As the observable to  monitor, we choose the particle's velocity,  $V(t) = \dot{X}(t)$. 
The average response to the applied force is then provided by
\begin{align}
\langle V(t) \rangle_f = \int_{-\infty}^t \mu(t-\bar{t}) f(\bar{t}) \diff \bar{t},
\end{align}
where the response function $\mu(t)$ is referred to as the time-dependent mobility. 
From the fluctuation-dissipation theorem, the mobility is related to the equilibrium velocity autocorrelation function
\begin{align}
\mu(t)  = \frac{1}{k_B T} \Theta(t) \langle V(t) V(0) \rangle . 
\end{align}
We denote the velocity autocorrelation function in equilibrium as $Z(t) = \langle V(t) V(0) \rangle$. This function is directly measurable in computer simulations. 

For a step force $f(t) = F\Theta(t)$, the  particle  reaches a steady state at long times with a constant drift velocity
\begin{align}\label{eq:lin_force}
\langle V(t\to \infty)\rangle_f = \frac{F}{k_B T} \int_0^\infty \langle V(t) V(0) \rangle \diff t.
\end{align}
The proportionality of the drift velocity to the applied force defines the stationary mobility $\mu$ via  $\langle V \rangle(t\to\infty) = \mu F$.

Rather than directly measuring the velocity-autocorrelation function, it is  often more practical to  evaluate the mean-square displacement (MSD), defined as
\begin{align}
\text{MSD}(t-t') \coloneq \langle [ X(t) - X(t')]^2 \rangle. 
\end{align} 
Here we already used that in the unperturbed equilibrium state the mean-square displacement depends only on the lag time $t-t'$. By taking the second derivative of the MSD, we can relate it to the velocity autocorrelation function
\begin{align}
\frac{\diff^2 }{\diff t^2} \text{MSD}(t-t') &= - \frac{\diff }{\diff t'} \frac{\diff }{\diff t} \langle [ X(t) - X(t')]^2 \rangle \nonumber \\
&=-2 \frac{\diff }{\diff t'} \langle V(t) [ X(t) - X(t')]\rangle \nonumber \\
&= 2 \langle V(t) V(t') \rangle.
\end{align}
At long times, the mean-square displacement is expected to grow linearly, $\text{MSD}(t) \to 2 D t$ as $t\to\infty$, where $D$ is the diffusion coefficient. It is customary to define the time-dependent diffusion coefficient as
\begin{align}
D(t) \coloneq \frac{1}{2} \frac{\diff}{\diff t}\text{MSD}(t) ,
\end{align}
such that $D(t) \to D$ as $t\to \infty$. Since $Z(t) = \diff D(t)/\diff t$, the Green-Kubo relation for the diffusion coefficient follows
\begin{align}
D = \int_0^\infty \langle V(t) V(0) \rangle \diff t. 
\end{align}
Comparing the Green-Kubo relation with the linear-response result in Eq.~\eqref{eq:lin_force}, we arrive at the celebrated Einstein relation
\begin{align}
D = k_B T \mu,
\end{align} 
which connects the mobility $\mu$, describing the linear response to an external force, to the diffusion coefficient $D$ which 
quantifies  the long-time growth of the mean-square displacement -- a quantifier of the fluctuations of the particle movement. 

 \begin{figure}[htp!]
 \includegraphics[width=0.8\linewidth]{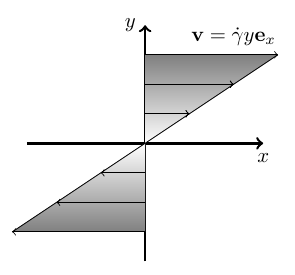}
\caption{Velocity profile for simple shear.
  \label{fig:Simple_shear}}
\end{figure}

As  a second example, consider a simple fluid subjected to time-dependent simple shear flow. The imposed velocity field is provided by $\vec{v}(\vec{r},t) = \dot{\gamma}(t) y \vec{e}_x$ where $\dot{\gamma}(t)$ is the time-dependent shear rate, see Fig~\ref{fig:Simple_shear} for an illustration. For a start-up flow, a constant shear rate is switched on at time $t=0$, i.e. $\dot{\gamma}(t) =\dot{\gamma} \Theta(t)$. The fluid will resist this shear developing a uniform stress in the plane of the shear flow. At large times, the system reaches a steady-state stress, $\langle \sigma_{xy}(t \to \infty) \rangle_{\dot{\gamma}}$. For small shear rates, this steady-state stress is proportional to the shear rate, as described by Newton's law of viscosity
\begin{align}
\langle \sigma_{xy}(t \to \infty) \rangle_{\dot{\gamma}} = \eta \frac{\partial v_x}{\partial y} = \eta \dot{\gamma}.
\end{align}
Here, the transport coefficient $\eta$ is known as the shear viscosity of the fluid. 

For a general time-dependent shear rate $\dot{\gamma}(t)$, the linear-response relation implies 
\begin{align}\label{eq:sigma_lin_resp}
\langle \sigma_{xy}(t) \rangle_{\dot{\gamma}} = \int_{-\infty}^t G(t-\bar{t}) \dot{\gamma}(\bar{t}) \diff \bar{t}.
\end{align} 
where  $G(t)$ is  the time-dependent shear modulus.

For the  start-up flow at long times, comparison with Newton's law of viscosity yields the connection between the shear modulus and  the shear viscosity
\begin{align}\label{eq:shear_viscosity}
\eta = \int_0^\infty G(t) \diff t.
\end{align}

In Appendix~\ref{Sec:Appendix_FDT}, a microscopic expression for general response coefficients is derived. For the time-dependent shear modulus,  Eq.~\eqref{eq:chi_J} reduces to $G(t) = - \Theta(t) \langle \sigma_{xy}(t) J \rangle /k_B T$, where $J$ is the dissipative flux. Heuristically, the dissipative flux can be identified from the dissipated work
\begin{align}
\Delta W_{\text{diss}} = \mathscr{V} \int  \dot{\gamma}(t) \langle \sigma_{xy}(t)\rangle_{\dot{\gamma}} \diff t,
\end{align} 
where $\mathscr{V}$ is the volume of the system. Combining this with the linear-response expression, Eq.~\eqref{eq:sigma_lin_resp}, we identify the dissipative flux as $ J(t) =  \mathscr{V}  \sigma_{xy}(t)$. Substituting this into the expression for $G(t)$, the time-dependent shear modulus becomes
\begin{align}\label{eq:shear_modulus}
G(t) = \frac{\mathscr{V}}{k_B T}\Theta(t) \langle \sigma_{xy}(t) \sigma_{xy} \rangle. 
\end{align}
A more rigorous derivation of this relation, based on equations  of motion with non-Hamiltonian couplings, can be found in Ref.~\cite{Tuckerman:Statistical_Mechanics:2023}. Combining Eq.~\eqref{eq:shear_viscosity} with Eq.~\eqref{eq:shear_modulus} leads to  the Green-Kubo relation for the shear viscosity
\begin{align}
\eta = \frac{\mathscr{V}}{k_B T}\int_0^\infty \langle \sigma_{xy}(t) \sigma_{xy} \rangle \diff t .
\end{align}

The examples provided establish formally exact relations for transport coefficients, which are essential for  characterizing  the properties of a fluid. In particular, 
the Green-Kubo relations enable the calculation of key transport properties, such as  the diffusion coefficient and the shear viscosity, directly from equilibrium simulations. This approach eliminates the need to apply external forces or strains, making it especially useful in computational studies.

From a theoretical physics perspective, the connection between response functions and equilibrium correlation functions suggests that the focus should be on developing theories for equilibrium correlation functions, as they are often simpler to analyze than the seemingly more complex response functions. This connection underscores the power of equilibrium statistical mechanics in predicting non-equilibrium behavior.

However, for interacting systems, it is virtually impossible to solve explicitly for the time-correlation functions. As a result, one must rely on theoretical models that aim to approximate the relevant properties of the system. These models must satisfy certain fundamental constraints to ensure physical consistency. For example, in the cases discussed above, the diffusion coefficient and the shear viscosity must be positive, reflecting the intuitive requirement that energy dissipation cannot be negative.

To further explore the role of energy dissipation and to gain deeper insight into the properties of response functions, we will examine two elementary models and analyze their associated response functions. These models serve as illustrative examples of how theoretical frameworks can capture the essential features of linear response while adhering to fundamental physical principles.

Consider the undamped harmonic oscillator, whose dynamics are governed by the constitutive equation
\begin{align}\label{eq:undamped_osci}
\ddot{X}(t) + \omega_0^2 X(t) = f(t)/m , 
\end{align}
where $\omega_0>0$ is the natural  frequency of the oscillator, $m$ the mass,  and $f(t)$ the external driving force. We assume that the oscillator is at rest in the infinite past, $t\to -\infty$, and that the driving force $f(t)$ vanishes rapidly for $t\to \pm \infty$.   The  causal solution 
of Eq.~\eqref{eq:undamped_osci} can be  obtained by elementary methods (e.g. by variation of the constant) and is  provided by
\begin{align}
X(t) = \int_{-\infty}^t \frac{\sin[\omega_0 (t-\bar{t})]}{m \omega_0} f(\bar{t}) \diff \bar{t} .
\end{align}
We  define the response function $\chi(t)$ for this particular system as 
\begin{align}
\chi(t) = \frac{\sin(\omega_0 t)}{m \omega_0} \Theta(t).
\end{align} 
Using this definition,  the response $X(t)$  is expressed as in Eq.~\eqref{eq:lin_resp}. 
The response $X(t)$ depends on the entire history of the force $\{ f(\bar{t}) : \bar{t} \leq t\}$, weighted by the response function $\chi(t-\bar{t})$. The weight depends only on the lag time $t-\bar{t}$ between the time of the driving $\bar{t}$ and the time $t$ at which  the response is evaluated. The Heaviside function ensures that $\chi(t) = 0$ for $t<0$, meaning  that the response is causal -- it depends  only on the driving prior to the observation time $t$. These properties are generic for systems in a stationary state. 

Although Eq.~\eqref{eq:undamped_osci} also  admits  an anticausal solution, we discard it as unphysical because it would imply a response that precedes the driving force. 

Now consider a system consisting of a linear superposition of undamped harmonic oscillators. In this case, the linear response relation, Eq.~\eqref{eq:lin_resp}, still holds, but the response function is given by\footnote{In this and in the following sections the reference to an underlying probability space and the random outcomes is only implicit and we shall use $\omega, \Omega$ as generic symbols for frequencies.  
}
\begin{align}
\chi(t) =  \int_0^\infty \frac{\sin(\Omega t)}{\Omega} \Theta(t)  \rho(\Omega) \diff \Omega ,
\end{align}
where $\rho(\Omega)\geq 0$ is the density of oscillators (per mass). This generalization demonstrates how the response function can encode the distribution of natural frequencies in a system.

To analyze the response function in the frequency domain, we introduce its Fourier transform
\begin{align}
\hat{\chi}(\omega) \coloneq \int e^{i \omega t} \chi(t) \diff t.
\end{align}
Since $\chi(t)$ vanishes for $t<0$,  the integral extends only over positive times, making it effectively  a one-sided Fourier transform. 
The dynamic susceptibility $\hat{\chi}(\omega)$ is complex-valued and is decomposed into a real part $\chi'(\omega) = \Real[\hat{\chi}(\omega)]$ and an imaginary part $\chi''(\omega) = \Imag[\hat{\chi}(\omega)]$. 
Because $\chi(t)$ is real-valued, 
the real and imaginary parts  are obtained directly as
\begin{align}
\chi'(\omega)  &= \int_0^\infty \cos(\omega t) \chi(t) \diff t ,\nonumber\\
\chi''(\omega)  &= \int_0^\infty \sin(\omega t) \chi(t) \diff t . 
\end{align}
From the above expression, we observe that
the real part is an even function, $\chi'(\omega) = \chi'(-\omega)$, whereas the imaginary part is an odd function, $\chi''(\omega) = - \chi''(-\omega)$. 

By inverting the Fourier-sine transform, the response function  can be represented as
\begin{align}
\chi(t) = \int_0^\infty \frac{2 \diff \omega }{\pi} \chi''(\omega) \sin(\omega t) \qquad \text{for } t> 0. 
\end{align}
This representation highlights how the imaginary part of the Fourier transform, $\chi''(\omega)$, determines the time-domain response function. Comparing this result with the example of a superposition of undamped harmonic oscillators, we identify the density of oscillators as $\rho(\omega) = 2\omega \chi''(\omega)/\pi$. 
In particular, we observe that $\omega \hat{\chi}(\omega)$ has a non-negative imaginary part. Later, we shall see that this property is a hallmark of complex susceptibilities and plays a fundamental role in ensuring the stability of matter. It reflects the fact that the system's response is dissipative and causal, with energy dissipation encoded in the imaginary part of the susceptibility.

\begin{figure}[htp!]
 \includegraphics[width=\linewidth]{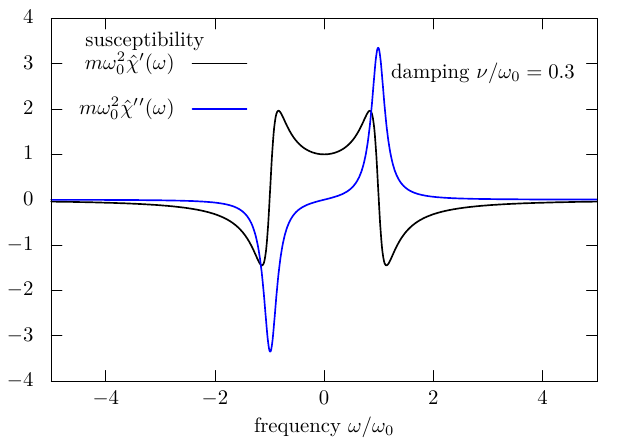}
\caption{Real $\chi'(\omega)$ and imaginary part $\chi''(\omega)$ of the complex susceptibility for a damped harmonic oscillator.
  \label{fig:susceptibility_damped_harmonic}}
\end{figure}

Now consider  the example of a damped harmonic oscillator driven by an external force. The constitutive equation is given by
\begin{align}
\ddot{X}(t) + \nu \dot{X}(t) + \omega_0^2 X(t) = f(t)/m ,
\end{align}
which differs from Eq.~\eqref{eq:undamped_osci} by the inclusion of a damping term characterized by the damping constant $\nu> 0$. In this discussion, we focus on the underdamped case, where $\nu/2 < \omega_0$. The response function encoding the causal solution for this system is
\begin{align}
\chi(t) = \frac{\sin(\omega_d t)}{m \omega_d} e^{-\nu t/2} \Theta(t),
\end{align}
where $\omega_d \coloneq \sqrt{\omega_0^2- (\nu/2)^2}$ is the damped characteristic frequency. Its Fourier transform is readily obtained as 
\begin{align}
\hat{\chi}(\omega) = \frac{1}{m}\frac{1}{\omega_0^2 - \omega^2 - i \omega \nu} ,
\end{align}
which describes the system's response in the frequency domain. The real and imaginary part of $\hat{\chi}(\omega)$ are given by
\begin{align}
\chi'(\omega) &= \frac{1}{m}\frac{\omega_0^2 - \omega^2}{(\omega_0^2 - \omega^2)^2 + (\nu \omega)^2}, \\
\chi''(\omega) &= \frac{1}{m} \frac{\nu \omega}{(\omega_0^2 - \omega^2)^2 + (\nu \omega)^2 }. 
\end{align}
These components are illustrated in Fig.~\ref{fig:susceptibility_damped_harmonic}. Since $\omega \chi''(\omega)\geq 0$ for all $\omega$,  the damped harmonic oscillator can be interpreted as a continuous superposition of a spectrum of undamped harmonic oscillators. In contrast to a finite superposition of harmonic oscillators, the continuous superposition allows for damping, which is essential for describing energy dissipation in the system. 

It is instructive to discuss the linear response also in the Fourier domain
\begin{align}
\hat{X}(\omega) \coloneq \int_{-\infty}^\infty e^{i \omega t} X(t) \diff t, 
\end{align}
and similarly for other quantities. The convolution theorem implies 
\begin{align}
\hat{X}(\omega) = \hat{\chi}(\omega) \hat{f}(\omega).
\end{align}
Recall that, by causality, $\hat{\chi}(\omega)$ is in fact the one-sided Fourier transform of $\chi(t)$. 

A key consequence of this result is that the response becomes local in the frequency domain. Specifically, the response $\hat{X}(\omega)$ at a given frequency $\omega$ depends only on the external force $\hat{f}(\omega)$ at the same frequency. This property is a hallmark of  linear response theory and reflects that there is no  frequency mixing in the linear regime. 

In contrast, higher-order responses exhibit phenomena such as overtones, frequency doubling and other forms of frequency mixing, where the response at a given frequency can depend on the external force at multiple frequencies. 

The locality of linear response in the frequency domain is a familiar concept in many physical systems. For example, in electrodynamics of continuous media, the dielectric function $\hat{\epsilon}(\omega)$ serves as the susceptibility that relates  the frequency-dependent electric field $\hat{E}(\omega)$ to the frequency-dependent displacement field $\hat{D}(\omega) =\hat{\epsilon}(\omega) \hat{E}(\omega)$. 

As an example, let us consider a pure sinusoidal external force $f(t) = \Real[ \hat{f}_\omega e^{-i\omega t}]$ where $\hat{f}_\omega >0$ is the amplitude of the driving force.  Strictly speaking, a pure sinusoidal force is not permitted, as the driving force is assumed to vanish rapidly for $t\to \pm \infty$. However, we assume that the  force is switched on and off  gradually at very distant times, allowing us to focus on the single frequency $\omega$ of the driving. 
Substituting the sinusoidal force in 
the linear-response relation, Eq.~\eqref{eq:lin_resp}, yields 
\begin{align}
X(t) &= \Real[\hat{\chi}(\omega) \hat{f}_\omega e^{-i \omega t} ] 
= |\hat{\chi}(\omega)| \hat{f}_\omega \cos(\omega t - \delta) ,
\end{align}
where we have expressed the complex susceptibility in polar form $\hat{\chi}(\omega) = |\hat{\chi}(\omega)| \exp(i \delta)$. 
This result shows  that the response  is again sinusoidal, $X(t) = \Real[ \hat{X}_\omega e^{-i \omega t} ] $,  with the same frequency $\omega$ as the driving force, 
but with a phase shift $\delta$.  The amplitude of the response  is the product of the  susceptibility  and the driving amplitude, $\hat{X}_\omega= \hat{\chi}(\omega) \hat{f}_\omega$.

The application of an external force leads to the dissipation of power in the system. Assume now that the monitored observable coincides with the observable conjugate to the force. Then the dissipated power is given by
\begin{align}
P(t) = f(t) \dot{X}(t).
\end{align}
This expression generalizes the familiar mechanical relation: \textbf{power = force $\times$ velocity}. Here, the observable $X(t)$ plays the role of a generalized coordinate, and $\dot{X}(t)$  represents its rate of change, analogous to velocity in classical mechanics. The microscopic justificaction  within classical dynamics can be found in Appendix~\ref{Sec:Appendix_FDT}. 

For the monochromatic excitation of the form $f(t) = \Real[\hat{f}_\omega e^{-i \omega t}]$, 
the power dissipated in the system, within the framework of linear response, is given by
\begin{align} 
P(t) &= \Real[\hat{f}_\omega e^{-i \omega t} ] \Real[ -i \omega \hat{X}_\omega e^{-i\omega t}] \nonumber\\
&= \frac{1}{4}[ i \omega \hat{f}_\omega \hat{X}_\omega^* - i \omega \hat{f}_\omega^* \hat{X}_\omega] + (\text{oscillatory terms}) ,
\end{align} 
where the oscillatory terms involve $\exp(\pm 2 i \omega t)$. These terms average out  over a cycle, leaving the time-averaged contribution. Substituting the linear response relation $\hat{X}_\omega = \hat{\chi}(\omega) \hat{f}_\omega$, the time-averaged dissipated power becomes
\begin{align}
\bar{P} = \frac{1}{4} \omega \chi''(\omega) |\hat{f}_\omega|^2. 
\end{align}
This result reveals that  it is the imaginary part of the complex susceptibility, $\chi''(\omega)$,  that governs the dissipation of energy in the system.  

A fundamental principle of physics is that no net  power can be extracted from a system at equilibrium by simply driving it. This principle, often referred to as the stability of matter, ensures that the system cannot  act as a perpetual energy source. From the expression for the averaged dissipated power, we infer the following condition
\begin{fundamentalbox}
\textbf{Stability of matter: }\qquad  $\omega \chi''(\omega) \geq 0$.
\end{fundamentalbox}
This condition guarantees that the dissipated power is always non-negative, regardless of the driving frequency $\omega$. 

To formalize this heuristic argument, let us consider a non-periodic driving force $f(t)$  that vanishes rapidly as $t\to \pm \infty$. The total dissipated work over the entire time interval is given by
\begin{align}
\Delta W_{\text{diss}} &= \int_{-\infty}^\infty  f(t) \dot{X}(t) \diff t = \frac{1}{2}\int \frac{\diff \omega}{2\pi} 
\hat{f}(\omega)^* \hat{\dot{X}}(\omega) + \text{c.c.} \nonumber \\
&= \int \frac{\diff \omega}{2\pi} \frac{1}{2} \hat{f}(\omega)^* (-i \omega) \hat{\chi}(\omega) \hat{f}(\omega) + \text{c.c.} ,
\end{align}
where in the first line Parseval's theorem has been used to express the dissipated work in the frequency 
domain and $\text{c.c.}$ means complex conjugate of the preceding term. In the second line, we have substituted the linear response relation $\hat{\dot{X}}(\omega) = (-i \omega) \hat{\chi}(\omega) \hat{f}(\omega)$. 

The integrand can be interpreted as the dissipated power per frequency interval $\diff \omega/2\pi$. Simplifying further, we find
\begin{align}
\Delta W_{\text{diss}} = \int \frac{\diff \omega}{2\pi}  \omega \chi''(\omega) |\hat{f}(\omega)|^2 . 
\end{align}
The stability of matter implies that, for any driving protocol $f(t)$ that vanishes rapidly  as $t\pm \infty$, the dissipated work $\Delta W_{\text{diss}}$ must be non-negative. This ensures that energy cannot be extracted from the system in a way that violates the second law of thermodynamics.

An interesting thought experiment arises here. Strictly speaking, the above analysis shows only that the total dissipated work in the distant future is non-negative. But what if we consider the dissipated work up to a finite time $T$? Specifically, let us define:
\begin{align}
\Delta W_{\text{diss}}(T) = \int_{-\infty}^T f(t) \dot{X}(t) \diff t.
\end{align}
We conclude this section with a thought-provoking question: 
Can one design a  driving protocol $f(t)$  such that $\Delta W_{\text{diss}}(T)$ becomes negative at some finite time $T$? If so, one could transiently  extract energy from an equilibrium system, with  the associated dissipation occurring  later -- effectively  "borrowing" energy from the system!!(?) We 
examine  this question and resolve the apparent paradox in Sec.~\ref{Sec:Linear_Response_revisited}.
 

\section{Scattering experiments}

\begin{figure}[tp!]
\begin{tikzpicture}[scale=1]
    \draw[->, thick, red, decorate, decoration={snake, amplitude=1mm, segment length=5mm}] (0,-0.3) -- (2,-0.3);
    \draw[->, thick, red, decorate, decoration={snake, amplitude=1mm, segment length=5mm}] (0,0.3) -- (2,0.3);
    \draw[->, thick, red, decorate, decoration={snake, amplitude=1mm, segment length=5mm}] (0,0.0) -- (2,0.0);
    \node[red] at (1,1) {plane wave};
    \node[red] at (1,-1) {$\vec{k}_\textrm{i}, \omega_\textrm{i}$};

    \fill[red!20] (4,0) -- (8,2.8) arc[start angle=30, end angle=60, radius=2.0cm] -- cycle;
     \draw[ultra thick, red]  (8,2.8) arc[start angle=30, end angle=60, radius=2.0cm];
     \node[red] at (8.5,3.7) {detector};  
    \node[red] at (8.5,3.3) {$\diff \Omega_\textrm{f}$};  
 
    \draw[ultra thick,fill=blue!20] (4,0.0) circle [radius=1cm];
    \node[blue] at (4,0) {sample};

     \draw[->, thick, red, decorate, decoration={snake, amplitude=1mm, segment length=5mm}] (5,0.85) -- (7.0,2.65);
     \node[red] at (7.0, 1.0) {scattered wave};
    \node[red] at (7.0, 0.5) {$\vec{k}_\textrm{f}, \omega_\textrm{f}$};
   
\end{tikzpicture}
\caption{Typical set-up of a scattering experiment. The sample is exposed to a plane-wave probe beam  of particles or waves characterized by a wave vector $\vec{k}_\textrm{i}$ and frequency $\omega_\textrm{i}$. The scattered particles in the direction of $\vec{k}_\textrm{f}$ are collected at the detector covering a solid angle $\diff \Omega_\textrm{f}$ and resolved according to their frequency $\omega_\textrm{f}$ or energy $\epsilon_{\textrm{f}}$.   }
\label{fig:scattering}
\end{figure}
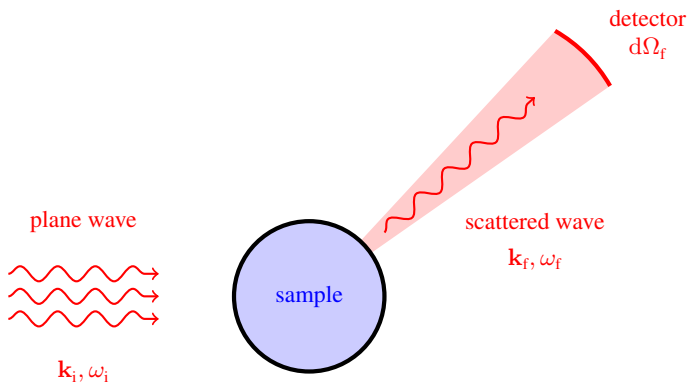

A second paradigm for probing material properties is scattering. In a typical experiment a sample is irradiated with a beam of particles or waves, and a fraction of the incident intensity is scattered and detected within a small solid angle (see Fig.~\ref{fig:scattering}). By analyzing the scattered intensity and its dependence on angle and frequency, one infers structural and dynamical information across a wide range of length and time scales.

A basic kinematic description uses the incident and scattered wave vectors $\vec{k}_\mathrm{i}$ and $\vec{k}_\mathrm{f}$ with magnitudes $k_\mathrm{i} = |\vec{k}_\mathrm{i}|, k_\mathrm{f} = |\vec{k}_\mathrm{f}|$.
 The scattering  vector is 
\begin{equation}
\vec{q} = \vec{k}_{\mathrm{i}} - \vec{k}_{\mathrm{f}},
\end{equation}
and $\vartheta = \angle (\vec{k}_\mathrm{f}, \vec{k}_\mathrm{i}) $ is the scattering angle. In a quantum picture, the waves correspond to  particles: photons in the case of light or X-rays, and massive particles  such as neutrons and electrons for matter waves. Associating a momentum $\hbar \vec{k}$ with each  probe particles via de Broglie's relation,  scattering transfers a momentum $\hbar \vec{q} = \hbar \vec{k}_\mathrm{i} - \hbar \vec{k}_\mathrm{f}$ from the probe beam to the sample. 
Likewise,  the frequency shift $\omega= \omega_\mathrm{i}-\omega_\mathrm{f}$ corresponds to an 
 energy transfer,  $\hbar \omega = \epsilon_\mathrm{i}-e_\mathrm{f}$, with $ \epsilon= \hbar^2 k^2 /2m$ for massive probes  and 
$\epsilon=\hbar c k$ for photons.

Spatial resolution requires that the probe wavelength be comparable to or smaller than the typical length scales of the sample. In practice, resolving a length scale $a$ typically demands $q \approx 2\pi/a$, hence $k_i \gtrsim 2\pi /a$. In hard condensed matter, thermal neutrons are ideal: their thermal wavelengths are of the order of an {\AA}ngstr{\"o}m, comparable to the lattice constants of atomic crystals, enabling the observation of Bragg peaks and atomic-scale dynamics. In soft condensed matter, where colloids play the role of effective constituents with sizes from hundreds of nanometers to micrometers, visible or near-infrared photons are the natural probe; their wavelengths are a fraction of a micrometer, well matched to colloidal length scales. In many situations
 the energy transfer is neglible compared to the incident energy, $|\omega| \ll \omega_\mathrm{i}$ (quasi-elastic scattering). Then $k_\mathrm{i}\approx k_\mathrm{f}$, such that the scattering angle fulfills $q = 2 k_\mathrm{i}\sin(\vartheta/2)$.  

We assume that the sample does not absorb the radiation. A perfectly uniform medium then acts only through  a (possibly frequency-dependent) refractive index, which changes the phase velocity of the wave but does not produce scattering (except in the strict forward direction, $\vec{q}=0$). Scattering arises from  microscopic spatial and temporal fluctuations of the material property that couples to the probe beam, here denoted as contrast field operator $\delta X(\vec{r},t)$ in the quantum problem and phase space function in the classical case.

In all these set-ups, the central observable is the (double) differential scattering cross section, which in many cases is 
proportional to the dynamic structure factor,
\begin{align}
\frac{\diff^{2}\sigma}{\diff \Omega_\mathrm{f}\,\diff \omega} \propto 
S(\vec{q},\omega) = \int \diff t\, e^{i\omega t}\,
\langle \delta X(\vec{q},t)\, \delta X(\vec{q},0)^\dagger \rangle,
\end{align}
where $\delta X(\vec{q},t)$ denotes the spatial Fourier transform of the fluctuating contrast field
\begin{align}
\delta X(\vec{q},t) = \int \diff \vec{r}\,  e^{-i \vec{q} \cdot \vec{r} } \delta X(\vec{r},t).   
\end{align}
In the quantum case $\delta X(\vec{q},t)$ becomes an operator with adjoint $\delta X(\vec{q},t)^\dagger$, whereas in the classical the adjoint is simply replaced by the complex conjugate $\delta X(\vec{q},t)^*$. 

For light, the relevant contrast is the dielectric susceptibility. In general, this is a tensor, with fluctuations written as 
  $\delta \bm{\chi}(\vec{r}, t)$. A general full quantum-mechanical derivation for the scattering cross section for light scattering is found in Appendix~\ref{Sec:Appendix_Light_Scattering}. 
In many soft-matter systems $\delta \bm{\chi}(\vec{r},t)$ is proportional  to number density fluctuations  or composition fluctuations in a mixture, and can also include  temperature fluctuations and orientational fluctuations in molecular liquids. For X-rays, the dominant contrast is the electron-density fluctuation. 
For thermal neutrons, the interaction is primarily with the nuclear potential~\cite{Lovesey:Theory_of_neutron_scatteringI:1984}, characterized by the scattering length. The corresponding contrast is the nuclear-scattering-length-density fluctuation, which reflects variations in nuclear positions (number density) and, where relevant, isotope or spin distributions. Neutrons also couple to magnetization, so magnetization fluctuations  give rise to magnetic scattering~\cite{Lovesey:Theory_of_neutron_scatteringII:1984}. For electrons, the contrast is set by fluctuations of the electrostatic potential, closely tied to charge-density variations. 

Other experimental techniques -- photon correlation spectroscopy
 (PCS, i.e., dynamic light scattering)~\cite{Berne:Dynamic_light_scattering:2000}, fluorescence correlation spectroscopy (FCS)~\cite{Rigler:Fluorescence:2012, Hofling:SM_7:2011, Hofling:RPP_76:2013}, 
fluorescence  recovery after photobleaching (FRAP)~\cite{Bancaud:fluorescence:2010, Hofling:RPP_76:2013}, differential dynamic microscopy (DDM)~\cite{Cerbino:PRL_100:2008}, and single-particle tracking (SPT)~\cite{Crocker:JCIS_179:1996} -- are likewise formulated
 in terms of time correlations, which serve as a unifying
 language for dynamical measurements. Their temporal correlators and 
Fourier transforms (via Wiener-Khinchin) offer complementary views of dynamics, 
enable direct comparison with theory through the intermediate scattering function and static/dynamic structure factors, and link to transport coefficients via Green-Kubo relations. 
This common framework explains their wide applicability to fluctuations across diverse length and time scales.

The link between measured signals and correlators naturally leads to the next section, where we develop the mathematical constraints on admissible correlation functions. In particular, the positive-definiteness of equilibrium correlators implies -- by Bochner's theorem -- the nonnegativity of the dynamic structure factor $S(\vec{q},\omega)$, consistent with the requirement that the detected intensity (and thus the inferred cross section) cannot be negative.

\section{Mathematical properties of correlation functions}

In this section we give a complete characterization of autocorrelation functions (henceforth simply correlation functions). We begin by establishing their structural properties (Hermiticity, positive-definiteness, boundedness/continuity conditions, and spectral nonnegativity).  In an outlook we then extend the discussion to multivariate settings, where one obtains matrix-valued correlation functions; cross-correlations appear as the off-diagonal entries of these matrices. Our guiding question is: given a candidate function $C(t)$, under what conditions does there exist a stationary stochastic process whose autocorrelation equals $C(t)$?

 \begin{figure}[htp!]
 \includegraphics[width=0.45\linewidth]{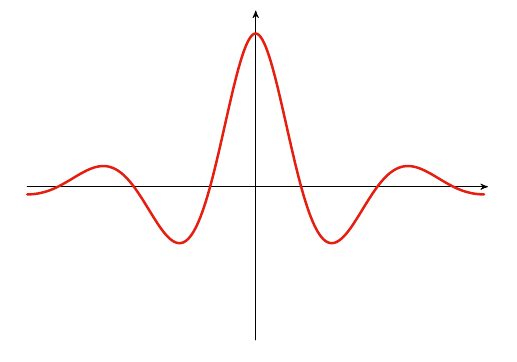}
 \includegraphics[width=0.45\linewidth]{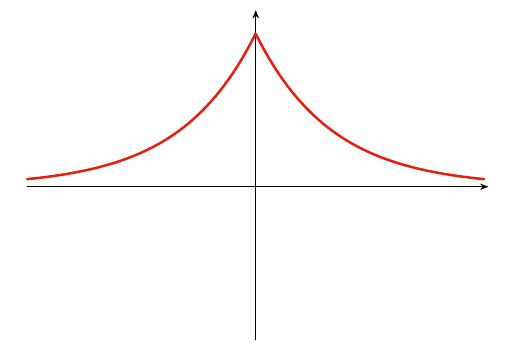} 
 \includegraphics[width=0.45\linewidth]{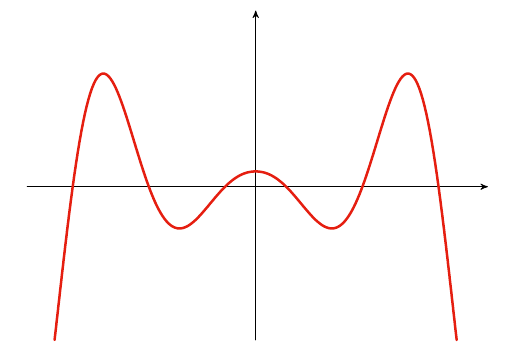}
 \includegraphics[width=0.45\linewidth]{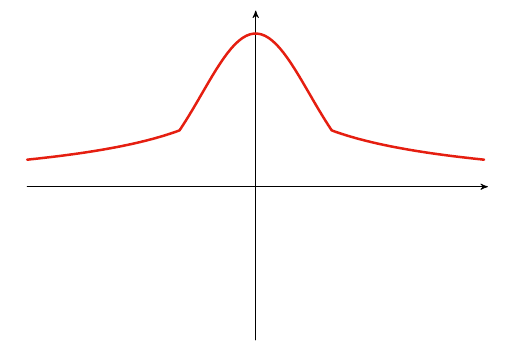} 
\caption{Which of these four functions corresponds to  a correlation function? Top left: damped harmonic oscillator. Top right: simple relaxator. Bottom left:  harmonic oscillator with negative damping. Bottom right: Gaussian coupling model.  
  \label{fig:candidates_correlation_functions}}
\end{figure}

Figure~\ref{fig:candidates_correlation_functions} illustrates four example functions. For simplicity, only real-valued and symmetric functions are shown. The first corresponds to the correlation function of a stochastically driven damped harmonic oscillator. The second represents a stochastically driven simple relaxator (a symmetrized relaxing exponential) and is thus also a correlation function.
The third has the same functional form as the first but with negative damping, which turns out not to be a valid correlation function. The fourth is a "coupling model," where a Gaussian is continuously concatenated with a stretched exponential. This, too, is not a valid correlation function.

The key insight into correlation functions is that they must respect a certain positivity property. 
\begin{definitionbox}
\textbf{Definition:}  A function $C: \mathbb{R}\to \mathbb{C}$ is \textbf{positive-semidefinite} if, for each $n\in \mathbb{N}$, any times $t_1,\ldots, t_n \in \mathbb{R}$, and complex numbers $\lambda_1,\ldots, \lambda_n\in \mathbb{C}$, the following inequality holds:
$$ \sum_{i=1}^n \sum_{j=1}^n \lambda_i C(t_i- t_j) \lambda_j^* \geq 0. $$
\end{definitionbox}
Equivalently, the $n\times n$ complex-valued matrix with matrix elements $C(t_i-t_j), i,j=1,\ldots n$ is positive-semidefinite. 

We shall show that all correlation functions fulfill this property. To do so,  we introduce the vector space of  square-integrable fluctuations of observables $\mathcal{H} \coloneq \{ \delta X : \Omega \to \mathbb{C}: \langle |\delta X|^2 \rangle< \infty \}$ equipped  with the Kubo scalar product:
\begin{definitionbox}
\textbf{Definition:}  The \textbf{Kubo scalar product} for  observables $X,  Y$ is  
$$ \langle X | Y \rangle \coloneq \langle \delta Y \delta X
^* \rangle.$$
\end{definitionbox}
Here the angular brackets represent the expectation value of the underlying stochastic process. 
One readily checks that $\langle X | Y \rangle$ is linear in the first argument and antilinear in the second. Furthermore, the Hermitian symmetry 
$\langle X | Y \rangle = [\langle Y | X \rangle]^*$ is obvious, and $\langle X | X \rangle = \langle |\delta X |^2 \rangle \geq 0$ with  equality  iff $\delta X = 0$ almost surely.  
The space $\mathcal{H}$ is complete, i.e. every Cauchy sequence converges, making $\mathcal{H}$  a Hilbert space. The definition is written such that it directly generalizes to the quantum case. In this case $X, Y$ become operators and  
$\langle X | Y \rangle = \langle \delta Y \delta X^\dagger \rangle$ where $\dagger$ indicates the adjoint. Here also the order of the operators matters. 

\begin{propositionbox}
\textbf{Proposition:} Autocorrelation functions $C(t) = \langle X | X(t) \rangle$ are positive-semidefinite. 
\end{propositionbox}
\textbf{Proof:} Using the Kubo scalar product, we write   $C(t) = \langle X | X(t) \rangle$. For any  $n\in \mathbb{N}$, times $t_1,\ldots, t_n$, and complex numbers $\lambda_1,\ldots, \lambda_n\in \mathbb{C}$, we compute
\begin{align*}
\sum_{i=1}^n \sum_{j=1}^n \lambda_i C(t_i-t_j) \lambda_j^* = \sum_{i=1}^n \sum_{j=1}^n \lambda_i \langle X(t_j) | X(t_i) \rangle  \lambda_j^* \\
=   \big\langle  \sum_{j=1}^n \lambda_j X(t_j) | \sum_{i=1}^n \lambda_i X(t_i) \big\rangle = \langle Z | Z \rangle \geq 0, 
\end{align*}
Here we used the time-translational invariance of the average in the first line, and  we have abbreviated $Z= \sum_{i=1}^n \lambda_i X(t_i)$ in the last line.  The proof also holds for the quantum case. 
\hfill $\square$ 

The proposition triggers the reverse question: Given a positive-semidefinite function $C: \mathbb{R}\to \mathbb{C}$, is there a stochastic process yielding $C$ as correlation function? The answer to this question is affirmative, in particular a  process with Gaussian random variables $\{ X(t) : t\in \mathbb{R}\}$ can be constructed.  The proof is deferred to Appendix~\ref{Sec:StochProcess}.

The positive-definiteness of correlation functions has a number of immediate consequences. 

The correlation function at time $t=0$ is non-negative: $C(0) \geq 0$.  
 This is clear for classical observables $C(0) = \langle |\delta X|^2 \rangle \geq 0$. In the quantum case this is also implied since $\delta X \delta X^\dagger\geq 0$ is a non-negative Hermitian operator. 
By stationarity we find additionally  that  autocorrelation functions are \emph{Hermitian}: $C(-t) = \langle X | X(-t) \rangle = \langle X(t) | X \rangle = C(t)^*$.

Correlation functions are bounded by their initial value: $|C(t)| \leq C(0)$. 

\textbf{Proof:} The proof follows the lines of the Cauchy-Schwarz inequality.  Choose $n=2, t_1=0, t_2= t, \lambda_1 =1, \lambda_2 = \lambda$. Then 
\begin{align*}
0 &\leq \sum_{i, j} \lambda_i C(t_i-t_j) \lambda_j^* \\ 
&= C(0) +  \lambda C(t) + C(-t) \lambda^* + |\lambda|^2 C(0)  \\
&= C(0) + 2 \Real[ \lambda C(t) ] + |\lambda|^2 C(0)  .
\end{align*}
If $C(t)=0$, nothing is to be shown. 
 Otherwise choose $\lambda = \mu C(t)^*/|C(t)|$ with $\mu \in \mathbb{R}$. Then the inequality becomes 
 \begin{align*}
 C(0) + 2 \mu |C(t)| + \mu^2 C(0) \geq 0. 
 \end{align*}
Since the latter has to hold for all real $\mu$, this is possible only if the quadratic equation has at most one root, i.e. the discriminant cannot be positive
\begin{align*}
(2 |C(t)| )^2 - 4  C(0)^2 \leq 0. 
\end{align*}
\hfill $\square$ 

 For two times $t,s\in \mathbb{R}$, a correlation function  satisfies
\begin{align}\label{eq:uniform_continuity}
|C(t) - C(s) | \leq 4 C(0) | C(0) - C(t-s) | .
\end{align}
The proof is left as an exercise but can be found in Refs.~\cite{Feller:Probability:1991,Shiryaev:Probability:2016}. 
The last relation implies that if $C(t)$ is continuous at the origin, it is uniformly continuous in $\mathbb{R}$.

A rather simple observation is that $e^{-i \Omega t}C(t)$ for $\Omega\in\mathbb{R}$ is positive-semidefinite iff $C(t)$ has the same property. 

\textbf{Proof:} 
\begin{align*}
\sum_{i,j} \lambda_i e^{-i \Omega (t_i-t_j)} C(t_i-t_j) \lambda_j^* = \sum_{i,j} \tilde{\lambda}_i C(t_i-t_j) \tilde{\lambda}_j^* \geq 0 ,
\end{align*}
where we have abbreviated $\tilde{\lambda}_i = e^{-i \Omega t_i} \lambda_i$.  The reverse direction follows by interchanging $\lambda_i$ and $\tilde{\lambda}_i$. \qquad $\square$. 

In particular, $C(t)= e^{-i \Omega t}, \Omega \in \mathbb{R}$ is positive-semidefinite. 
By linearity we also find that for $C_1(t), \ldots, C_n(t)$ positive-semidefinite and positive weights $p_1,\ldots, p_n>0$, the linear combination $\sum_{m=1}^n p_m C_m(t)$ is again positive-semidefinite. 
Combining the last two properties, we find that a finite sum
\begin{align}\label{eq:Fourier_discrete}
C(t) = \sum_{m=1}^n p_m e^{- i \Omega_m t}, 
\end{align}
with $ p_m > 0, \Omega_m \in \mathbb{R}, m=1,\ldots, n$
is positive-semidefinite. 
The continuous counterpart is
\begin{align}\label{eq:Fourier_continuous}
C(t)  = \int_\mathbb{R} e^{-i \Omega t} f(\Omega) \diff \Omega, 
\end{align}
with an integrable function $f(\Omega) \geq 0$. 
Both are special cases of the Fourier transform of a finite Lebesgue-Stieltjes measure
\begin{align}\label{eq:Fourier_Borel}
C(t) = \int_\mathbb{R} e^{-i \Omega t}\diff F(\Omega).
\end{align}
Physicists are often not familiar with the concepts of a finite Lebesgue-Stieltjes measure. Precise mathematical definitions can be found in Refs.~\cite{Feller:Probability:1991,Shiryaev:Probability:2016}. Here we just mention that it derives from a bounded non-decreasing right-continuous function  
 $F:\mathbb{R} \to [0,\infty)$, assigning  the measure $F(a,b] \coloneq F(b)- F(a)$ to the interval $(a,b]$. Taking $F$ as $F(\Omega) = \sum_{m=1}^n p_m \Theta(\Omega-\Omega_m)$ with the Heaviside function $\Theta(x) = 1$ for $x \geq 0$ and $0$ otherwise, recovers Eq.~\eqref{eq:Fourier_discrete}. Heuristically, the physicist would write $\diff F(\Omega) = \sum_{m=1}^n p_m \delta(\Omega- \Omega_m)\diff \Omega$ invoking delta-distributions. The case of Eq.~\eqref{eq:Fourier_continuous} 
is retrieved for differentiable functions $F(\Omega)$ with $\diff F(\Omega) = f(\Omega) \diff \Omega$, i.e. $f(\Omega)$ is the derivative of $F(\Omega)$. 

In measure theory it is shown that every Lebesgue-Stieltjes measure can be decomposed into a discrete part  comprising the jumps of $F(\Omega)$ and a continuous part. The latter can consist of an absolutely continuous part such that a derivative can be defined and a singular one. The singular part is concentrated on a set of Lebuesgue measure zero, for example the fractal Cantor set. Typically this third contribution does not arise such that we can view Eq.~\eqref{eq:Fourier_Borel} as an elegant notation to write both cases Eqs.~\eqref{eq:Fourier_discrete}, \eqref{eq:Fourier_continuous} and positive linear combinations thereof.

Let's repeat the direct proof of positive-semidefiniteness for the case of a Fourier transform of a Lebesgue-Stieltjes measure
\begin{align*}
\lefteqn{\sum_{i=1}^n \sum_{j=1}^n \lambda_i C(t_i-t_j) \lambda_j^* = } \\
 &= \int \big(\sum_{i=1}^n \lambda_i e^{-i \Omega_{t_i} }\big) \big(\sum_{j=1}^n \lambda_j e^{-i \Omega_{t_j} }\big)^* \diff F(\Omega) \\ 
&= \int \big|\sum_{i=1}^n \lambda_i e^{-i \Omega_{t_i}}\big|^2 \diff F(\Omega) \geq 0 .
\end{align*}

The remarkable insight by Bochner is that Fourier transforms of finite  Lebesgue-Stieltjes measures essentially exhaust the class of positive-semidefinite functions.
\begin{propositionbox}
\textbf{Theorem (Bochner):} A function $C: \mathbb{R} \to \mathbb{C}$ that is continuous at $t=0$ is positive-semidefinite 
iff there exists a finite Lebesgue-Stieljes measure $F(\Omega)$ such that
$$ C(t) = \int_{\mathbb{R}} e^{-i \Omega t} \diff F(\Omega).$$
\end{propositionbox}

The easy direction has been shown above. Recall that if $C(t)$ is continuous at the origin, it is already uniformly continuous by the positive-semidefiniteness. 
The proof of the reverse direction is not difficult in principle but invokes in its final step  another non-trivial theorem of probability theory (Levy's continuity theorem, Ref.~\cite{Shiryaev:Probability:2016}, III \S 3).  We have deferred the proof to Appendix~\ref{Sec:Appendix_Bochner}.

 The sign in the exponential is a question of convention. Here, we have chosen the sign such that the physicist readily thinks about $\Omega$ as frequency and $t$ as time. If we had chosen as variables $x$ for spatial position and $k$ as a wavenumber, the physicist would choose the opposite sign. 

Let us make a side remark. The notion of the Fourier transform of a probability measure is fundamental in probability theory. 
If $F(\Omega) = \textsf{Prob} [X\leq \Omega]$ corresponds to a probability measure for a real-valued random variable $X$, the Fourier transform 
\begin{align*}
C(t) = \int_\mathbb{R} e^{-i \Omega t} \diff F(\Omega) = \langle e^{-i t X} \rangle,
\end{align*}
is known as the characteristic function. From the discussion above, one infers that $C(t)$ is positive-semidefinite and Bochner's theorem guarantees that each such function is the Fourier transform of a probability measure if it is normalized, $C(0)=1$.

One can show that characteristic functions determine the probability distributions of real-valued random variables uniquely~\cite{Shiryaev:Probability:2016}, in other words, different probability distributions yield different characteristic functions. We cite from Ref.~\cite{Shiryaev:Probability:2016} the following properties of characteristic functions.
 If the moment $\langle | X|^n  \rangle$ exists for some $n\geq 1$ then $C^{(r)}(t)$, the $r$-th derivative of $C(t)$,   exists for every $r\leq n, r\in \mathbb{N}$ and 
\begin{align}\label{eq:probability_moments}
C^{(r)}(t) &= \int_{\mathbb{R}} (-i \Omega)^r e^{- i\Omega t} \diff F(\Omega), \nonumber \\
\langle X^r \rangle &= i^r C^{(r)}(0), \nonumber \\
C(t) &= \sum_{r=0}^n \frac{(-i t )^r}{r!} \langle X^r \rangle +o(t^n) \qquad \text{for } t\to 0. 
\end{align}

For correlation functions in stochastic processes, we can think of $C(t)/C(0)$ as the characteristic function of a random frequency drawn from the distribution $F(\Omega)$. This notion becomes useful later once we discuss sum rules. 

Next we discuss some properties following from the representation theorem by Bochner. Assume that $F(\Omega)$ only consists  of finitely many jumps, yielding a correlation function of the form in Eq.~\eqref{eq:Fourier_discrete}. In this case the correlation function keeps oscillating forever such that correlations never die out. Rather even at large lag times the correlation comes infinitely often close to its initial value. In particular, the correlation function is not integrable with respect to the Lebuesgue measure. Similar observations hold whenever $F(\Omega)$ contains infinitely many jumps.  

Let's assume that the continuous correlation function is integrable, $C \in L^1(\mathbb{R})$, i.e., $\int_{-\infty}^\infty | C(t)| \diff t < \infty$. Then we define the \emph{spectrum}
\begin{align}\label{eq:spectrum}
S(\omega) = \int_{\mathbb{R}} e^{i \omega t} C(t) \diff t.
\end{align}
Clearly, for the spectrum to exist, $C(t\to \pm \infty) =0$. 
An immediate consequence of the Hermitian property of correlation functions, $C(-t) = C(t)^*$, is that the spectrum is real (and even):
\begin{align*}
S(\omega)^* &= \int_{\mathbb{R}} e^{-i\omega t} C(t)^* \diff t \stackrel{t\mapsto -t}{=} \int_{\mathbb{R}} e^{i \omega t} C(-t)^* \diff t = S(\omega).
\end{align*}
 In Feller~\cite{Feller:Probability:1991} (Chapter XIX, 2, Lemma 1)
it is shown that the Lebesgue-Stieltjes measure corresponding to the correlation function is then provided by $\diff F(\Omega) = S(\Omega) \diff \Omega/2\pi$.  A pedestrian version is contained in the proof of Bochner's theorem in Appendix~\ref{Sec:Appendix_Bochner}.
Thus, $S(\Omega)/2\pi$ corresponds to the density of the measure and satisfies, in particular, $S(\Omega)\geq 0$. In conclusion, continuous correlation functions display non-negative spectra (if they exist). 

For the physicist ignoring the existence of jumps and the singular continuous part in the Lebesgue-Stieltjes measure and the conditions for Fourier transforms, this fact follows by inverse Fourier transform from
\begin{align*}
C(t) &= \int e^{-i \Omega t} S(\Omega) \frac{\diff \Omega}{2\pi}, \qquad S(\Omega)\geq 0, \\
\Rightarrow \qquad S(\omega) &= \int e^{i \omega t} C(t) \diff t \geq 0 .
\end{align*}

Let us revisit the examples shown in  Fig.~\ref{fig:candidates_correlation_functions}. 
The upper left panel illustrates the solution of an underdamped harmonic oscillator, described by the equation 
\begin{align}\label{eq:oscillator}
m\ddot{X}(t) + \zeta \dot{X}(t) + k X(t) =f(t),
\end{align}
where the oscillator is 
 driven by a stochastic force $f(t)$. 
 Here $X(t)$ is the displacement of the oscillator from its equilibrium position, $m$ is the mass of the particle, $\zeta>0$ is the  friction coefficient,    and $k> 0$ the spring constant. We impose the equilibrium assumption, $\langle X(0) f(t) \rangle = 0$ for $t>0$, which states that the stochastic  force is uncorrelated with the displacement of the oscillator at earlier times.
Correlating Eq.~\eqref{eq:oscillator} with $X(0)$ yields for the correlation function $C(t) = \langle X(t) X \rangle$ the equation of motion for $t>0$
\begin{align}\label{eq:eom_oscillator}
\ddot{C}(t) +\nu \dot{C}(t) + \omega_0^2 C(t) = 0.
\end{align}
Here $\omega_0 = \sqrt{k/m}$ is the undamped natural frequency of the oscillator and $\nu = \zeta/m$ is the damping constant. 
To connect to statistical physics in equilibrium, we  impose the initial conditions $C(0) = \langle X^2 \rangle = k_B T / k$  as dictated by the equipartition theorem, and $\dot{C}(0) = \langle \dot{X} X \rangle =0$ 
which follows from the factorization of the equilibrium probability density into a Maxwellian for  velocities and Gaussian for  displacements. The solution of Eq.~\eqref{eq:eom_oscillator} for  $t\geq 0$ is 
\begin{align}\label{eq:correlation_harmonic_oscillator}
C(t) = \frac{k_B T}{k} e^{-\nu t/2} \left[ \cos(\omega_d t) + \frac{\nu}{2\omega_d} \sin(\omega_d t) \right] ,
\end{align}
where  $\omega_d = ( \omega_0^2 - \nu^2/4)^{1/2} > 0$ is the characteristic  frequency of the underdamped oscillator. Since $C(-t) = C(t)^*$, 
this completely prescribes the correlation function. 

Importantly, no explicit assumption on the noise autocorrelation $\langle f(t)f(0)\rangle$
 (nor Gaussianity) is needed to obtain $C(t)$. The two-point function is fixed by the  equation of motion, Eq.~\eqref{eq:oscillator}, the equilibrium assumption for the noise,  and the equilibrium initial conditions,  while detailed  assumptions on $f(t)$  affect higher-order moments.

 The correlation function $C(t)$ is integrable, and its  Fourier transform is given by
\begin{align*}
S(\omega) =  \frac{2   k_B T \zeta /m^2  }{\nu^2 \omega^2 + (\omega^2- \omega_0^2)^2}>  0. 
\end{align*}
The non-negativity of the spectrum confirms that $C(t)$ is a valid correlation function. For reference, 
the white-noise spectrum  $S_f(\omega) = 2k_B T \zeta$ for  the random forces follows from the spectrum of the positional autocorrelation function using the equation of motion, Eq.~\eqref{eq:oscillator}. An inverse Fourier transform of Eq.~\eqref{eq:spectrum} then yields $\langle f(t) f(0) \rangle = 2 k_B T \zeta \delta(t)$. 

 In contrast, for an oscillator with negative damping the 'spectrum' becomes negative, 
indicating that no stationary stochastic process can produce such a function as its correlation function. More directly, the condition $|C(t)| \leq C(0)$ is violated, ruling out physical realizability.

We now consider the simple relaxator, which can be viewed as the limiting case of the damped harmonic oscillator, Eq.~\eqref{eq:oscillator}, for vanishing mass $m\to 0^+$.
The  constitutive equation for the relaxator is
\begin{align}
\zeta \dot{X}(t) + k X(t) =f(t).
\end{align} 
Imposing  the same  equilibrium assumption $\langle X(0) f(t) \rangle =0$ for $t>0$, the equation of motion (e.o.m.) for the correlation function $C(t) = \langle X(t) X\rangle$ becomes
\begin{align}\label{eq:eom_relaxator}
 \dot{C}(t) + \gamma C(t) = 0,
\end{align}
with the relaxation rate $\gamma = k/\zeta \eqcolon \tau^{-1}$. Using the equipartition theorem,  $C(0) = \langle X^2 \rangle = k_B T/k$, the solution  is 
\begin{align}\label{eq:relaxator}
C(t) = \frac{k_B T}{k}  e^{-\gamma |t|}. 
\end{align}

The Fourier transform of this correlation function is a  centered Lorentzian
\begin{align}\label{eq:relaxator_spectrum}
S(\omega)  =  \frac{2k_BT \zeta }{1  + \omega^2 \tau^2} > 0, 
\end{align}
The non-negativity of the spectrum confirms that  the simple relaxator is a valid correlation function. 

Last, the Gaussian coupling model is piecewise defined by a Gaussian $\exp(- \gamma^2 t^2)$ for small times and a stretched exponential $\exp[- (|t|/\tau)^\beta]$ with $0< \beta < 1$ for large times such that they are continuously concatenated. Performing a numerical Fourier transform shows that the 'spectrum' becomes  negative at certain frequencies. The reason is the kink in the derivative which leads to slowly oscillating contribution. Interestingly, replacing the Gaussian by a relaxing exponential $\exp(-\gamma |t|)$ yields positive spectra. More generally, by Polya's theorem~\cite{Shiryaev:Probability:2016}, any even continuous convex function $C(t) \geq 0$ satisfying $C(t) \to 0$ as $t\to \infty$ is a correlation function. 

Another interesting example is a Lorentzian in the temperal domain, $C(t) = 1/[1+ (\gamma t)^2 ], \gamma> 0$ \cite{Straube:Comm_Phys_3:2020} with associated exponentially decay power spectral density $S(\omega) =(\pi/\gamma) \exp( - |\omega|/\gamma) $.

To conclude this section we make a few remarks on cross-correlation functions $\langle \delta X(t) \delta Y^* \rangle$ and matrix-valued correlation functions. Let $(X_1(t), \ldots, X_n(t))$ be a finite family of  time-dependent observables. Then we define the matrix-valued correlation function
\begin{align}
C_{ij}(t) &= \langle \delta X_i(t) \delta X_j^* \rangle, \qquad i,j=1,\ldots,n. 
\end{align}
Clearly the diagonal elements of the matrix are autocorrelation functions and thereby are positive-semidefinite. The off-diagonal elements are cross-correlations.  One readily infers that the auto-correlation function corresponding to the linear combination $Z(t) = \sum_{i=1}^n \mu_i X_i(t)$ with arbitrary complex numbers, $\mu_i\in \mathbb{C}, i=1,\ldots, n$ is given by $\sum_{i,j} \mu_i C_{ij}(t) \mu_j^*$.  By Bochner's theorem (assuming continuity) it can be represented as the Borel transform of a suitable finite Lebesgue-Stieltjes measure. The generalization to the matrix case~\cite{Gesztesy:MathMeth_218:2000} is now
\begin{align}
C_{ij}(t) = \int_{\mathbb{R}} e^{-i \Omega t} \diff F_{ij}(\Omega) ,
\end{align}
such that $(F_{ij}(\Omega) )_{i,j=1,\ldots,n}$ is a self-adjoint matrix-valued measure, i.e. $F_{ij}(\Omega)$ is a complex finite Borel measure on the real line $\Omega \in \mathbb{R}$ and a positive-semidefinite matrix for fixed $\Omega$. The latter property guarantees that every contraction $\sum_{i,j} \mu_i \diff F_{ij}(\Omega) \mu_j^*$ is a finite Lebesgue-Stieltjes measure and its Fourier transform corresponds to the autocorrelation of $Z(t)$.

\section{Fourier-Laplace domain}
In this section, we introduce the Fourier-Laplace transform of correlation functions as equivalent representation of these functions. This transform is defined for complex frequencies in the upper half-plane and exhibits specific analytic properties known as \emph{Nevanlinna functions} (sometimes also referred to as Herglotz or Pick functions). The goal is to provide the necessary and sufficient conditions for such functions to represent the Fourier-Laplace transform of a correlation function. 

It is defined as the Fourier-Laplace transform of a function $C: [0,\infty) \to \mathbb{C}$ via 
\begin{align}\label{eq:Fourier-Laplace-transform}
\hat{C}(z) \coloneq i \int_0^\infty e^{i z t} C(t) \diff t, \qquad z \in \mathbb{C}_+,
\end{align}
whenever the integral exists. Here $\mathbb{C}_+ = \{ z \in \mathbb{C}: \Imag[z] > 0 \}$ denotes the complex upper half-plane, and $z$ is referred to as a complex frequency. Throughout the remaining chapters, we will indicate 
functions defined on the upper complex half-plane by a hat (e.g., $\hat{C}(z)$). 

The factor $i$ in front of the integral is a matter of convention. Up to this factor, $\hat{C}(z)$ is the one-sided Fourier transform extended to complex frequencies. The Fourier-Laplace transform is also closely related to the standard Laplace transform, with the substitution $s = - i z$ and an additional prefactor $i$. Working in the Fourier-Laplace domain is particularly convenient because, in many cases, the limit as $z$ approaches a real frequency $\omega$ can be performed. 

Let's revisit the example of a simple relaxator, Eq.~\eqref{eq:relaxator}. The Fourier-Laplace transform yields 
\begin{align}
\hat{C}(z) = \frac{- k_B T/k}{z+ i \gamma} ,
\end{align}
which is complex analytic in the complex upper half-plane. 
Furthermore the imaginary part 
\begin{align}
\Imag[\hat{C}(z) ] =\frac{k_B T}{k}  \frac{ \Imag[z] + \gamma}{\Real[z]^2  + (\Imag[z]+ \gamma)^2} > 0,
\end{align}
is non-negative for $z\in \mathbb{C}_+$. We shall see that these properties transfer to the general case.

In this and the following sections, we assume that all correlation functions are continuous, meaning they admit a representation via Bochner's theorem. Under this assumption, the Fourier-Laplace transform is well-defined for all $z\in \mathbb{C}_+$ since the integrand $e^{i z t} C(t)$ is continuous and dominated by an integrable function. Specifically, 
the boundedness property of correlation functions $|C(t)| \leq C(0)$ ensures that $|e^{i z t} C(t) | \leq \exp(-\Imag[z] t) C(0)$. 
Since $\exp(-\Imag[z] t)$ is integrable on $[0,\infty)$ for $\Imag[z]> 0$, the Fourier-Laplace transform exists.   
Furthermore, the following inequality holds:
\begin{align}
|\hat{C}(z) | \leq \int_0^\infty e^{- \Imag[z] t}| C(t) | \diff t \leq \frac{C(0)}{\Imag[z]}. 
\end{align}

The function $\hat{C} : \mathbb{C}_+ \to \mathbb{C}$ is \emph{complex analytic} in $\mathbb{C}_+$. This follows from the dominated convergence theorem, which ensures that differentiation under the integral is permitted. 
This argument is analogous to the one used for the conventional Laplace transform. Additionally, by the uniqueness property of the (Fourier-)Laplace transform, two different correlation functions 
cannot have the same (Fourier-)Laplace transform. Thus, the Fourier-Laplace transform uniquely characterizes the correlation function.  

Bochner's theorem for continuous correlation functions implies that the Fourier-Laplace transform can be expressed as
$\hat{C}(z) = i \int_0^\infty \diff t \, e^{i z t} \int e^{-i \Omega t}\diff F(\Omega)$, where $F(\Omega)$ is the finite Lebesgue-Stieltjes measure associated with the correlation function. By interchanging the order of integration (justified by Fubini's theorem), 
we arrive at the \emph{Riesz-Herglotz representation}
\begin{align}\label{eq:Riesz-Herglotz}
\hat{C}(z) = \int \frac{\diff F(\Omega)}{\Omega- z}, \qquad z\in \mathbb{C}_+. 
\end{align}

Collecting results, we find that the Fourier-Laplace transform $\hat{C}(z)$ is a complex-analytic function in the upper half-plane $\mathbb{C}_+$, and it is uniquely associated with the measure $F(\Omega)$. The transform $\hat{C}(z)$ is also known as the Borel transform of the measure $F(\Omega)$.

Let us decompose $\hat{C}(z)$ into real and imaginary parts. Using the identity $1/(\Omega- z) = (\Omega- z^*) / |\Omega- z|^2$, we find the following expressions for the real and imaginary parts of $\hat{C}(z)$:
\begin{align}
\Real[\hat{C}(z) ] &= \int \frac{\Omega- \Real[z]}{|\Omega-z|^2} \diff F(\Omega), \nonumber \\
\Imag[\hat{C}(z) ] & = \int \frac{\Imag[z]}{|\Omega-z|^2} \diff F(\Omega).
\end{align}
Remarkably, the imaginary part is always non-negative for $z\in\mathbb{C}_+$ (the upper half-plane). More precisely, $\Imag[\hat{C}(z)]>0$ for all $z\in \mathbb{C}_+$ unless the measure is identically zero, $\diff F(\Omega)=0$, in which case $\hat{C}(z)=0$. This observation motivates the definition of a new class of functions. 
\begin{definitionbox}
\textbf{Definition:}  
A function $\hat{\sigma} : \mathbb{C}_+ \to \mathbb{C}_+ \cup \mathbb{R}$ is called a \textbf{Nevanlinna function} if it is complex analytic in the upper half-plane $\mathbb{C}_+$ and satisfies $\Imag[\hat{\sigma}(z)] \geq 0$ for all $z\in \mathbb{C}_+$.  
If, in addition, the real part of $\hat{\sigma}(z)$ vanishes on the positive imaginary axis, i.e., $\Real[\hat{\sigma}(i y)] = 0$ for $y>0$,
then $\hat{\sigma}(z)$ is called a \textbf{positive-real function (PR function)}.
\end{definitionbox}

The term 'positive-real' originates in electrical network synthesis, where one uses the classical Laplace transform
$
\tilde{C}(s)=\int_0^\infty e^{-s t}\,C(t)\,dt$, $\Real[ s] >0$ of a real-valued function $C(t)$.
With our Fourier-Laplace convention [Eq.~\eqref{eq:Fourier-Laplace-transform}] one has $
\tilde{C}(s)=i \hat{C}(i s)$ and $\Real[ s]>0$. 
Thus the PR condition $\Real[ \tilde{C}(s)] \geq 0$ for $\Real[s]>0$ is equivalent to $\Imag[ \hat{C}(z)]\geq 0$ for $\Imag[ z]>0$. If $C(t)$ is real-valued, then $\tilde{C}(y)\in\mathbb{R}$ for $y>0$. We therefore retain the term 'positive-real', although in our upper-half-plane convention the corresponding inequality involves the imaginary part.

Nevanlinna functions map the upper complex half-plane into itself and possibly to the reals. However, the latter case is not particularly interesting since it turns out to be trivial. Since the imaginary part $\Imag[\hat{\sigma}(z)]$ of a Nevanlinna function is a harmonic function (i.e., it fulfills the Laplace equation), it obeys the minimum principle: either $\Imag[\hat{\sigma}(z)]$ is constant, or it has no minimum in the connected open set $\mathbb{C}_+$. Now suppose there exists a point $z_0\in \mathbb{C}_+$ such that $\Imag[\hat{\sigma}(z_0)] = 0$. Then the point $z_0$ corresponds to a minimum of $\Imag[\hat{\sigma}(z)]$ implying $\Imag[\hat{\sigma}(z)] = 0$ for all $z\in \mathbb{C}_+$. 
By the Cauchy-Riemann equations this entails that $\hat{\sigma}(z) = \text{const.}\in\mathbb{R}$. Thus Nevanlinna functions either map the upper complex half-plane into itself or to a real constant. We may safely exclude the trivial case. 

The Riesz-Herglotz representation clearly shows that the Fourier-Laplace transform of a correlation function is a Nevanlinna function. Let us now examine the conditions under which it is also a positive-real (PR) function. For $y> 0$, we evaluate the real part of $\hat{C}(iy)$ as follows:
\begin{align*}
\Real[\hat{C}(i y) ] = \int \frac{\Omega}{\Omega^2 + y^2} \diff F(\Omega) =  -\int_0^\infty e^{-y t} \Imag[C(t)] \diff t . 
\end{align*} 
Thus $\Real[\hat{C}(i y)]=0$ whenever the correlation function is real. Conversely, if $\Real[\hat{C}(i y)]=0$ for all $y>0$, then $\Imag[C(t)]=0$ for all $t\geq 0$ by the uniqueness of the Laplace transform~\cite{Zemanian:Distribution:1987}, which implies that $C(t)$ is real-valued. Likewise, $C(t)$ is real-valued iff the measure $\diff F(\Omega)$ is symmetric about $\Omega=0$. Collecting these facts, we conclude that $\hat{C}(z)$ is a PR function iff $C(t)$ is real-valued. In classical physics, we will see that most autocorrelation functions in equilibrium are indeed real-valued.  

Next we infer the asymptotic behavior for large frequencies. Here, by $z\to \infty$ we mean $|z| \to \infty$ with $\delta < \arg z < \pi -\delta$ for any fixed $\delta >0$ (i.e., uniformly in sectors),  and we use $O(1/z)$ as shorthand for $O(|z|^{-1})$; likewise, $o(1/z)$ means $o(|z|^{-1})$, and more generally $O(z^{-k})$ and $o(z^{-k})$ denote $O(|z|^{-k})$ and $o(|z|^{-k})$, respectively. From Eq.~\eqref{eq:Riesz-Herglotz} we may write
\[
\hat{C}(z) \;=\; -\frac{C(0)}{z} \;+\; \int \frac{\Omega}{\Omega-z}\, \frac{\diff F(\Omega)}{z}.
\]
For $z=iy$ with $y\to\infty$, one has $y\,\Imag[\hat{C}(iy)]\to C(0)$ by monotone convergence 
\[
\hat{C}(iy) \;=\; -\,\frac{C(0)}{iy} \;+\; o(1/y).
\]
More generally, the same first-order asymptotics
\[
\hat{C}(z) \;=\; -\,\frac{C(0)}{z} \;+\; o(z^{-1}),
\qquad z\to\infty,
\]
holds uniformly in sectors (nontangentially to infinity) for Nevanlinna functions arising from finite measures $F$, in accordance with the Herglotz representation discussed above.

Interestingly, the two properties we have identified for Borel transforms of correlation functions characterize them completely. 
\begin{propositionbox}
\textbf{Theorem (Herglotz representation):} Let $\hat{C} : \mathbb{C}_+ \to \mathbb{C}_+$ be a Nevanlinna function that satisfies either
\begin{enumerate}
\item[1.] the growth condition
\[|\hat{C}(z)| \le \frac{M}{\Imag[ z]}\quad \text{for all } z\in\mathbb{C}_+ \text{ and some } M<\infty,
\]
\end{enumerate}
or
\begin{enumerate}
\item[2.] the sectorial growth condition
$\hat{C}(z)=O(1/|z|)$ uniformly as  $|z|\to\infty$  for   $0<\delta<\arg z<\pi-\delta$.
\end{enumerate}
Then there exists a (finite) Lebesgue-Stieltjes measure $F$ on $\mathbb{R}$ such that
\[
\hat{C}(z)=\int_{\mathbb{R}} \frac{1}{\Omega-z}\,\diff F(\Omega),\qquad z\in\mathbb{C}_+.
\]
Moreover, in case \emph{(1)} one has $\int_{\mathbb{R}} \diff F(\Omega)\le M$.
\end{propositionbox}
In case (1), contour integration with the bound $|\hat{C}(z)|\leq M/\Imag[z]$ yields the pure Cauchy form and $F(\mathbb{R})\le M$ (Appendix~\ref{Sec:Appendix_Herglotz}). In case (2), sectorial decay and the Nevanlinna representation force the affine and cutoff terms to vanish, giving the same Cauchy form with finite $F$ (Appendix~\ref{Sec:Appendix_Representation_Nevanlinna}).

Note that the sectorial growth condition is satisfied if there exists $C(0)\in \mathbb{R}$ such that, uniformly for $0<\delta =\text{arg } z=\pi-\delta$,
\[
\hat{C}(z)= -\frac{C(0)}{z}+o\!\left(1/z\right)\quad\text{as }|z|\to\infty,
\]
since this implies $\hat{C}(z)=O(1/|z|)$ in that sector.

As a side result of the proof, Eq.~\eqref{eq:analytic_representation}, we find also the analytic representation
\begin{align}
\hat{C}(z) = \int_\mathbb{R}  \frac{ \Imag[\hat{C}(\Omega+i\epsilon)] \diff \Omega}{\Omega+ i \epsilon - z} ,
\end{align}
for $\Imag[z]> \epsilon> 0$. 
This is similar to a Kramers-Kronig relation and implies that knowledge of the imaginary part parallel and arbitrarily close to the real line of a $O(z^{-1})$ Nevanlinna function is enough to reconstruct the complete function.

We now assume that the measure has a density $\diff F(\Omega) = f(\Omega) \diff \Omega $. Writing $z = \omega+ i \epsilon, \epsilon>0$, 
\begin{align}\label{eq:inversion_density}
\Imag[\hat{C}(\omega+i \epsilon)] = \int \frac{ \epsilon}{(\Omega-\omega)^2 + \epsilon^2} f(\Omega) \diff \Omega \to \pi f(\Omega), 
\end{align}
as $\epsilon \to 0^+$
at all points of continuity of $f(\Omega)$. We use the following notation
\begin{align}
C'(\omega) + i C''(\omega) &= \lim_{\epsilon\to 0^+} \hat{C}(\omega+ i \epsilon) \nonumber \\
&= \lim_{\epsilon\to 0^+} i \int_0^\infty e^{i \omega t- \epsilon t}C(t) \diff t ,
\end{align}
with real and imaginary part $C'(\omega), C''(\omega)$
provided the limits exist. By the Nevanlinna property $C''(\omega) \geq 0$, and we shall see later that this condition  is related to the non-negativity of power spectral densities and the 
 stability of matter. The previous consideration shows that the density is (up to a factor of $\pi$) the imaginary part of $\hat{C}(z)$ continued to real frequencies. In this case, the  Kramers-Kronig relation simplifies to
\begin{align}
\hat{C}(z) = \int \frac{C''(\Omega) }{\Omega- z } \frac{\diff \Omega}{\pi}.
\end{align}
In other words, the Fourier-Laplace transform of a correlation function can be reconstructed knowing only the imaginary part for real frequencies.

We conclude this section by stating the Stieltjes inversion formula
\begin{align}
\lefteqn{\frac{1}{2} \int_{[a,b]} \diff F(\Omega) + \frac{1}{2}  \int_{(b,a)} \diff F(\Omega)  = } \nonumber \\
&= \lim_{\epsilon\to 0^+} \frac{1}{\pi} \int_{a}^{b} \Imag[ \hat{C}(\Omega+i\epsilon) \diff \Omega. 
\end{align}

Hence, the measure $F(\Omega)$ can be recovered from the imaginary part of $\hat{C}(z)$. For the special case of a density, this relation reduces to  Eq.~\eqref{eq:inversion_density}. A proof of the Stieltjes inversion formula and many more useful relations can be found in Ref.~\cite{Teschl:Mathematical_methods:2014}. The generalization to matrix-valued correlation functions is discussed in Ref.~\cite{Gesztesy:MathMeth_218:2000}:

\section{Power spectral density}

In this section, we introduce the concept of the power spectral density (PSD)  and discuss its relationship with correlation functions and their Fourier-Laplace transform. The PSD is a fundamental concept in signal processing, physics, and statistics, as it provides a frequency-domain representation of the power distribution of a signal. Formally it is defined as follows:
\begin{definitionbox}
\textbf{Definition:}  For a continuous-time stationary stochastic process $( X(t) : t\in \mathbb{R})$ with fluctuations $\delta X(t) = X(t) - \langle X \rangle$, consider the finite-time Fourier transform  for $T>0$
$$
\delta \hat{X}_T(\omega) \coloneq \int_{-T/2}^{T/2} \delta X(t) e^{i \omega t} \diff t, \qquad \omega \in \mathbb{R}. 
$$ 
The \textbf{power spectral density (PSD)} is then defined as
$$ S(\omega) \coloneq \lim_{T\to \infty} \frac{1}{T} \langle |\delta \hat{X}_T(\omega)|^2 \rangle, $$ 
whenever the limit exists. 
\end{definitionbox}
The intuition behind the definition is that the finite-time Fourier transform $\delta \hat{X}_T(\omega)$ 
represent the signal filtered over a finite observation time $T$, isolating frequencies  in the interval $[\omega, \omega+\diff\omega]$. The quantity  $|\delta \hat{X}_T(\omega)|^2$ corresponds to the energy of the signal $X(t)$ in the time interval $[-T/2, T/2]$ within the frequency band $\diff \omega/2\pi$. Dividing by the observation time $T$ yields the corresponding power. Since the signal is random, an ensemble average $\langle \ldots \rangle$ is taken to account for fluctuations, and the observation time $T$ is assumed to be large. 

One might argue that the ensemble average is unnecessary and  that the PSD could be obtained directly from the finite-time Fourier transform of a single long trajectory. However, even for simple processes such as Brownian motion, the result remains random for large $T$~\cite{Krapf:NJP_20:2018}. Instead, a practical approach is to divide the long trajectory into several large sub-trajectories and average their PSD estimates to reduce randomness.

From the definition of the PSD, one infers immediately that it is non-negative, $S(\omega)\geq 0$, in accordance of our interpretation of  power in the signal. For real-valued processes, it is also symmetric, $S(\omega) = S(-\omega)$. 

We choose the same symbol $S(\omega)$ for the PSD as for the spectrum defined in Eq.~\eqref{eq:spectrum}. 
This is justified by the following fundamental result, which reveals that the PSD and the spectrum are indeed the same quantity.  

\newpage

\begin{propositionbox}
\textbf{Theorem (Wiener-Khinchin):} 
For a stationary stochastic process $( X(t): t\in \mathbb{R})$ with integrable 
 correlation function $C(t) = \langle \delta X(t) \delta X^* \rangle$, the power spectral density is given by 
$$
S(\omega) = \int e^{i \omega t} C(t) \diff t.
$$
\end{propositionbox}

\textbf{Proof:} Expanding the squared modulus in the definition of the PSD yields 
\begin{align*}
& S(\omega) = \lim_{T\to \infty}\frac{1}{T} \int_{-T/2}^{T/2} \diff t_1 \int_{-T/2}^{T/2} \diff t_2 e^{i \omega (t_1- t_2) }\langle \delta X(t_1) \delta X(t_2)^* \rangle .
\end{align*}
The angular brackets yield $C(t_1-t_2)$ by stationarity. Substitute $t=t_1-t_2$
\begin{align*}
\ldots&= \lim_{T\to \infty} \frac{1}{T} \int_{-T}^{T} \diff t \, C(t) e^{i \omega t} \left( \int_{t_2 \in [-T/2, T/2], t+t_2 \in [-T/2, T/2] } \!\!\!\!\! \diff  t_2
\right)  \\
&= \lim_{T\to \infty}  \int_{-\infty}^{\infty} \diff t \, C(t) e^{i \omega t} (1- |t|/T )  \Theta(T^2- t^2).
\end{align*}
The integrand converges pointwise to $C(t) e^{i \omega t}$ and is dominated by the integrable function $|C(t)|$. By the dominated convergence theorem, we can take the limit inside the integral
\begin{equation*}
 S(\omega)= \int_{-\infty}^\infty \diff t \, C(t) e^{i \omega t}. \tag*{$\square$} 
\end{equation*}
A  direct consequence for integrable correlation functions is the following identity for the power spectral density
\begin{align}
S(\omega) = 2 \lim_{\epsilon\to 0^+} \Imag[\hat{C}(\omega+ i \epsilon) ] = 2 C''(\omega).
\end{align}
This result follows  by splitting the integral for the PSD into two parts 
\begin{align*}
S(\omega) =& \int_{-\infty}^0 e^{i \omega t} C(t) \diff t + \int_0^\infty e^{i \omega t} C(t) \diff t  .
\end{align*}
Using the identity valid for correlation functions $[e^{i \omega (-t) }C(-t)]^* = e^{i \omega t} C(t)$ and substituting $t\mapsto - t$ in the first integral, we obtain
\begin{align*}
S(\omega) &=\lim_{\epsilon\to 0^+} \left[ \frac{1}{i} \hat{C}(\omega+ i \epsilon) \right]^* + \lim_{\epsilon\to 0^+} \left[ \frac{1}{i} \hat{C}(\omega+ i \epsilon) \right] \\
&= 2 C''(\omega).
\end{align*}
This result highlights the connection between the power spectral density and the imaginary part of the Fourier-Laplace transform of the correlation function. 

\section{Time-Reversible Processes}

An important subclass of stochastic processes consists of those for which it is impossible to determine, by inspecting their trajectories, whether time flows forward or backward. For a given stochastic process, the time-reversed process is obtained by flipping the sign of time, i.e., $( X(-t) : t\in \mathbb{R} )$.

\begin{definitionbox}
\textbf{Definition:}
A stochastic process $ (X(t) : t\in \mathbb{R})$
 is  \textbf{time-reversible} if the time-reversed process $( X(-t) : t\in \mathbb{R})$ obeys the same probabilistic law as the original process.
\end{definitionbox}

The most prominent examples of time-reversible processes in physics are observables in thermal equilibrium. For instance, monitoring the trajectories of individual particles in real space does not allow one to distinguish the future from the past. In such cases, the time-reversed trajectory has the same probability as the original one.

This concept extends naturally to all observables with a well-defined time parity. In classical physics, we can associate with each observable $X=X(q,p)$ 
 (where $q$ represents positions and $p$ represents momenta) a new observable under time reversal, defined as 
$X^T = X^T(q,p)\coloneq X(q,-p)$.  An observable is said to displays positive time parity if $X^T = +X$, and negative one if $X^T = - X$. It is important to note that not all observables exhibit positive or negative time parity; some observables may not have a well-defined behavior under time reversal.

For observables with a definite time parity in a time-reversible process, one readily infers that the correlation function is even under time reversal:
$C(t) = \langle X | X(t) \rangle = \langle X^T | X^T(t) \rangle = \langle X | X(-t) \rangle = C(-t)$. 
 Combining this with the general property of a correlation function $C(-t) = C(t)^*$, reveals that such correlation functions are necessarily real-valued. 

Bochner's  theorem for continuous correlation functions that are even under time reversal then implies that the Lebesgue-Stieltjes measure is symmetric $ \int_{(a,b] }\diff F(\Omega) = \int_{[-b,-a)} \diff F(\Omega)$. For the case of a density $\diff F(\Omega) = f(\Omega) \diff \Omega$, this simply means that $f(\Omega) = f(-\Omega)$. 
In particular, the representation 
simplifies to 
\begin{align}\label{eq:Bochner_real}
C(t) = \int_{\mathbb{R}} \cos(\Omega t) \diff F(\Omega). 
\end{align}

For the Fourier-Laplace transform, we then find
\begin{align*}
\hat{C}(z) &= \int_{\mathbb{R}} \frac{\diff F(\Omega)}{\Omega-z} = \frac{1}{2} \int_{\mathbb{R}} \diff F(\Omega)  \left( \frac{1}{\Omega- z} + \frac{1}{-\Omega-z}\right)   ,
\end{align*} 
or 
\begin{align}\label{eq:PR_representation}
\hat{C}(z) = \int_{\mathbb{R}} \frac{z \diff F(\Omega)}{\Omega^2 - z^2} .
\end{align}
In particular, $\Real[ \hat{C}(iy )] = 0$ for $y>0$, thus $\hat{C}(z)$ is a positive-real function.  
A consequence of Eq.~\eqref{eq:PR_representation} is 
\begin{align}
\hat{C}(-z^*) = - \hat{C}(z)^* \qquad \text{for } z\in \mathbb{C}_+.
\end{align}
Note that $-z^* \in \mathbb{C}_+$ such that $\hat{C}(-z^*)$ is well-defined. Conversely, the latter condition implies that if $\hat{C} : mathbb{C}_+ \to \mathbb{C}_+ \cup \mathbb{R}$ is a Nevanlinna function with $|\hat{C}(z)| \leq M/\Imag[z]$, then $C(t) = C(-t)$. 

Writing the complex frequency as $z = \omega+ i \epsilon$ where  $\omega\in \mathbb{R}$ and $ \epsilon> 0$, we can express the Fourier-Laplace transform of the correlation function as
\begin{align}
\hat{C}(\omega+ i \epsilon) = i \int_0^\infty [ \cos(\omega t) + i \sin(\omega t) ] e^{-\epsilon t } C(t) \diff t .
\end{align}
This expression
reveals that for real-valued correlation functions
\begin{align}
\Real[\hat{C}(\omega+ i \epsilon) ] &= -\int_0^\infty \sin(\omega) e^{-\epsilon t} C(t) \diff t, \nonumber \\
\Imag[\hat{C}(\omega+ i \epsilon) ] &= \int_0^\infty \cos(\omega) e^{-\epsilon t} C(t) \diff t.
\end{align}
Furthermore, one infers that 
$\Real[ \hat{C}(\omega+ i \epsilon) ]$ is an odd function in $\omega$ and $\Imag[\hat{C}(\omega+ i \epsilon) ]> 0$ is even in $\omega$. 
In the special case where  the Lebesgue-Stieltjes measure has a  density, $\diff F(\Omega) = C''(\omega) \diff \omega/\pi$, the representation formula for the correlation function simplifies to 
\begin{align}
C(t) = \frac{2}{\pi} \int_0^\infty C''(\omega) \cos(\omega t) \diff \omega . 
\end{align}

\section{Short-time expansion and frequency sum rules}
In many applications, only the short-time behavior of a correlation function is known such that one has to conceive a model for the full time dependence of the correlation function. Of course, these models should not violate probability theory, therefore certain constraints must be fulfilled. Most importantly, Bochner's representation theorem should hold. Specializing to time $t=0$ yields
\begin{align}
C(0) = \int_{\mathbb{R}} \diff F(\Omega) .
\end{align}
Such a relation is known as a \emph{sum rule}. Whatever measure is considered as candidate for a model, the total variation is fixed by the initial value of the correlation function. For the special case of a density
\begin{align}\label{eq:sum_rule}
C(0) = \int_{-\infty}^\infty S(\Omega) \frac{\diff \Omega}{2\pi},
\end{align}
this implies that the total area under $S(\Omega)/2\pi \geq 0$ has to match the initial value of the correlation function. 

This observation can be generalized to low-order moments of the frequency. Suppose the short-time expansion of the correlation function is known
\begin{align}\label{eq:short-time-expansion}
C(t) = \sum_{r=0}^n C^{(r)}(0) \frac{t^r}{r!} + o(t^n) \qquad \text{for } t\to 0,
\end{align}
where $C^{(r)}(0)$ is the $r$-th derivative of the correlation function evaluated at time $t=0$. Equivalently, the Fourier-Laplace transform displays the high-frequency expansion
\begin{align}
\hat{C}(z) = -\sum_{r=0}^n \frac{i^r}{z^{1+r}} C^{(r)}(0) + o(z^{-1-n}) \qquad \text{as } z\to \infty,
\end{align}
uniformly in every sector $0<\delta<\arg z<\pi-\delta$. Taking derivatives in Bochner's representation theorem provides the sum rules for the Taylor coefficients, $r=0,\ldots, n$,
\begin{align}
C^{(r)}(0) &= \int_{-\infty}^\infty (-i \Omega)^r \diff F(\Omega) 
=  \int_{-\infty}^\infty (-i \Omega)^r S(\Omega) \frac{\diff \Omega}{2\pi} ,
\end{align}
where the second identity holds for the case that $F$ has a density. 

The sum rules allow for a neat probabilistic interpretation. Since $C(t)/C(0)$ corresponds to a characteristic function of a probability density, we can consider the previous relation as the $r$-th moment  
\begin{align}
C^{(r)}(0)/ C(0 ) = (-i )^r\langle  \Omega^r \rangle ,
\end{align}
of the 'random' frequency $\Omega$. Employing this analogy reveals that the short-time expansion, Eq.~\eqref{eq:short-time-expansion}, exists precisely if the moments $\langle |\Omega|^r \rangle$ exist and are finite, compare Eq.~\eqref{eq:probability_moments}.   

By Lebesgue's dominated convergence theorem and Bochner's representation theorem, the correlation function $C(t)$ is $r$ times differentiable for all $t\in \mathbb{R}$ if the moments $\langle |\Omega|^r \rangle$ exist and are finite.

Let's illustrate this for the simple relaxator $C(t) = (k_BT /k) e^{-  |t| /\tau}$ with $\tau>0$. The short-time expansion
\begin{align*}
C(t)/C(0) = 1 -   | t| /\tau  + o(t) \qquad \text{for } t\to 0,  
\end{align*}
is singular already to first order in $t$ due to the jump in the left- and right-derivative at $t=0$. The spectrum corresponds to the centered Lorentzian, Eq.~\eqref{eq:relaxator_spectrum}. For the Lorentzian, the integral in  
\begin{align}
- i \langle \Omega \rangle \stackrel{?}{=} \int_{-\infty}^\infty \frac{2 \Omega \tau }{1 +\Omega^2 \tau^2 } \frac{\diff \Omega}{2\pi} ,
\end{align}
does not converge. This is consistent with the observation that the short-time expansion does not exist beyond the zeroth order. 

As a second example, we consider the damped harmonic oscillator, Eq.~\eqref{eq:correlation_harmonic_oscillator} with
 short-time expansion
\begin{align}\label{eq:harmonic_short_time}
C(t) /C(0) =  1- \omega_0^2 t^2/2 + o(t^2) \qquad \text{for }t\to 0.
\end{align}
The sum rule associated with the second moment is then
\begin{align}
\langle \Omega^2 \rangle = \int_{-\infty}^\infty \Omega^2 \frac{2 \gamma\omega_0^2}{\gamma^2 \Omega^2 + (\Omega^2- \omega_0^2)^2}   \frac{\diff \Omega}{2\pi} =  \omega_0^2.
\end{align}

Last, let us provide an example of a correlation function such that moments of the power spectral density exist, but the radius of convergence for the short-time expansion is finite: choose for $\gamma>0$
\begin{equation}
C(t) = \frac{1}{1+ \gamma^2 t^2 } = \sum_{r=0}^\infty (- \gamma^2 t^2)^r, \qquad |t| < 1/\gamma.
\end{equation}

\section{Representation in terms of memory kernels}

In this section, we discuss representation theorems that allow the formulation of exact equations of motion for correlation functions, albeit at the cost of introducing memory kernels. These exact identities often serve as a convenient starting point for modeling: instead of attempting to construct the correlation function directly, one models the associated memory kernel. Even simple approximations for the memory kernel can yield non-trivial results for the correlation function, and, in particular, the short-time behavior is correctly encoded. 

Let us assume that the correlation function exhibits a short-time expansion similar to that of a relaxator
\begin{align}\label{eq:high}
C(t)/C(0) &= 1- \gamma | t| + o(t) \qquad \text{for } t\to 0, \nonumber \\
\hat{C}(z)/C(0) &=  -\frac{1}{z} + \frac{i \gamma}{z^2} + o(z^{-2}) \qquad \text{for } z\to \infty,
\end{align}
uniformly in the sector $0<\delta<\arg z<\pi-\delta$. Here $\gamma>0$ characterizes the short-time behavior of the correlation function. 

\begin{propositionbox}
\textbf{Theorem:} If $\hat{C}: \mathbb{C}_+ \to \mathbb{C}_+ \cup \mathbb{R}$ is a Nevanlinna function with high-frequency expansion \eqref{eq:high} (uniformly in a sector), then it can be represented as
\[
\hat{C}(z) = \frac{-C(0)}{z+ i \gamma + \hat{K}(z)} ,
\]
where $i\gamma + \hat{K}$ is again a Nevanlinna function and $\hat{K}(z) \to 0$ as $|z|\to \infty$ uniformly in every sector $0<\delta<\arg z<\pi-\delta$.
\end{propositionbox}

\textbf{Proof:} Since $\hat{C}$ is a Nevanlinna function and the high-frequency expansion in \eqref{eq:high} holds uniformly in a sector $0<\delta<\arg z<\pi-\delta$, the Nevanlinna representation yields a finite Lebesgue-Stieltjes measure $F$ on $\mathbb{R}$ such that
\[
\hat{C}(z) = \int_{\mathbb{R}} \frac{\diff F(\Omega)}{\Omega- z}\,.
\]
Normalizing by $C(0)=1$ (the case $C(0)=0$ is trivial) gives $\int_{\mathbb{R}} \diff F=1$. Taking imaginary parts,
\[
\Imag[\hat{C}(z)] = \int_{\mathbb{R}} \frac{\Imag[z]}{|\Omega-z|^2} \,\diff F(\Omega) > 0 \qquad (z\in\mathbb{C}_+),
\]
so $\hat{C}$ has no zeros in $\mathbb{C}_+$. Define
\[
i\gamma + \hat{K}(z) \coloneq - z -\frac{1}{\hat{C}(z)}\,,
\]
which is analytic on $\mathbb{C}_+$. From \eqref{eq:high}, the right-hand side converges to $i\gamma$ as $|z|\to\infty$ in sectors, so $\hat{K}(z)\to 0$ accordingly. To verify that $i\gamma+\hat{K}$ is Nevanlinna, observe
\[
\gamma+ \Imag[\hat{K}(z)] = - \Imag[z] - \Imag\!\big[\hat{C}(z)^{-1}\big]
= -\Imag[z] + \frac{\Imag[\hat{C}(z)]}{|\hat{C}(z)|^2}.
\]
Define $x(\Omega) \coloneq (\Omega-z)^{-1} \hat{C}(z)^{-1} -1$. Then by Cauchy-Schwarz,
\[
0 \le \int_{\mathbb{R}} |x(\Omega)|^2 \,\diff F(\Omega)
= \frac{1}{|\hat{C}(z)|^2}\int_{\mathbb{R}} \frac{\diff F(\Omega)}{|\Omega-z|^2} - 1.
\]
Multiplying by $\Imag[z]>0$ gives
\[
0 \le \frac{1}{|\hat{C}(z)|^2}\int_{\mathbb{R}} \frac{\Imag[z]}{|\Omega-z|^2}\,\diff F(\Omega) - \Imag[z]
= \frac{\Imag[\hat{C}(z)]}{|\hat{C}(z)|^2} - \Imag[z],
\]
whence $\gamma+\Imag[\hat{K}(z)] \ge 0$ for $z\in\mathbb{C}_+$. Thus $i\gamma+\hat{K}$ is a Nevanlinna function. \hfill $\square$

The representation of $\hat{C}(z)$ in terms of the Nevanlinna function $i \gamma + \hat{K}(z)$ can be rearranged to obtain the following equation of motion in the frequency domain:
\[
[ z + i \gamma + \hat{K}(z) ] \hat{C}(z) = - C(0).
\]
Assume in addition that $\hat{K}: \mathbb{C}_+ \to \mathbb{C}_+ \cup \mathbb{R}$ is itself a Nevanlinna function and satisfies the sectorial growth condition
\[
\hat{K}(z) = O(|z|^{-1}) \quad \text{uniformly as } |z|\to\infty,
\]
for  $0<\delta<\arg z<\pi-\delta$. 
This implies that $\hat{K}(z)$ is the Fourier-Laplace transform of a correlation function $K(t)$. Equivalently, we assume the high-frequency expansion for $\hat{C}(z)$ and the corresponding short-time expansion for $C(t)$:
\begin{align}\label{eq:high_freq}
\hat{C}(z)/C(0) &= \frac{-1}{z} + \frac{i \gamma}{z^2} + \frac{\gamma^2- K(0)}{z^3} + o(z^{-3}) , \nonumber\\
C(t) /C(0) &= 1 - \gamma | t|  + \frac{\gamma^2- K(0)}{2}\, t^2 + o(t^2),
\end{align}
where $\gamma> 0$ and $K(0) \geq 0$. 

Using these assumptions, the equation of motion can be transformed into an integro-differential equation in the time domain. First, the Fourier-Laplace transform of the derivative is
\begin{align*}
i \int_0^\infty \dot{C}(t) e^{iz t} \diff t &= i \left( \left.  C(t) e^{iz t} \right|_{t=0}^\infty - i z \int_0^\infty e^{iz t} C(t) \diff t  \right), \\
\Rightarrow\quad \hat{\dot{C}}(z) &= -i C(0) - i z \hat{C}(z) ,
\end{align*}
using boundedness of $C(t)$ so that the boundary term at $t=\infty$ vanishes for $z\in\mathbb{C}_+$. Second, by the convolution theorem, for $(K \ast C)(t)=\int_0^t K(t-\bar{t})C(\bar{t})\diff \bar{t}$,
\begin{align*}
& i \int_0^\infty e^{i z t} (K \ast C)(t) \diff t \\
&= i \left( \int_0^\infty e^{i z t_1} K(t_1) \diff t_1 \right)
   \left( \int_0^\infty e^{i z t_2} C(t_2) \diff t_2 \right) \\
&\Rightarrow\quad \widehat{ K \ast C}(z)  = - i \hat{K}(z) \hat{C}(z) .
\end{align*}
Combining these identities with the frequency-domain equation yields the desired memory-kernel representation.

\begin{propositionbox}
\textbf{Theorem (Memory kernel): }
The correlation function $C:\mathbb{R} \to \mathbb{C}$ with high-frequency expansion \eqref{eq:high_freq} satisfies the integro-differential equation
\[
\dot{C}(t) + \gamma C(t) + \int_0^t K(t-\bar{t})\, C(\bar{t}) \,\diff \bar{t} = 0,
\]
where the memory kernel $K:\mathbb{R} \to \mathbb{C}$ is the correlation function associated with the Nevanlinna function $\hat{K}:\mathbb{C}_+ \to \mathbb{C}_+ \cup\mathbb{R}$, which obeys the sectorial growth condition $\hat{K}(z) = O(|z|^{-1})$ uniformly for $0<\delta<\arg z<\pi-\delta$.
\end{propositionbox}

The equivalence between the integro-differential equation in the time domain and the equation of motion in the frequency domain shows that the solution is unique once the initial condition $C(0)$ is specified. When the memory kernel is absent, i.e., $K(t)\equiv 0$, the equation reduces to the simple relaxator of Eq.~\eqref{eq:eom_relaxator}.

The term $\gamma C(t)$ can be interpreted as an instantaneous (time-local) friction term. In contrast, the convolution with $K$ represents a retarded friction that depends on the entire past of the system; this history dependence motivates the term memory kernel for $K$.

In the frequency domain, $\hat{K}$ acts as a frequency-dependent friction; in particular, only $\Imag[\hat{K}]$ directly quantifies dissipation. In other areas of physics, especially quantum field theory and condensed matter physics, the memory kernel is often referred to as the self-energy or mass operator.

As an alternative representation, introduce $\hat{M}(z)$ via
\begin{align}
i \gamma + \hat{K}(z) &= \frac{-1}{\,i \gamma^{-1} + \hat{M}(z)\,} \nonumber \\
&= \frac{-\Real[\hat{M}(z)] + i \gamma^{-1} + i\,\Imag[\hat{M}(z)]}{\big|i \gamma^{-1} + \hat{M}(z)\big|^2}.
\end{align}
Since $i\gamma+\hat{K}$ is a Nevanlinna function, it has no zeros in $\mathbb{C}_+$, so the right-hand side is well-defined and analytic on $\mathbb{C}_+$. Equivalently,
\[
\hat{M}(z) \;=\; -\frac{1}{\,i\gamma+\hat{K}(z)\,} - i\gamma^{-1}.
\]
Because $z\mapsto -1/z$ maps $\mathbb{C}_+$ into $\mathbb{C}_+$, it follows that $i\gamma^{-1}+\hat{M}=-1/(i\gamma+\hat{K})$ is a Nevanlinna function. From the high-frequency behavior $\hat{K}(z)\to 0$ uniformly in sectors, we obtain $\hat{M}(z)\to 0$ as $|z|\to\infty$ uniformly in every sector $0<\delta<\arg z<\pi-\delta$.

For the correlation function $C(t)$, this yields the following frequency-domain representation.

\newpage 

\begin{propositionbox}
\textbf{Theorem:} If $\hat{C}:\mathbb{C}_+ \to \mathbb{C}_+ \cup \mathbb{R}$ is a Nevanlinna function with high-frequency expansion \eqref{eq:high} (uniformly in a sector), then
\[
\hat{C}(z) = \frac{-C(0)}{\,z- \big[i \gamma^{-1} + \hat{M}(z)\big]^{-1}} ,
\]
where $i\gamma^{-1} + \hat{M}$ is a Nevanlinna function and $\hat{M}(z) \to 0$ as $|z|\to \infty$ uniformly in sectors.
\end{propositionbox}

If, in addition, $\hat{K}(z)=O(|z|^{-1})$ uniformly for $0<\delta<\arg z<\pi-\delta$, then the same sectorial decay holds for $\hat{M}(z)=O(|z|^{-1})$. Hence $\hat{M}$ is the Fourier-Laplace transform of a correlation function $M$, and the representation in terms of $\hat{M}$ translates to a time-domain Volterra equation.

\begin{propositionbox}
\textbf{Theorem (Memory kernel): }
The correlation function $C: \mathbb{R}\to \mathbb{C}$ with high-frequency expansion \eqref{eq:high_freq} satisfies
\[
\gamma^{-1}\,\dot{C}(t) + C(t) + \int_0^t M(t-\bar{t})\,\dot{C}(\bar{t})\, \diff \bar{t} \;=\; 0,
\]
where the memory kernel $M:\mathbb{R}\to \mathbb{R}$ is the correlation function associated with the Nevanlinna function $\hat{M}:\mathbb{C}_+ \to \mathbb{C}\cup \mathbb{R}$, which obeys the sectorial growth condition $\hat{M}(z)=O(|z|^{-1})$ uniformly for $0<\delta<\arg z<\pi-\delta$.
\end{propositionbox}

As a second example, consider a short-time/high-frequency expansion of the correlation function
\begin{align}
C(t)/C(0) &= 1- \frac{\omega_0^2}{2}\, t^2 + o(t^2), \\
\hat{C}(z)/C(0) &= -\frac{1}{z} - \frac{\omega_0^2}{z^3} + o(z^{-3}) \qquad (z\to\infty),
\end{align}
uniformly for $0<\delta<\arg z<\pi-\delta$. This coincides with the expansion of a harmonic oscillator, Eq.~\eqref{eq:harmonic_short_time}. The representation theorem of the previous example applies with $\gamma=0$, yielding
\[
\hat{C}(z) \;=\; \frac{-C(0)}{\,z+ \hat{K}(z)\,}, 
\]
where $\hat{K}$ is a Nevanlinna function with $\hat{K}(z)\to 0$ as $|z|\to\infty$ uniformly in sectors. Comparing with the high-frequency expansion above shows
\[
\hat{K}(z) \;=\; -\,\frac{\omega_0^2}{z} + o(z^{-1}) \qquad (|z|\to\infty),
\]
indicating that $\hat{K}$ is the Fourier-Laplace transform of a correlation function $K$ with $K(0)=\omega_0^2$.

Further progress can be made by assuming that $K$ itself exhibits a short-time/high-frequency expansion analogous to \eqref{eq:high}:
\begin{align}
K(t)/ \omega_0^2  &= 1 - \nu |t| + o(t), \\
\hat{K}(z)/\omega_0^2 & = -\frac{1}{z} + \frac{i \nu}{z^2} + o(z^{-2}) \qquad (|z|\to\infty),
\end{align}
uniformly for $0<\delta<\arg z<\pi-\delta$, with $\nu\ge 0$. Equivalently, this yields the refined expansions
\begin{align}
C(t)/ C(0) &= 1- \frac{\omega_0^2}{2}\, t^2 + \frac{\omega_0^2 \nu}{6}\, t^3 + o(t^3), \label{eq:short_damped_ho} \\
\hat{C}(z)/C(0) &= -\frac{1}{z} - \frac{\omega_0^2}{z^3} + \frac{i \omega_0^2 \nu}{z^4} + o(z^{-4}) \qquad (|z|\to\infty),
\end{align}
uniformly in sectors. By the representation theorem, $\hat{K}(z)$ can be written as
\[
\hat{K}(z) \;=\; \frac{-\omega_0^2}{\,z+ i \nu + \hat{M}(z)\,},
\]
where $i\nu+\hat{M}$ is a Nevanlinna function and $\hat{M}(z)\to 0$ as $|z|\to\infty$ uniformly in sectors. Substituting this into $\hat{C}(z)$ gives
\begin{align}\label{eq:damped_ho}
\hat{C}(z) \;=\; \frac{-C(0)}{\,z- \omega_0^2/[\,z+ i \nu + \hat{M}(z)\,]}\,.
\end{align}

If $\hat{M}(z) = O(|z|^{-1})$ uniformly for $0<\delta<\arg z<\pi-\delta$, then $\hat{M}$ is the Fourier-Laplace transform of a correlation function $M$. Transforming \eqref{eq:damped_ho} back to the time domain yields:

\begin{propositionbox}
\textbf{Theorem (Memory kernel): }
The correlation function $C: \mathbb{R} \to \mathbb{C}$ with short-time expansion \eqref{eq:short_damped_ho} satisfies the integro-differential equation
\[
\ddot{C}(t) + \nu \dot{C}(t) + \omega_0^2 C(t) + \int_0^t M(t-\bar{t})\, \dot{C}(\bar{t}) \,\diff \bar{t} \;=\; 0,
\]
where the memory kernel $M:\mathbb{R} \to \mathbb{C}$ is the correlation function associated with the Nevanlinna function $\hat{M}:\mathbb{C}_+\to \mathbb{C}_+ \cup \mathbb{R}$, which obeys the sectorial growth condition $\hat{M}(z)=O(|z|^{-1})$ for $|z|\to \infty$ uniformly for $0<\delta<\arg z<\pi-\delta$.
\end{propositionbox}

This integro-differential equation is reminiscent of a damped harmonic oscillator. Setting $M(t)\equiv 0$ recovers the conventional damped harmonic oscillator equation. The convolution with $M(t)$ acts as a retarded friction, introducing memory effects that depend on the system's history; equivalently, in the frequency domain, $\hat{M}(z)$ represents a frequency-dependent friction that modifies the damping at different frequencies. 

Equations of motion of this type have been derived for Markov processes using the Mori-Zwanzig projection-operator formalism~\cite{Kubo:Statistical_physics_II:2012, Hansen:Theory_of_simple_liquids:2013}. The formalism provides exact microscopic expressions for memory kernels; for instance, for the velocity-autocorrelation function, the associated memory kernel is a force-autocorrelation function evolving under projected dynamics. While these expressions are seldom directly computable, they serve as a convenient starting point for approximations, including perturbative schemes and mode-coupling theories~\cite{Goetze:Complex_Dynamics}.

\section{Linear Response revisited}\label{Sec:Linear_Response_revisited}

In this section, we elaborate a more rigorous characterization of linear-response functions. The fluctuation-dissipation theorem, combined with the mathematical properties of correlation functions,  imposes constraints on  the class of admissible response function. Here, we take a broader perspective and derive the constraints from three phenomenological principles:  passivity, stability of matter, and causality.  

This section assumes some familiarity with the theory of distributions. The presentation closely follows the excellent textbook by Zemanian~\cite{Zemanian:Distribution:1987}.

We begin by reconsidering the class of  force protocols $f=f(t)$ that should be admissible. In the language of signal theory, the force corresponds to the input signal, while the observable $X= X(t)$ we monitor is  the output signal. The input signal $f$ must satisfy the following conditions. First, it should be a smooth function, i.e., it is infinitely differentiable, in order to allow for a broad class of response kernels. Second,  along with all its derivatives, it should vanish
 rapidly for distant times, meaning faster than any power law. Formally this expressed as  
\begin{align}
f\in C^\infty(\mathbb{R}), \quad \text{and} \quad  t^n \partial_t^m f(t) \to 0 \quad \text{as } t\to \pm \infty ,
\end{align} 
for $n,m\in \mathbb{N}_0$. The space  of such functions, equipped with a certain translationally invariant locally convex topology, is known  as the Schwartz space, denoted by $\mathcal{S}(\mathbb{R})$, or the space of test functions of rapid descent. 

The output signal $X=X(t)$ should satisfy the following properties.
\begin{enumerate}
\item The output signal $X$ must be a linear functional of the input signal. 
\item The mapping from $f$ to $X$ must be continuous, meaning that    small changes in the input signal result  in small changes in $X$. 
\item A time shift in the input signal should result in a corresponding time shift in the output signal. 
\end{enumerate}
The first condition is basically the definition of linear response. Implicitly, we are assuming small signals such nonlinear contributions are neglible. The second condition on continuity is natural and rules out pathologically sensititve responses and ensures well-posedness and robustness to noise.  The third condition reflects homogeneity of time. 

As shown in Ref.~\cite{Zemanian:Distribution:1987}, the most general mapping satisfying these conditions is given by the convolution integral:
\begin{align*}
X(t) = (\chi \ast f)(t) = \int \chi(t - \bar{t}) f(\bar{t}) \diff \bar{t},
\end{align*}
where $\chi$ is the response kernel. 
However, here $\chi$ is  not necessarily a function but rather a  tempered distribution, $\chi \in \mathcal{S}'(\mathbb{R})$. 
The space of tempered distributions $\mathcal{S}'(\mathbb{R})$ is the dual space to the Schwartz space $\mathcal{S}(\mathbb{R})$. The theory of distributions guarantees  that, under this convolution, the  output signal is  infinitely differentiable, $X\in \mathbb{C}^\infty$, and $X(t)$ and all its derivatives grow at most polynomially.  Consequently,  the output itself belongs to the space of tempered distributions, $X \in \mathcal{S}'(\mathbb{R})$. 

To proceed further, we assume that there is a functional that quantifies the dissipated work exerted on the system by the input signal. The dissipated work up to time $T$ is given by
\begin{align}
\Delta W_{\text{diss}}(T) = \int_{-\infty}^T f(t) \dot{X}(t) \diff t.
\end{align}
This integral is well-defined because $f$ is smooth and vanishes faster than any power law at distant times, while $\dot{X}$ is smooth and grows at most polynomially.

In Sec.~\ref{Sec:Linear_Response}, we introduced the concept of stability of matter, which implies that the dissipated work in the distant future must be non-negative. A natural question arises: can one design a clever force protocol $f\in \mathcal{S}(\mathbb{R})$ to extract work from a system at a finite time $T$? Experience tells us that this is impossible unless the system is actively sustained by an external energy source. This insight motivates the following definition
\begin{definitionbox}
\textbf{Definition (Passivity):}   A system is called passive if the dissipated work is non-negative for all times $T$:
$$ \Delta W_{\text{diss}}(T) = \int_{-\infty}^T f(t) \dot{X}(t) \diff t \geq 0. $$
\end{definitionbox}

So far, we have identified three fundamental principles relevant to the linear response of a system.

\begin{enumerate}
\item    Causality: The output signal at any time $t$ depends only on the input signal at earlier times.
 \item  Stability of Matter: The dissipated work in the distant future is non-negative.
 \item   Passivity: The dissipated work is non-negative at all times.
\end{enumerate}

One might wonder which of these principles is the most fundamental. Clearly, passivity implies stability of matter, as $T$ can become arbitrarily large. Causality, on the other hand, is universal in physics and not restricted to linear response. Surprisingly, the following theorem~\cite{Zemanian:Distribution:1987} establishes a direct link between passivity and causality.
\begin{propositionbox}
\textbf{Theorem:} Passivity implies causality.
\end{propositionbox}

\textbf{Proof:} Let $f,f_1 \in \mathcal{S}(\mathbb{R})$ be two test functions (input signals), with corresponding responses $X= \chi \ast f, X_1 = \chi \ast f_1$. Both responses are smooth functions of slow growth $X, X_1 \in C^\infty(\mathbb{R}) \cap \mathcal{S}'(\mathbb{R})$. Assume  that the first signal vanishes for negative times, $f(t) =0$ for $t<0$. We aim to show that the corresponding response satisfies $\dot{X}(t) =0$ for $t<0$.  

Define a new input signal $f_2(t) \coloneq f_1(t) + \alpha f(t)$ where $\alpha\in \mathbb{R}$ is arbitrary. For $t<0$, this implies $f_2(t) = f_1(t)$. The response to this new input signal is $X_2=  \chi \ast f_2 = X_1+ \alpha X$. By passivity, the corresponding dissipated work is non-negative
\begin{align*}
\int_{-\infty}^T \dot{X}_2(t) f_2(t) \diff t \geq 0,
\end{align*}
 for all times $T\in \mathbb{R}$. Specializing to negative times, $T<0$, we have
 \begin{align*}
 \int_{-\infty}^T [ \dot{X}_1(t) + \alpha \dot{X}(t) ] f_1(t) \diff t \geq 0.
 \end{align*}
Since $\alpha$ is arbitrary, this inequality can hold only if  
\begin{align*}
\int_{-\infty}^T \dot{X}(t) f_1(t) \diff t = 0. 
\end{align*}
Now, suppose the smooth function $\dot{X}(t)$ is non-zero for some $t<0$. In that case, we could construct $f_1\in \mathcal{S}(\mathbb{R})$ as a non-negative, smooth, and narrow pulse centered around this time. Such a choice would make the integral non-zero, leading to a contradiction. Hence, we conclude that  $\dot{X}(t) = 0$ for all negative times, $t< 0$. \hfill $\square$  

We associate  the complex susceptibility $\hat{\chi}$ with the response kernel $\chi \in \mathcal{S}'(\mathbb{R})$ via the Laplace  transform with convention
\begin{align}\label{eq:Laplace_response}
\hat{\chi}(z) \coloneq \int_0^\infty e^{i z t} \chi(t) \diff t,
\end{align} 
for $z\in \mathbb{C}$ whenever the integral exists. Note that we use a different convention for response kernels, Eq.~\eqref{eq:Laplace_response}, and correlation functions, Eq.~\eqref{eq:Fourier-Laplace-transform}, the latter includes an extra $i$ as prefactor. Using two different conventions for these quantities will turn out useful to unify the discussion. 

We demonstrate  the existence of the integral  for $z \in \mathbb{C}_+$, given that
 the response kernel is a tempered distribution, $\chi \in \mathcal{S}'(\mathbb{R})$. Let $z= \omega+ i \epsilon$, where $ \omega\in\mathbb{R}$ and $ \epsilon>0$. Choose a smooth cutoff function  $\lambda \in C^\infty(\mathbb{R})$ such that  $\lambda(t) = 0$ for $t<-1$ and $\lambda(t) =1 $ for $t>0$. Then, the function $\lambda(t) e^{i z t} = \lambda(t) e^{-\epsilon t} e^{i \omega t}$ belongs to $ \mathcal{S}(\mathbb{R})$ and is an admissible test function. Consequently  the integral
\begin{align*}
\int_{-\infty}^\infty \chi(t) \lambda(t) e^{i z t} = \int_0^\infty \chi(t) e^{iz t} \diff t = \hat{\chi}(z), 
\end{align*}
exists. The equality follows from causality since $\chi(t) =0$ for $t<0$. 

The complex susceptibility $\hat{\chi}$ is an analytic function in the complex upper half-plane because differentiation under the integral is permitted. To see this, consider the difference quotient 
\begin{align*}
\Phi_{\Delta z}(t) \coloneq \lambda(t) \frac{e^{i (z+ \Delta z) t}- e^{iz t}}{\Delta z} \stackrel{\Delta z\to 0}{\longrightarrow} \lambda(t) t e^{iz t}.
\end{align*}
where the  convergence is in $\mathcal{S}(\mathbb{R})$. This justifies  differentiation under the integral, ensuring that $\hat{\chi}(z)$ is analytic in $\mathbb{C}_+$. 

It is favorable to allow also for complex input signals $f(t) = \Real[f(t)] + i \Imag[f(t)]$. The response is then defined as 
\begin{align*}
X(t) = (\chi \ast f)(t) = (\chi \ast \Real[f])(t) + i (\chi \ast \Imag[f])(t).
\end{align*}
Since $\chi$ is real-valued, this implies that the input signal can be decomposed into two real-valued components, $(\Real[f], \Imag[f])$, with corresponding output signals $(\Real[X], \Imag[X]$. The associated dissipated work is then expressed as
\begin{align*}
\Delta W_{\text{diss}}(T) = \int_{-\infty}^T \{& \Real[\dot{X}(t)] \Real[f(t)]  \\ 
&+ \Imag[\dot{X}(t)] \Imag[f(t)] \} \diff t \geq 0.
\end{align*}
The latter result can be compactly rewritten in terms of the complex conjugate of the input signal
\begin{align*}
\Delta W_{\text{diss}}(T) = \int_{-\infty}^T \Real[ \dot{X}(t) f(t)^* ] \geq 0.
\end{align*}

The next property provides a mathematically precise formulation of the concept of stability of matter, which was  introduced heuristically earlier as $\omega \chi''(\omega) \geq 0$. This is formalized in the following theorem: 
\begin{propositionbox}
\textbf{Theorem:} Passivity implies $\Imag[z \hat{\chi}(z)] \geq 0$ for $z\in \mathbb{C}_+$, i.e.,  $z\hat{\chi}(z)$ is a positive real (PR) function. 
\end{propositionbox}

\textbf{Proof:} We have already established that $\hat{\chi}$ is analytic in $\mathbb{C}_+$, and therefore  $z\hat{\chi}(z)$ is also analytic in $\mathbb{C}_+$. Furthermore, since $\chi$ is real-valued, it follows directly  that for $z=i \epsilon, \epsilon>0$
\begin{align*}
\Real[ i \epsilon \int_0^\infty e^{-\epsilon t} \chi(t) \diff t] = 0.
\end{align*}
It remains to  show that $\Imag[z \hat{\chi}(z)] \geq 0$. To do so, we use the property of passivity, which implies that for any complex-valued signal $f(t)$, the dissipated work satisfies
\begin{align*}
0 \leq & \Real \int_{-\infty}^T \diff t f(t)^* \dot{X}(t) \\ =& \Real \int_{-\infty}^T \diff t f(t)^* \int_{-\infty}^\infty \diff \bar{t} \dot{\chi}(t-\bar{t}) f(\bar{t}).
\end{align*}
Now, choose as input signal $f(t) = e^{-i z t}, z\in \mathbb{C}_+$ defined for $t< t_1$ (where $T<t_1 < \infty$) and extended smoothly such that $f\in\mathcal{S}(\mathbb{R})$. Then 
\begin{align*}
0 \leq & \Real \int_{-\infty}^T \diff t\, e^{i z^* t} \int_{-\infty}^t \diff \bar{t}\, \dot{\chi}(t-\bar{t}) e^{-i z \bar{t}} \\
=&  \Real \int_{-\infty}^T \diff t\, e^{2 \Imag[z] t} \int_{-\infty}^t \diff \bar{t} \,\dot{\chi}(t-\bar{t}) e^{iz(t-\bar{t})} \\
=& \int_{-\infty}^T \diff t\, e^{2 \Imag[z] t} \Real \int_0^\infty \diff t' \,\dot{\chi}(t') e^{i z t'} \\
=& \int_{-\infty}^T \diff t \, e^{2 \Imag[z] t} \Real[ -i z \hat{\chi}(z) ] .
\end{align*}
Here we used in the first line causality to restrict the inner integral to times $\bar{t}\leq t$ and in the third line we substituted $t' = t-\bar{t}$. In the last line, we used the rules of differentiation in the distributional sense. Rearranging, this yields
\begin{align*}
0 \leq \Imag[ z \hat{\chi}(z) ] \int_{-\infty}^T \diff t\, \exp( 2 \Imag[z]t ) .
\end{align*}
Since the integral is strictly positive, this implies  $\Imag[z\hat{\chi}(z)]\geq 0$. \hfill $\square$

In our heuristic considerations, we concluded that the response kernel can be written as 
\[
\chi(t) =\Theta(t) \int_{0}^\infty \frac{2\diff \Omega}{\pi} \chi''(\Omega) \sin(\Omega t).
\]
By the stability of matter $\Omega \chi''(\Omega) \geq 0$, this implied that the kernel is a positive superposition of oscillating sine functions. Using the symmetry $\chi''(-\Omega) = -\chi''(\Omega)$, this representation is equivalent to
\begin{align}\label{eq:time-representation}
\chi(t) =\Theta(t) \int_{-\infty}^\infty \frac{\diff \Omega}{\pi} \chi''(\Omega) \sin(\Omega t).
\end{align}
By applying a Laplace transform, we infer the following representation for the complex susceptibility
\begin{align}\label{eq:Kramers-Kronig}
\hat{\chi}(z) = \int_{-\infty}^\infty \frac{\diff \Omega}{\pi} \frac{\Omega\chi''(\Omega)}{\Omega^2- z^2} .
\end{align}
 Taking the real part of Eq.~\eqref{eq:Kramers-Kronig} and considering the limit $\epsilon \to 0^+$ in the complex frequency $z =\omega+ i \epsilon$ yields a Kramers-Kronig relation, which intertwines the real and imaginary part of the complex susceptibility.  
The following powerful representation theorem for positive-real (PR) functions  makes these heuristic considerations rigorous.
\begin{propositionbox}
\textbf{Theorem (Cauer): } A PR function $z \hat{\chi}(z)$ can be represented as 
$$ z \hat{\chi}(z) = z \chi_\infty + \int \frac{z}{\Omega^2- z^2} (1+ \Omega^2) \diff H(\Omega),$$
with $\chi_\infty \geq 0$ and a real-valued, odd, bounded, nondecreasing function $H:\mathbb{R} \to \mathbb{R}$.   
\end{propositionbox}
The proof is a consequence of a more general representation theorem for Nevanlinna functions and is deferred to Appendix~\ref{Sec:Appendix_Representation_Nevanlinna}. The integral is  understood in the Lebesgue-Stieltjes sense. Comparing this with our heuristic considerations in Eq.~\eqref{eq:Kramers-Kronig}, we identify the replacement rule $\Omega \chi''(\Omega) \diff \Omega/\pi \to (1+ \Omega^2) \diff H(\Omega)$. 
Using  Cauer's representation theorem, we can now provide the analog of Eq.~\eqref{eq:time-representation} by performing a formal inverse Laplace transform
\begin{align*}
\chi(t) = \chi_\infty \delta(t) + \Theta(t) \int \frac{\sin(\Omega t)}{\Omega} (1+ \Omega^2) \diff H(\Omega), 
\end{align*}
where this representation is  understood in the distributional sense, $\chi \in \mathcal{S}'(\mathbb{R})$. 
The response function reduces to an ordinary function if and only if $\chi_\infty =0$ and  the first absolute moment of the Lebesgue-Stieltjes measure is finite, $\int |\Omega| \diff H(\Omega) < \infty$. In this case the response function is  continuous in time. An atom of $\diff H(\Omega)$ at $\Omega=0$ corresponds to a contribution increasing linearly in time. 

The output signal, given by the convolution of the response kernel with an input signal $f \in \mathcal{S}(\mathbb{R})$, is
\begin{align*}
\lefteqn{ X(t) =  (\chi \ast f)(t) =} \nonumber \\ 
 &= \chi_\infty f(t) 
 +   \int (1+\Omega^2)  \diff H(\Omega)
\int_{-\infty}^t \frac{\sin[\Omega(t-\bar{t})]}{\Omega} f(\bar{t}) \diff \bar{t},
\end{align*}
 where in the case of $\chi(t)$ being a distribution the time integration is performed before the integral over the measure $dH(\Omega)$. Performing a change of variables $\bar{t}\mapsto t-\bar{t}$ in the inner integral reveals that it and all its time derivatives are $O(\Omega^{-2})$. Therefore, the outer integral converges since $H(\Omega)$ is bounded. 

It is instructive to compare this result with the linear response  derived from the fluctuation-dissipation theorem,  
$\chi(t) = - (k_B T)^{-1} \Theta(t) \diff C(t)/\diff t$,  combined with Bochner's representation theorem of  real-valued correlation functions 
\begin{align*}
C(t) = \int \cos(\Omega t) \diff F(\Omega) .
\end{align*}
This yields the following representation for the response kernel
\begin{align}
\chi(t) &= \frac{1}{k_B T} \Theta(t) \int \sin(\Omega t) \Omega \diff F(\Omega) ,
\end{align}
and in the frequency domain
\begin{align}\label{eq:chi_rep_F}
\hat{\chi}(z) &= \frac{1}{k_B T}\Theta(t) \int \frac{\Omega^2 \diff F(\Omega)}{\Omega^2- z^2} .
\end{align}
Comparison with our phenomenological approach based on passivity, two differences become apparent. 
 First, the term $\chi_\infty \delta(t) $ 
accounts for an instantaneous response, which physically represents degrees of freedom with dynamics much faster than the time scales of interest and therefore are not included in the description of the fluctuation-dissipation theorem. Second, the integral contribution becomes formally identical upon identifying $\diff F(\Omega) =k_B T (1/\Omega^2 + 1) \diff H(\Omega)$. However, the corresponding function $F(\Omega)$ may not be bounded anymore due to an abundance of modes at low frequencies. Note that Eq.~\eqref{eq:chi_rep_F} reduces to the heuristic result, Eq.~\eqref{eq:Kramers-Kronig} only for the case where $\diff F(\Omega)$ has a density.

The next theorem closes the loop and answers the provocative question we raised earlier: Is it possible to find a smart protocol $f\mathcal{S}(\mathbb{R})$ for the input signal such that work can be extracted at finite time, despite the fact that stability of matter requires that at infinite times work can only be dissipated? 

\newpage

\begin{propositionbox}
\textbf{Theorem: } Stability of matter, i.e. $z\hat{\chi}(z)$ is a positive real (PR) function, implies passivity
$$ \Delta W_{\text{diss}}(T) = \int_{-\infty}^T \diff t  \dot{X}(t) f(t) \geq 0, $$ 
for real-valued signals $f\in \mathcal{S}(\mathbb{R})$ and all times $T\in\mathbb{R}$.  
\end{propositionbox}
\textbf{Proof:} Using Cauer's representation, the dissipated work at finite time $T$ is expressed as
\begin{align*}
\Delta W_{\text{diss}} =& \int_{-\infty}^T \diff t f(t) \chi_\infty \dot{f}(t) 
+ \int_{-\infty}^T \diff t f(t) \int (1+ \Omega^2) \diff H(\Omega) \\ 
& \times \int_{-\infty}^t \diff \bar{t} \cos[ \Omega(t-\bar{t}) ]f(\bar{t}) .
\end{align*}  
The first term corresponds to the reversibly stored energy in the system and is nonnegative
\begin{align*}
\frac{1}{2} \chi_\infty \int_{-\infty}^T \diff t \frac{\diff}{\diff t} f(t)^2 = \frac{1}{2} \chi_{\infty} f(T)^2 \geq 0 ,
\end{align*}
where we used that $\chi_\infty\geq 0$. 

The second term involves a triple integral. Since the  integral over $\diff \bar{t}$ in the second term is $O(\Omega^{-2})$, the   integral over $\diff \Omega$  exists such that  
the two outer integrals can be interchanged
\begin{align*}
\ldots =&  \int (1+ \Omega^2) \diff H(\Omega) \\
& \times  \int_{-\infty}^T \diff t \int_{-\infty}^t  
\diff \bar{t} f(t)
\cos[ \Omega (t-\bar{t}) ] 
f(\bar{t}) .
\end{align*}
The term $f(t) \cos[\Omega(t-\bar{t}) ]f(\bar{t})$ is symmetric under the exchange $\bar{t} \leftrightarrow t$. Therefore, we can symmetrize and extend the domain of integration of the two inner integrals
\begin{align*}
\ldots &= \frac{1}{2} \int_{-\infty}^T \diff t \int_{-\infty}^T \diff \bar{t} f(t) \cos[ \Omega (t-\bar{t}) ] f(\bar{t}) \\
&=\frac{1}{2} \Real\left[ \int_{-\infty}^T \diff t f(t) e^{-i \Omega t} \int_{-\infty}^T \diff \bar{t}  f(\bar{t}) e^{i \Omega \bar{t} } \right] \\
&= \frac{1}{2} \left| \int_{-\infty}^T \diff t f(t) e^{-i\Omega t} \right|^2.
\end{align*}
Collecting results, we find for the second term
\begin{align*}
\frac{1}{2} \int (1+ \Omega^2) \diff H(\Omega) \left| \int_{-\infty}^T \diff t f(t) e^{-i\Omega t} \right|^2 \geq 0 .
\end{align*}
Note that the integral over $\diff H(\Omega)$ is finite since $\int_{-\infty}^T f(t) \exp(-i \Omega t) \diff t= O(\Omega^{-1})$. This proves that the dissipated work is nonnegative for all times. \hfill $\square$

It may appear that we could have simply taken a truncated signal $f(t) \Theta(T-t)$, i.e., instantaneously switching off the input signal at the finite time $T$. However such a signal would not fulfill the requirement of being a test function of rapid descent.

In this section, we derived representation theorems for general linear response functions based solely on the principle of passivity. It was shown that passivity is equivalent to the seemingly weaker condition of stability of matter. Furthermore, passivity, and therefore stability of matter, implies causality, ensuring that the response of a system cannot precede the applied stimulus.

For cases where the response kernel is derived from a correlation function via the fluctuation-dissipation theorem, a simpler representation is obtained. In this case, no instantaneous response arises, reflecting the physical constraint that the system's response is inherently delayed. However, the Kramers-Kronig relations are valid only when the associated Lebesgue-Stieltjes measure possesses a density. This highlights the additional mathematical structure required for the full applicability of these relations.

\section{Summary and Outlook}

In these lecture notes, we have developed a rigorous framework for understanding linear-response theory, correlation functions, and their mathematical properties. The material bridges fundamental physical principles such as causality, passivity, and stability with powerful mathematical tools, including representation theorems and the analytic structure of response functions.

Linear response theory provides the foundation for describing how physical systems respond to small external perturbations. The response is characterized by a kernel that encodes the system's dynamical properties. A key result is the connection between causality and the analytic structure of the response function, which ensures that the system's response cannot precede the applied stimulus. This principle is rigorously captured by the Kramers-Kronig relations, which link the real and imaginary parts of the response function. However, these relations hold only when the associated Lebesgue-Stieltjes measure has a density, highlighting the importance of additional mathematical structure.

Correlation functions, which quantify  fluctuations in stationary processes, were examined in depth. Their defining structural feature is positive-definiteness, captured rigorously by Bochner's theorem: a function is a correlation function precisely when it is the Fourier transform of a finite positive measure. This yields a robust spectral characterization. 
In the physically common case of an integrable function, the characterization simplifies substantially: the Fourier transform exists as a continuous function and the candidate is a valid correlation function exactly when its spectrum  is non-negative. In classical equilibrium, the spectrum is additionally even, while in quantum settings it generally need not be even reflecting the  asymmetry  between emission and absorption  of quanta.  
By the Wiener-Khinchin theorem, this spectrum coincides with the power spectral density, which is often directly accessible experimentally, for example in a scattering experiment. The fluctuation-dissipation theorem further links equilibrium correlations to response, showing how fluctuations determine dissipative behavior and excluding instantaneous responses, thereby reinforcing causal delay in the system's dynamics.

Representation theorems were a central focus of the lecture notes, providing a systematic framework for characterizing response and correlation functions. The Riesz-Herglotz and Nevanlinna representation theorems were derived, offering a rigorous way to represent functions analytic in the upper half-plane. These theorems were further specialized to positive-real functions, which are essential in the study of physical systems governed by passivity and causality. The Cauer representation theorem, derived as a consequence, provides a systematic way to represent such functions in terms of a linear term and an integral over a Lebesgue-Stieltjes measure. This result underscores the deep connection between physical principles such as passivity and the mathematical structure of response functions.

The lecture notes also emphasized the equivalence between passivity and the stability of matter. Passivity, a fundamental principle ensuring that systems do not generate energy, was shown to imply causality and stability. This equivalence provides a unifying perspective on the physical constraints governing response functions and their mathematical properties.

To illustrate these concepts, several examples were discussed, with particular emphasis on the harmonic oscillator. The harmonic oscillator serves as a paradigmatic example of a system where the response function, correlation function, and their spectral properties can be explicitly computed. It demonstrates how the fluctuation-dissipation theorem connects the oscillator's equilibrium fluctuations to its dissipative dynamics. The harmonic oscillator also provides a concrete example of the Kramers-Kronig relations and the analytic structure of response functions, making it an invaluable tool for understanding the general theory.

Looking ahead, the results presented in these notes form the basis for further exploration of advanced topics in physics. Extensions of linear-response theory to nonlinear regimes, the role of memory effects in complex systems, and the application of representation theorems to quantum systems are promising directions for current research and future study. The mathematical tools developed here, such as Bochner's theorem and the representation theorems, are also central to the study of quantum field theory, condensed matter physics, and nonequilibrium statistical mechanics. These areas offer rich opportunities to deepen our understanding of physical systems and their fundamental properties.

\begin{acknowledgments}
I am grateful to Rolf Schilling, Matthias Meiners, Tanja Schilling, Gerhard Jung, Walter Schirmacher, Thomas Voigtmann, and Felix H\"ofling for many fruitful discussions and helpful comments on the manuscript. I particularly thank Matthias Meiners for the proof presented in Appendix~\ref{Sec:StochProcess} on constructing a stochastic process for a given positive-semidefinite function. I also wish to commemorate my PhD adviser, the late Wolfgang G\"otze ($\dagger$ 2021), whose guidance during my time in his group taught me much about correlation functions and their properties.
This research was funded in part by the Austrian Science Fund
(FWF) No. 10.55776/P35673.
\end{acknowledgments}

\appendix

\section{Fluctuation-Dissipation Theorem}\label{Sec:Appendix_FDT}

In this appendix we provide a derivation of the fluctuation-dissipation theorem in the context of classical statistical physics. 
The derivation is standard and can be found for example in Ref.~\cite{Hansen:Theory_of_simple_liquids:2013}. The extension to the non-Hamiltonian coupling is adapted from Ref.~\cite{Tuckerman:Statistical_Mechanics:2023}. 

The time-independent Hamilton function $\mathcal{H}_0 = \mathcal{H}_0(q,p)$ generates the dynamics in equilibrium. Recall that $q= (q^1, \ldots, q^f)$ are the generalized coordinates with $f$ the number of degrees of freedom, and $p =(p_1, \ldots, p_f)$ the associated generalized momenta. The 
time evolution in equilibrium of an observable, a phase-space function, $X=X(q,p)$,  is determined by $\dot{X} = i \mathcal{L}_0 X$ where the Liouville operator $\mathcal{L}_0$ is defined via the Poisson bracket 
\begin{align*}
i \mathcal{L}_0  = \{ \cdot, \mathcal{H}_0 \} = \frac{\partial \mathcal{H}_0}{\partial p_i } \frac{\partial }{\partial q^i} - \frac{\partial \mathcal{H}_0}{\partial q^i} \frac{\partial}{\partial p_i}.
\end{align*} 
Here and in the following  the Einstein summation convention  is used. Applying an external force $f=f(t)$, the Hamiltonian becomes time-dependent $\mathcal{H}_0(q,p) \mapsto \mathcal{H}(q,p,t) = \mathcal{H}_0(q,p) - f(t) Y(q,p)$. The real-valued phase-space function  $Y=Y(q,p)$ to which the force couples in the Hamiltonian is referred to as conjugate to the applied force.  The perturbed dynamics is then provided by an explicitly time-dependent Liouville operator $i\mathcal{L}(t) = i \mathcal{L}_0 + i \delta \mathcal{L}(t)$ where the perturbation enters as
\begin{align*}
i \delta \mathcal{L}(t) = - f(t) \{ \cdot , Y \} = - f(t) \left( \frac{\partial Y}{\partial p_i} \frac{\partial }{\partial q^i} - \frac{\partial Y}{\partial q^i } \frac{\partial}{\partial p_i } \right) .
\end{align*} 
More generally, one considers also non-Hamiltonian couplings 
\begin{align*}
i \delta \mathcal{L}(t) = f(t) \left( C^i \frac{\partial }{\partial q^i} + D_i \frac{\partial}{\partial p_i} \right) ,
\end{align*}
with phase functions $C^i = C^i(q,p), D_i=D_i(q,p), i=1,\ldots, f$. The function should not spoil the property that the flow in phase space is incompressible. Therefore we impose the incompressibility condition
\begin{align}\label{eq:incompressible}
\frac{\partial}{\partial q^i } C^i + \frac{\partial}{\partial p_i} D_i =0.
\end{align}
Clearly, for a Hamiltonian flow this is automatically fulfilled. For the equations of motion in phase space  this implies
\begin{align}\label{eq:eom_nonHamiltonian}
\dot{q}^i &= i \mathcal{L}(t) q^i = \frac{\partial \mathcal{H}_0}{\partial p_i } + f(t) C^i, \nonumber \\ 
\dot{p}_i & = i \mathcal{L}(t) p_i = -\frac{\partial \mathcal{H}_0}{\partial q^i} + f(t) D_i.
\end{align}

The system is initially considered to be in thermal equilibrium before it is exposed to the external force. The force is switched on only for positive times, i.e., $f(t) = 0$ for $t<0$. The goal of the following calculation is to obtain a formally exact expression for the induced shift of the expectation in observable $X$ to linear order in the force. With the external force the ensemble is described by a time-dependent phase space density $\rho = \rho(t) = \rho(q,p,t)$ that allows calculating averages 
\begin{align*}
\langle X(t) \rangle_f = \int X(q,p) \rho(q,p,t) \diff q \diff p.
\end{align*}
For simplicity, we assume that the average vanishes in equilibrium $\langle X \rangle$, otherwise switch from $X$ to the fluctuation $\delta X=  X - \langle X\rangle$.

The equation of motion of the phase space density follows from the local conservation of probability
\begin{align*}
\frac{\diff }{\diff t}\rho = 
\frac{\partial}{\partial t} \rho + \frac{\partial }{\partial q^i } (\dot{q}^i \rho ) + \frac{\partial }{\partial p_i} (\dot{p}_i \rho) = 0.
\end{align*}
Employing the imcompressibility of the phase-space flow, this can be rearranged to obtain the Liouville equation
\begin{align*}
\frac{\diff}{\diff t} \rho &= \frac{\partial}{\partial t} \rho + \left( \dot{q}^i \frac{\partial }{\partial q^i} + \dot{p}_i \frac{\partial}{\partial p_i }\right) \rho =0 \\
&= \frac{\partial }{\partial t} \rho + i \mathcal{L}(t) \rho =0  .
\end{align*}
Solving the Liouville equation exactly is of course hopeless. Here we need to solve it only to linear order in the perturbation. The strategy is then to decompose the phase-space density $\rho(t) = \rho_0 + \delta \rho(t)$ into its equilibrium density $\rho_0 = Z_0^{-1} \exp(-\beta \mathcal{H}_0)$ and its deviation $\delta \rho(t)$.  Here $\beta = 1/k_BT $ is the inverse temperature and $Z_0$ the partition function. By the set-up, there is no deviation before the force acts on the system, $\delta \rho(t) = 0$ for $t<0$. The Liouville equation linearized in the force $f(t)$ then reads
\begin{align*}
\frac{\partial }{\partial t} \delta \rho(t) &= - i \mathcal{L}_0 \delta \rho(t) - f(t) \left( C^i \frac{\partial}{\partial q^i} + D_i \frac{\partial}{\partial p_i} \right) \rho_0 \\
&=  - i \mathcal{L}_0 \delta \rho(t) + \frac{f(t)}{k_B T}  \left( C^i \frac{\partial \mathcal{H}_0}{\partial q^i} + D_i \frac{\partial\mathcal{H}_0}{\partial p_i} \right) \rho_0.
\end{align*}
The term linear in the force plays an important role in the following and will be referred to as the dissipative flux\footnote{We reverse the sign of the dissipative flux relative to Ref.~\cite{Tuckerman:Statistical_Mechanics:2023}.}
\begin{align}\label{eq:flux}
J =J(q,p)  \coloneq  \left( C^i \frac{\partial \mathcal{H}_0}{\partial q^i} + D_i \frac{\partial \mathcal{H}_0}{\partial p_i} \right).
\end{align}
In the case of a Hamiltonian coupling, it simplifies to 
\begin{align*}
J = \{ Y, \mathcal{H}_0 \} =  i\mathcal{L}_0 Y =  \dot{Y}. 
\end{align*}
In terms of the dissipative flux, the linearized Liouville equation can be cast in the form
\begin{align*}
\frac{\partial}{\partial t} \delta \rho(t) + i \mathcal{L}_0 \delta \rho(t) =  \frac{f(t)}{k_B T} \rho_0 J.
\end{align*}
The solution without the inhomogeneity is $\delta \rho(t) = e^{-i \mathcal{L}_0 t} K$ with a phase-space function $K=K(q,p)$. To find a particular solution of the full equation, the method of the variation of the constant is used. The solution for the deviation fulfilling the initial condition is then found to 
\begin{align*}
\delta \rho(t) =  \frac{1}{k_B T} \rho_0 \int_0^t f(\bar{t}) e^{-i \mathcal{L}_0 (t-\bar{t})}J  \diff \bar{t}.
\end{align*}
Here we used that the Liouville operator commutes with the equilibrium density, $\mathcal{L}_0 ( \rho_0 J) = \rho_0 \mathcal{L}_0 J$. 

The equation of motion for observables in equilibrium
\begin{align*}
\frac{\diff}{\diff t} X(t) = \{ X(t) , \mathcal{H}_0 \} = i \mathcal{L}_0 X(t),
\end{align*}
is solved formally by $X(t) = e^{i \mathcal{L}_0 t} X$ (recall the convention $X= X(0)$). This observation allows writing the solution for the deviation in the phase-space density as
\begin{align*}
\delta \rho(t) = \frac{1}{k_B T} \rho_0 \int_0^t f(\bar{t}) J(\bar{t}-t) \diff \bar{t}. 
\end{align*}
Multiplying with $X$ and performing the phase space integral entails 
for the shift in the observable $X$
\begin{align*}
 \langle X(t) \rangle_f= \frac{1}{k_B T} \int_0^t f(\bar{t}) \langle X J(\bar{t}-t ) \rangle \diff \bar{t}.
 \end{align*}
 Using the time-translational invariance of the equilibrium correlation function, an alternative form is
 \begin{align*}
\langle X(t) \rangle_f = \frac{1}{k_B T} \int_0^t f(\bar{t}) \langle X(t-\bar{t})  J \rangle \diff \bar{t}. 
 \end{align*}
 This previous result is already in the form of a convolution of a response function $\chi_{XJ}(t)$ with the external force
 \begin{align}\label{eq:Appendix_LinResp}
 \langle X(t) \rangle_f = \int_0^\infty \chi_{XJ}(t-\bar{t}) f(\bar{t}) \diff \bar{t}. 
 \end{align}
Comparison yields the response function 
 \begin{align}\label{eq:chi_J}
 \chi_{X J}(t) = 
\frac{1}{k_B T} \Theta(t) \langle X(t) J \rangle.
 \end{align}
  The subscripts indicate that $X$ is the observable that is monitored and $J$ is the dissipative current to which the external force couples. 
 The Heaviside function $\Theta(t)$ ensures that the integral in Eq.~\eqref{eq:Appendix_LinResp} extends only up to time $t$ such that the response depends only on the history of the force not on its future. 
 
 The integral in Eq.~\eqref{eq:Appendix_LinResp} starts at time $t=0$, the time when the force is switched on. However, we could have chosen any other reference time $t_0$ after which the force acts on the system. In particular, shifting this reference time to the infinite past, yields
 \begin{align*}
  \langle X(t) \rangle_f = \int_{-\infty}^\infty \chi_{XJ}(t-\bar{t}) f(\bar{t}) \diff \bar{t},
 \end{align*}
provided the force is vanishing rapidly at distant times, $f(t\to \infty) =0$.

 In the case of a Hamiltonian coupling,  the response function can be related to the correlation function $\langle \delta X(t) \delta Y \rangle$ by observing 
\begin{align*}
\chi_{XY}(t) &=\frac{1}{k_B T}  \langle  X(t) \dot{Y} \rangle = \frac{1}{k_B T} \langle X \dot{Y}(-t) \rangle \\
  &= -\frac{1}{k_B T} \frac{\diff }{\diff t}\langle \delta X \delta Y(-t) \rangle = -\frac{1}{k_B T} \frac{\diff }{\diff t}\langle \delta X(t) \delta Y \rangle,   
 \end{align*}
 for $t>0$. The subscripts indicate the monitored observable and the variable conjugate to the force. 

So far we have considered only real-valued observables $X,Y$ for the linear-response set-up. However, as familiar from electrodynamics, it is advantageous to complexify the theory. Then the coupling to the complex force is achieved via the substitution $\mathcal{H}_0 \mapsto \mathcal{H}_0 - \frac{1}{2} [ f(t) Y^* + c.c.] = \mathcal{H}_0 -\Real[ f(t) Y^* ]$. Going through the steps again, shows that all that needs to be changed is to replace $Y \mapsto Y^*$ in the entire calculation. In the complexified version, the fluctuation dissipation theorem assumes the form
\begin{align}\label{eq:FDT_Appendix}
\chi_{XY}(t) = -\frac{1}{k_B T} \Theta(t) \frac{\diff}{\diff t} C_{XY}(t) ,
\end{align}
with the correlation function $C_{XY}(t) = \langle \delta X(t) \delta Y^*\rangle$. 

It is often useful to recast the response in terms of Poisson brackets, which makes the classical-quantum correspondence explicit. 
We use Hamiltonian dynamics, $\dot X(t)=\{X(t),\mathcal H_0\}$, and equilibrium stationarity to relate the time derivative of the correlation to a Poisson bracket. With respect to the canonical  equilibrium measure and employing Liouville's theorem (phase-space volume preservation) together with integration by parts, one obtains
\begin{align*}
\frac{\mathrm{d}}{\mathrm{d}t}\,C_{XY}(t)
=\big\langle \{X(t),\mathcal H_0\}\,\delta Y^{*}\big\rangle
=-\,\big\langle \{X(t), Y^{*}\}\big\rangle.
\end{align*}
Substituting into the fluctuation-dissipation relation, Eq.~\eqref{eq:FDT_Appendix}, yields the classical Kubo formula in Poisson-bracket form,
\begin{align*}
\chi_{XY}(t) \;=\; \frac{1}{k_B T}\,\Theta(t)\,\big\langle \{\,X(t),\,Y^{*}\,\}\big\rangle.
\end{align*}
This representation transfers directly to the quantum case by replacing the Poisson bracket with the commutator and $Y^{*}$ with the adjoint $Y^{\dagger}$. In the Heisenberg picture and for an equilibrium (Gibbs) state,
\begin{align*}
\chi_{XY}(t) \;=\; \frac{i}{\hbar}\,\Theta(t)\,\big\langle \,[\,X(t),\,Y^{\dagger}\,] \,\big\rangle.
\end{align*}
In the classical limit $\hbar\to 0$, $(i/\hbar)[\cdot,\cdot]\to\{\cdot,\cdot\}$, recovering the Poisson-bracket formula above.

We conclude  by deriving explicit expressions for the dissipated power $P(t) := (\diff / \diff t) \langle \mathcal{H}_0(t) \rangle_f$. Since $\dot{\mathcal{H}}_0 = \{ \mathcal{H}_0, \mathcal{H}_0 \} =0$, the linear-response formula yields zero (as for any other conserved quantity). We therefore have to go back to the exact relation
\begin{align*}
\frac{\diff}{\diff t} \langle \mathcal{H}(t) \rangle_f =& \frac{\diff}{\diff t} \int \mathcal{H}_0(q,p) \rho(q,p,t) \diff q \diff p  \\
=& -\int \mathcal{H}_0  \left( \dot{q}^i \frac{\partial}{\partial q^i}\rho(t) + \dot{p}_i \frac{\partial}{\partial p_i} \rho(t) \right) \diff q \diff p \\
=& \int \rho(t) \Big[ \frac{\partial}{\partial q^i} \left( \mathcal{H}_0 \left( \frac{\partial \mathcal{H}_0}{\partial p_i}  + f(t) C^i \right) \right) \\
&  + \frac{\partial}{\partial p_i} \left( \mathcal{H}_0 \left( -\frac{\partial \mathcal{H}_0}{\partial q^i} + f(t) D_i \right) \right) \Big] \diff q \diff p,
\end{align*}
where the last line uses  integration by parts in phase space (surface terms vanish) and  the equations of motion, Eq.~\eqref{eq:eom_nonHamiltonian}. Invoking the incompressibility condition, 
Eq.~\eqref{eq:incompressible}, and the defintion of the flux, Eq.~\eqref{eq:flux}, we find the exact epression for the dissipated power
\begin{align}
P(t) =  f(t) \langle J(t) \rangle_f. 
\end{align}
In the case of a Hamiltonian coupling this reduces to 
\begin{align}
P(t) =f(t) \langle \dot{Y}(t) \rangle_f.
\end{align}

\section{Quasi-elastic light scattering}\label{Sec:Appendix_Light_Scattering}

We consider light scattering  by a medium exhibiting microscopic spatio-temporal fluctuations. Neglecting magnetic effects (no magnetization), the electromagnetic coupling is captured by the  polarization field $\vec{P}(\vec{r},t)$. 
On time scales much slower than electronic ones (Born--Oppenheimer-type approximation)  a local, instantaneous linear constitutive relation holds~\cite{Hellwarth:Progress_in_Quantum_Electronic:1977}
\begin{align*}
\vec{P}(\vec{r},t) = \bm{\chi}(\vec{r},t) \cdot \vec{E}(\vec{r},t),
\end{align*}
where $\bm{\chi}(\vec{r}, t)$ is  the (possibly anisotropic) fluctuating local susceptibility tensor. In an isotropic liquid, we decompose the susceptibility  into a  stationary average $\langle \bm{\chi}(\vec{r},t) \rangle = \chi \bm{I}$ with $\bm{I}$ the unit tensor and a fluctuation
  \begin{align*}
 \bm{\chi}(\vec{r},t) = \chi \bm{I} + \delta \bm{\chi}(\vec{r},t).
 \end{align*}
 The average $\chi>0$ determines  the refractive index of the uniform medium via $n= \sqrt{\epsilon}, \epsilon=1+4\pi\chi$, while  $\delta \bm{\chi}(\vec{r},t)$ captures fluctuations due to particle motion, composition, temperature, or orientation. These fluctuations occur at frequencies  small compared to the optical carrier frequency, so that the scattered light undergoes only a small frequency shift (quasi-elastic light scattering, QELS). 

In a full quantum-mechanical derivation, the starting point is the Hamilton operator
\begin{align*}
\mathcal{H}_\text{tot} &= \mathcal{H}_\text{field}  + \mathcal{H}_\text{int} + \mathcal{H}_0, \\
\mathcal{H}_\text{field} &= \frac{1}{8\pi} \int_\mathscr{V} \diff \vec{r} \left[ \epsilon \vec{E}(\vec{r}) ^2 + \vec{B}(\vec{r})^2 \right], \\
\mathcal{H}_\text{int} &= \frac{-1}{2} \int_\mathscr{V} \diff \vec{r}\,  \vec{E}(\vec{r}) \cdot \delta\bm{\chi}(\vec{r}) \cdot \vec{E}(\vec{r}) ,
\end{align*}
where $\mathcal{H}_\text{field}$ describes the energy of the electromagnetic fields in the uniform medium of volume $\mathscr{V}$, $\mathcal{H}_\text{int}$ the light-matter interaction in the Born-Oppenheimer-type approximation, and $\mathcal{H}_0$ the Hamiltonian of the sample. The electric field $\vec{E}(\vec{r})$, the magnetic field $\vec{B}(\vec{r})$ as well as the fluctuation of the susceptibility $\delta \bm{\chi}(\vec{r})$  become operators with time evolution driven by the  Hamiltonian. 

The electric field in the uniform medium is quantized 
\begin{align*}
\vec{E}(\vec{r}) = \sum_{\vec{k}, \lambda} \mathcal{E}_k \left[ a_{\vec{k}\lambda } \vec{e}_{\vec{k}\lambda} e^{i \vec{k} \cdot \vec{r}} + \text{h.c.} \right] , \quad \mathcal{E}_k = \sqrt{\frac{2 \pi \hbar \omega_k}{\mathscr{V}}} ,
\end{align*}
with $\omega_k = c k /\sqrt{\epsilon}$, and polarization vectors $\vec{e}_{\vec{k}\lambda} \perp \vec{k}$. Here $a_{\vec{k}\lambda}$ denotes the destruction operator of a photon of wave vector $\vec{k}$ and polarization vector $\vec{e}_{\vec{k}\lambda}$, and h.c. means Hermitian conjugate. 

For photon scattering, take as initial state a single photon with wave vector $\vec{k}_\text{i}$ (frequency $\omega_\text{i}$) and polarization $\lambda_\text{i}$, and the sample in state $|I \rangle$ of energy $\epsilon_I$. The product state is $| i \rangle = |1_{\vec{k}_\text{i} \lambda_\text{i} }\rangle \otimes | I \rangle$. After scattering, $ | f \rangle = |1_{\vec{k}_\text{f} \lambda_\text{f} }\rangle \otimes | F \rangle$ 
with photon  $(\vec{k}_\text{f}, \lambda_\text{f})$, $\omega_{\text{f}} = c k_\text{f}/\sqrt{\epsilon}$
and sample energy $\epsilon_F$. 

The interaction  $\mathcal{H}_\text{int}$ induces transitions between these states. Using the plane-wave expansion, the photon part of the transition matrix element becomes
\begin{align*}
\langle f | \mathcal{H}_\text{int}| i \rangle =& -  \mathcal{E}_{k_\text{i}} \mathcal{E}_{k_\text{f}} \langle F | \delta \hat{\chi}_{\text{fi}}(-\vec{q}) | I \rangle \nonumber \\
=& -  \mathcal{E}_{k_\text{i}} \mathcal{E}_{k_\text{f}} \langle I | \delta \hat{\chi}_{\text{fi}}(\vec{q}) | F \rangle^*,
\end{align*}
with the spatial Fourier transform of the projection of the fluctuating susceptibility tensor on the polarization of the incoming and outgoing wave 
\begin{align*}
\delta \hat{\chi}_{\text{fi}}(-\vec{q}) = \int_V \diff \vec{r} \, e^{i\vec{q}\cdot \vec{r}} \vec{e}_{\vec{k}_\text{f}\lambda_\text{f}}^* \cdot \delta \bm{\chi}(\vec{r},t) \cdot \vec{e}_{\vec{k}_\text{i}\lambda_\text{i}} ,
\end{align*}
where $\vec{q} = \vec{k}_\text{i}-\vec{k}_\text{f}$ is the scattering vector. The transition rate  follows from Fermi's golden rule
\begin{align*}
\Gamma_{\text{i}\to \text{f}} = \frac{2\pi}{\hbar} \left| \langle \text{f}| \mathcal{H}_\text{int} | \text{i} \rangle \right|^2  \delta(\epsilon_I - \epsilon_F + \hbar\omega),
\end{align*}
with  frequency shift $\omega = \omega_\text{i}-\omega_\text{f}$. Since the initial sample states $|I\rangle$ are not known individually, we average with weights $p_I$, and sum over all unresolved final sample states $|F\rangle$.

For the flux of the scattered light in the solid angle $\diff \Omega_\text{f}$ with final wavenumber in the interval $[k_\text{f}, k_\text{f}+ \diff k_\text{f} ]$, 
we sum over all final states of the photon compatible with the observation (indicated by a prime at the summation symbol) which generates the rule
\begin{align*}
\sum_\text{f}{}' (\ldots )\mapsto & \mathscr{V} \frac{\diff \vec{k}_\text{f}}{(2\pi)^3} (\ldots) = \mathscr{V} \frac{k_\text{f}^2 \diff k_\text{f} \diff \Omega_\text{f}}{(2\pi)^3}(\ldots) \nonumber \\
 &= \mathscr{V} \frac{\omega_\text{f}^2 \diff\omega \diff \Omega_\text{f}}{  (2\pi c)^3} (\ldots),
\end{align*}
where we transformed the wave-vector element into spherical coordinates and then used $k_\text{f} = \omega_\text{f}/c$ for propagation in vacuum outside the sample.

The scattering cross section is defined by the ratio
\begin{align*}
\diff \sigma = \frac{\text{flux in  } \diff \Omega_\text{f} \text{ and frequency interval }  \diff \omega }{\text{incident flux density}} .
\end{align*} 
For box normalization with one incident photon in the medium, the  incident flux is $ c / \mathscr{V} \sqrt{\epsilon} $. In quasi-elastic scattering, we approximate $\omega_\text{i} \approx \omega_\text{f}$. Collecting results,
\begin{align*}
&\frac{\diff \sigma}{\diff \Omega_\text{f} \diff \omega} =  
 \left( \frac{c}{\mathscr{V} \sqrt{\epsilon}} \right)^{-1}  \frac{\mathscr{V} \omega_\text{f}^2 }{(2\pi c)^3 }  \sum_{I, F} p_I \Gamma_{i\to f}  \nonumber  \\
=&  \frac{\sqrt{\epsilon}}{2\pi} \left(\frac{ \omega_\text{f}}{c} \right)^4    \sum_{I, F} p_I \left|  \langle I | \delta \hat{\chi}_{\text{fi}}(\vec{q}) | F \rangle \right|^2 2\pi \hbar \delta(\epsilon_I - \epsilon_F + \hbar\omega) .
\end{align*}
The delta-function reflects energy conservation: the photon  transfers  $\hbar \omega$ to the sample, $\epsilon_F= \epsilon_I + \hbar \omega$. 

We can make further progress by expressing the delta-function as Fourier transform 
\begin{align*}
\lefteqn{\frac{\diff \sigma}{\diff \Omega_\text{f} \diff \omega} = } \nonumber \\
=&  \frac{\sqrt{\epsilon}}{2\pi} \left(\frac{ \omega_\text{f}}{c} \right)^4  \int \diff t e^{i\omega t}   \sum_{I, F} p_I e^{i (\epsilon_I-\epsilon_F) t/\hbar} \left|  \langle I | \delta \hat{\chi}_{\text{fi}}(\vec{q}) | F \rangle \right|^2 \nonumber  \\
=&  \frac{\sqrt{\epsilon}}{2\pi} \left(\frac{ \omega_\text{f}}{c} \right)^4  \int \diff t e^{i\omega t} 
\sum_{I, F}
p_I \langle I | \delta \hat{\chi}_{\text{fi}}(\vec{q},t) | F\rangle \langle F | \delta \hat{\chi}_{\text{fi}}(\vec{q})^\dagger | I\rangle .
\end{align*}
Here we introduced the Heisenberg operator $\delta \hat{\chi}_{\text{fi}}(\vec{q},t) = e^{i \mathcal{H}_0 t/\hbar} \delta \hat{\chi}_{\text{fi}}(\vec{q}) e^{-i \mathcal{H}_0 t/\hbar}$ driven by the matter Hamiltonian and used that $|I \rangle, | F\rangle$ are eigenstates of $\mathcal{H}_0$ with energies $\epsilon_I, \epsilon_F$.
The sum over all final states can be eliminated using the completeness relation, while the averaging over the initial states merely yields an  ensemble average. For the scattering cross section we obtain the final form
\begin{align}
\frac{\diff \sigma}{\diff \Omega_\text{f} \diff \omega}
=&  \frac{\sqrt{\epsilon}}{2\pi} \left(\frac{ \omega_\text{f}}{c} \right)^4  \int \diff t e^{i\omega t} \langle  \delta \hat{\chi}_{\text{fi}}(\vec{q},t)  \delta \hat{\chi}_{\text{fi}}(\vec{q})^\dagger\rangle .
\end{align}
The scattering cross section is essentially the Fourier transform of a correlation function of fluctuations. With the present sign conventions  for the scattering vector $\vec{q}$ and the frequency shift $\omega$, the time-dependent operator is always to the left, consistent  with our definition of correlation functions and the Kubo scalar product. The classical analog follows by  treating $\delta \hat{\chi}(\vec{q},t)$ as a phase space variable, making the  operator ordering immaterial.

\section{Construction of a stochastic process for given positive-semidefinite function}\label{Sec:StochProcess}

In the main text, we showed that the correlation function $C(t) = \langle \delta X(t) \delta X \rangle$ for a stochastic process $\{ X(t) : t\in \mathbb{R}\}$ is positive-semidefinite. Here we show that the correlation functions exhaust the set of  real-valued positive-semidefinite functions, i.e., to every real-valued positive-semidefinite function $C:\mathbb{R} \to \mathbb{R}$ there exists a stochastic process $( X(t) : t\in \mathbb{R})$ that yields $C$ as its correlation function~\footnote{We thank  Matthias Meiners (Universit\"at Gie{\ss}en) for a sketch of the proof.}.  

Assume $C$ to be real-valued and positive-semidefinite. Then $C(t) = C(-t)^* = C(-t)$ is symmetric. By definition it fulfills 
\begin{align*}
\sum_{i,j=1}^n a_i C(t_i-t_j) a_j^* \geq 0 ,
\end{align*}
for any choice $n\in \mathbb{N}, t_1,\ldots, t_n, a_1,\ldots, a_n \in \mathbb{C}$. The $n\times n$ matrix $C_{ij} \coloneq C(t_i-t_j) = C_{ji}$ is then real-valued, symmetric, and positive-semidefinite. 

Step 1: There is a Gaussian vector (i.e. a tuple of Gaussian random variables) $(X_1,\ldots, X_n) $ with $\langle X_i \rangle =0, \langle X_i X_j \rangle = C_{ij}$. This can be seen as follows. Diagonalize $C= O \Lambda O^T$ with an orthogonal matrix $O$: $O O^T = \mathbbm{1}$ and the diagonal matrix $\Lambda =\text{diag} (\lambda_1,\ldots, \lambda_n)$ with non-negative entries. Then the square root of $C$ can be defined as $A \coloneq O \sqrt{\Lambda} O^T$ with $\sqrt{\Lambda}= \text{diag} (\sqrt{\lambda_1}, \ldots, \sqrt{\lambda_n})$. Clearly, $AA^T = O \Lambda O^T = C\geq 0$. 

Let now $Y_1, \ldots, Y_n$ be independently identically distributed Gaussian variables with $\langle Y_i\rangle =0$ and $\langle Y_i^2 \rangle =1$. Define $X_i  = \sum_{j=1}^n A_{ij} Y_j$. Then $\langle X_i \rangle = 0$ and $\langle X_i X_k \rangle = \sum_{j=1}^n\sum_{l=1}^n \langle A_{ij} A_{kl} \langle Y_j Y_l \rangle = \sum_{j=1}^n A_{ij} A_{kj} = C_{ij}$. Thus, $(X_1,\ldots, X_n)$ is a Gaussian vector with covariance matrix $C_{ij}$.  

Step 2: This construction yields a family of finite-dimensional distributions: for each $t_1< t_2 < \ldots < t_n $ the (unique) distributions of the Gaussian vectors build above. These should  correspond to $(X(t_1) , \ldots, X(t_n) )$ of a stochastic process $( X(t) : t\in \mathbb{R})$. The existence of such a process follows from Kolmogorov's extension (consistency) theorem~\cite{Kolmogorov:Grundbegriffe:1933, Kallenberg:Foundations_of_modern_probability:1997}: 
 There is a stochastic process with given finite-dimensional distributions if the family is consistent. 

Consistency means that one can marginalize by leaving out certain elements. However, for our Gaussian vector this is clear since this corresponds merely to deleting columns and rows of the matrix $(C_{ij})_{i,j=1,\ldots,n}$. Note that it is not necessary that $C$ is continuous. 
  \hfill $\square$
  
  The proof can be extended to cover also the case of complex-valued positive-semidefinite functions~\cite{Meiners}.

\section{Proof of Bochner's theorem}\label{Sec:Appendix_Bochner}

In this section we prove Bochner's theorem, stating that any function $C: \mathbb{R} \to \mathbb{C}$ that is positive-semidefinite and continuous at $t=0$ can be represented as $C(t) = \int e^{-i \Omega t}\diff F(\Omega)$ with a finite Lebuesgue-Stieltjes measure $F:\mathbb{R}\to \mathbb{R}$. 

First, we recall that since $C$ is positive-semidefinite and continuous at $t=0$, it is uniformly continuous by Eq.~\eqref{eq:uniform_continuity}.

As a first step, we assume that the function is integrable, $C \in L^1(\mathbb{R})$, i.e. $\int_{\mathbb{R}} |f(t) | \diff t < \infty$. Then by Lebuesgue's dominated convergence theorem, the Fourier transform
\begin{align*}
S(\Omega) = \int  e^{-i \Omega t} C(t) \diff t,
\end{align*}
exists and is uniformly continuous and bounded. We show that $S(\Omega) \geq 0$.  The inequality 
\begin{align*}
0 \leq \frac{1}{T}\int_0^T \diff t \int_0^T \diff s\,  e^{-i \Omega t } e^{i \Omega s} C(t-s) ,
\end{align*}
for $T>0$ is obtained  as continuous limit of  positive definiteness with $\lambda_i = \lambda(t_i) = e^{-i \Omega t_i}/\sqrt{T}$. Substituting $\tau= t-s$ yields 
\begin{align*}
0 \leq &  \frac{1}{T}\int_{-T}^T C(\tau) e^{-i \Omega t} \left( \int_{s\in [0,T], s+\tau \in [0,T]} \diff s \right) \diff \tau \\ 
=& \int_{-\infty}^\infty C(\tau) e^{-i \Omega t}  (1-|\tau| / T) \Theta(T^2-\tau^2) \diff \tau . 
\end{align*}
The limit $T\to \infty$ follows from the dominated convergence theorem: the integrand converges pointwise and is dominated by an integrable function, since $C \in L^1(\mathbb{R})$.  Therefore,
\begin{align*}
0 \leq \int e^{-i \Omega t} C(t) \diff t = S(\Omega).
 \end{align*}
 Next we show that $S\in L^1(\mathbb{R})$  and the correlation function is recovered by inverse Fourier transform.

Introduce approximants  
\begin{align*}
S_\sigma(\Omega) \coloneq S(\Omega) \exp(- \sigma^2 \Omega^2 /2), \qquad \sigma>0. 
\end{align*}
Clearly, $S_\sigma\in L^1(\mathbb{R})$, such that the inverse Fourier transform exists
\begin{align*}
\int e^{i \Omega t}S_\sigma(\Omega) \frac{\diff \Omega}{2\pi} = \int e^{i \Omega t} e^{-\sigma^2 \Omega^2/2} \left( \int C(s) e^{-i \Omega s} \diff s\right) \frac{\diff \Omega}{2\pi} .
\end{align*}
 By Fubini's theorem
 \begin{align*}
 \ldots = \frac{1}{\sqrt{2\pi \sigma^2}} \int C(s) \left(  \int e^{i \Omega (t-s)} 
\frac{e^{-\sigma^2 \Omega^2 /2}}{\sqrt{2\pi/\sigma^2}} \diff \Omega \right) \diff s .
 \end{align*}
 The inner integral is the Fourier transform of a Gaussian and evaluates to $e^{-(t-s)^2/2 \sigma^2}$. Substituting $t-s = \sigma u$ yields
 \begin{align}\label{eq:Fourier_Gaussian}
 \ldots = \frac{1}{\sqrt{2\pi}} \int C(t+ \sigma u) e^{-u^2/2} \diff u. 
 \end{align}
 In particular, specializing to $t=0$ and substituting $\tau = \sigma u$ implies
 \begin{align*}
 \int S_\sigma(\Omega) \frac{\diff \Omega}{2\pi} =& \int C(\tau) \frac{e^{-\tau^2/2\sigma^2}}{\sqrt{2\pi \sigma^2} } \diff \tau  \\ 
\leq & \text{sup}_{\tau\in \mathbb{R}} | C(\tau) | = C(0),
 \end{align*}
 where the last inequality follows from positive semidefiniteness. 
 
Since $S_\sigma(\Omega) \uparrow S(\Omega)$ as $\sigma \downarrow 0$, the monotone-convergence theorem implies
\begin{align*}
\int S(\Omega) \frac{\diff \Omega}{2\pi} \leq C(0),
\end{align*}
 which reveals that $S\in L^1(\mathbb{R})$. Furthermore, by dominated convergence, we conclude from Eq.~\eqref{eq:Fourier_Gaussian} as $\sigma\downarrow 0$
 \begin{align*}
\int e^{i \Omega t} S(\Omega) \frac{\diff \Omega}{2\pi} =  \frac{1}{\sqrt{2\pi}} \int C(t) e^{-u^2/2} \diff u = C(t).
 \end{align*}
 
 If $C$ is not integrable, approximate
 \begin{align*}
 C_\epsilon(t) \coloneq C(t) e^{-\epsilon^2 t^2/2} = \int C(t) e^{-i \omega t } \frac{e^{-\omega^2/2\epsilon^2}}{\sqrt{2\pi \epsilon^2}} \diff \omega, 
 \end{align*}
 with $\epsilon>0$. 
 Since $C_\epsilon(t)$ is uniformly continuous, integrable, $C_\epsilon(0) = C(0) \geq 0$, and positive linear combination of the positive-semidefinite functions $C(t) e^{-i \omega t}, \omega\in \mathbb{R}$, it is again positive-semidefinite. According to the first part of this proof, it can be represented as
 \begin{align*}
 C_\epsilon(t) = \int e^{-i \Omega t} \diff F_\epsilon(\Omega),
 \end{align*}
 with $\diff F_\epsilon(\Omega) = S_\epsilon(\Omega) \diff \Omega/2\pi$. In other words, each $C_\epsilon$ is [up to a factor $C(0)$] the characteristic function of some measure. Since $C_\epsilon(t) \to C(t)$ pointwise and $C(t)$ is continuous at $t=0$, 
$F_\epsilon \to F$ weakly to some measure $F$ 
(Levy's continuity theorem, Ref.~\cite{Shiryaev:Probability:2016}, III \S 3), in particular, 
 \begin{align*}
 C(t) = \int e^{-i\Omega t} \diff F(\Omega).  \qquad \square
 \end{align*}

\section{Proof of Herglotz's representation theorem}\label{Sec:Appendix_Herglotz}
In this appendix we sketch the proof of the Herglotz representation for a Nevanlinna function $\hat{C} : \mathbb{C}_+ \to \mathbb{C}_+$ with inequality $| \hat{C}(z) | \leq M / \Imag[z]$. The proof is adapted from Ref.~\cite{Teschl:Mathematical_methods:2014}. 

 \begin{figure}[htp!]
\includegraphics[width=\linewidth]{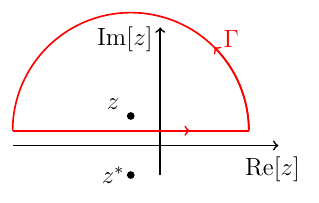}
\caption{Contour $\Gamma$ encircling $z$ in the complex upper half-plane..
  \label{fig:Contour}}.
\end{figure}

Fix $\epsilon>0$ and  consider complex frequencies 
 $z = x+ i y$, $y > \epsilon, x\in \mathbb{R}$ and write  $v(z) = \Imag[\hat{C}(z)]$.
Define  the contour $\Gamma = \{ x+ i\epsilon + \Omega : \Omega\in [-R, R] \} \cup \{ x+ i \epsilon + R e^{i \varphi} : \varphi \in [0,\pi) \} $ for $R>0$
encircling $z$, see Fig.~\ref{fig:Contour}. Then by Cauchy's formula
\begin{align*}
\hat{C}(z) =& \frac{1}{2\pi i} \int_{\Gamma} \left( \frac{1}{w-z} - \frac{1}{w- (z^* + 2 i \epsilon)} \right) \hat{C}(w) \diff w \nonumber \\
=&  \int_{-R}^R \frac{y-\epsilon}{\Omega^2 +(y-\epsilon)^2 } \hat{C}(x+ i \epsilon+ \Omega) \frac{\diff \Omega}{\pi} \nonumber \\
&+ \int_0^\pi \frac{y-\epsilon}{R^2 e^{2i \varphi} + (y-\epsilon)^2}  \hat{C}(x+ i \epsilon+ Re^{i\varphi}) R i e^{i\varphi} \frac{\diff \varphi}{\pi} .
\end{align*} 
Here we used in the first line that  $z^* + 2 i \epsilon$ lies outside the contour $\Gamma$ such that the second integral does not contribute. 
By the bound $| \hat{C}(x+i \epsilon+Re^{i\varphi}) | \leq M / \epsilon$, the integral over the semicircle vanishes in the limit $R\to \infty$. After shifting $\Omega\mapsto \Omega- x$, we  arrive at
\begin{align*}
\hat{C}(z) = \int_\mathbb{R} \frac{y-\epsilon}{(\Omega-x)^2 + (y-\epsilon)^2}\hat{C}(\Omega+ i \epsilon) \frac{\diff \Omega}{\pi}.
\end{align*}
Taking imaginary parts yields the analytic representation for $\Imag[z]> \epsilon$
\begin{align}\label{eq:analytic_representation}
v(z) = \int_\mathbb{R} \Imag\left[ \frac{1}{\Omega+ i \epsilon - z} \right] v_\epsilon(\Omega) \diff \Omega,
\end{align}
 with $v_\epsilon(\Omega) = \Imag[\hat{C}(\Omega+i \epsilon)]/\pi > 0$. Since $\hat{C}$ is analytic in $\mathcal{C}_+$ with $\hat{C}(z) \to 0$ as $z\to \infty$, we can extend Eq.~\eqref{eq:analytic_representation} uniquely to a representation 
\begin{align*}
\hat{C}(z) = \int_\mathbb{R}  \frac{ v_\epsilon(\Omega) \diff \Omega}{\Omega+ i \epsilon - z} .
\end{align*}

Next we show $\int_\mathbb{R} v_{\epsilon}(\Omega) \diff \Omega \leq M$, in particular, $v_\epsilon$ is integrable. Since $y>\epsilon$, 
\begin{align*}
v(i y) = \int_\mathbb{R} \frac{y-\epsilon}{\Omega^2 + (y-\epsilon)^2 } v_{\epsilon}(\Omega) \diff \Omega.
\end{align*}
From the growth condition $|\hat{C}(z)| \leq M /\Imag[z]$ we conclude
\begin{align*}
M \geq y v(i y) \geq \int_\mathbb{R} \frac{(y-\epsilon)^2}{\Omega^2 + (y-\epsilon)^2} v_{\epsilon}(\Omega) \diff \Omega.
\end{align*}
The fraction approaches unity for large $y$. More precisely, for $m>0$,  $(y-\epsilon)^2/[\Omega^2 +(y-\epsilon)^2] \geq 1/(1+m^2)$ if $|\Omega| \leq m (y-\epsilon)$. Therefore
\begin{align*}
M \geq \frac{1}{1+ m^2} \int_{-m (y-\epsilon)}^{ m (y- \epsilon)} v_{\epsilon}(\Omega) \diff \Omega.
\end{align*}
Now let first $y\to \infty$, then $m\to 0$ to find the desired bound.

The analytic representation, Eq.~\eqref{eq:analytic_representation}, holds for all $\epsilon>0$. Since the left-hand side is independent of $\epsilon$, we can take the limit
\begin{align*}
v(z) = \lim_{\epsilon \to 0^+} \int_\mathbb{R} \Imag\left[ \frac{1}{\Omega+ i \epsilon - z} \right] v_\epsilon(\Omega) \diff \Omega. 
\end{align*}
Yet, $\Imag[ 1/(\Omega+ i \epsilon- z)] \to \Imag[1/(\Omega-z)]$ pointwise as $\epsilon\to 0^+$ for all $z\in \mathbb{C}_+$. Then
\begin{align*}
v(z) = \lim_{\epsilon \to 0^+} \int_\mathbb{R} \Imag\left[ \frac{1}{\Omega- z} \right] \diff F_\epsilon(\Omega) . 
\end{align*}
with $\diff F_{\epsilon}(\Omega) = v_{\epsilon}(\Omega) \diff \Omega$. Since $F_{\epsilon}(\mathbb{R} ) = \int_\mathbb{R} v_{\epsilon}(\Omega) \diff \Omega \leq M$, the measures are uniformly bounded. 

In the last step, we choose a positive sequence $\epsilon_n\to 0^+$ and write $F_n(\Omega) = F_{\epsilon_n}(\Omega)$. Since $F_n(\mathbb{R}) \leq M$,  there is a subsequence that converges weakly to some Lebesgue-Stieltjes measure $F:\mathbb{R}\to \mathbb{R}$ (selection theorem, Ref.~\cite{Feller:Probability:1991}, VIII.6 ). Therefore
\begin{align*}
v(z) = \int_\mathbb{R} \Imag\left[ \frac{1}{\Omega-z} \right] \diff F(\Omega) .
\end{align*}
Now $\hat{C}(z)$ and $\int_{\mathbb{R}} (\Omega-z)^{-1} \diff F(\Omega)$ have the same imaginary part and therefore differ at most by a real constant. By our bound this constant must be zero. \hfill $\square$

\section{Representation theorem for Nevanlinna functions}\label{Sec:Appendix_Representation_Nevanlinna}

In this section, we prove a representation theorem for Nevanlinna functions. Then we include growth conditions to arrive at the Herglotz representation theorem of the main text. Last,
by specializing to positive-real (PR) functions, we subsequently derive Cauer's representation. 

\subsection{Harmonic functions on the disk}
The first result is a representation theorem for positive harmonic functions on the unit disk. The presentation follows closely the one of Koosis~\cite{Koosis:Hp_spaces:1998}.

Consider the power series expansion of an analytic function for $|z|< 1$, $F(z) = \sum_{n=0}^\infty a_n z^n$. The real part of $F(z)$ can be expressed as
\begin{align*}
U(z= r e^{{\rm i} \vartheta}) &= \frac{1}{2} \sum_{n=0}^\infty r^n \left(a_n e^{i n \vartheta} + a_n^* e^{-i n \vartheta} \right)\\
&\equiv \sum_{n=-\infty}^\infty A_n r^{|n|} e^{i n \vartheta} .
\end{align*}
Hence $U(z)$ is harmonic and we assume that it is  positive, $U(z)>0$ for $|z|<1$.

Using the orthogonality of the exponentials on the unit circle, the coefficients $A_n$ can be obtained by projection
\begin{equation*}
A_n r^{|n|} = \frac{1}{2\pi} \int_{-\pi}^{\pi} U(r e^{{\rm i} t}) e^{-{\rm i} n t} \diff t . 
\end{equation*}
In particular, the positivity of $U(z)$ implies that the zeroth coefficient satisfies $A_0>0$. Thus, the measure $\diff F_r(t) = (2\pi A_0)^{-1} U(r e^{{\rm i} t}) \diff t$ constitutes a probability distribution on the circle $[-\pi,\pi)$ with Fourier coefficients $A_n r^{|n|} /A_0$. 
Now consider a sequence $r_k \to 1$ as $k\to \infty$. 
By Helly's selection theorem~\cite{Feller:Probability:1991,Shiryaev:Probability:2016}, there exists a probability distribution function $F$ such that a subsequence $(F_{r_{k_i}})_{i\in \mathbb{N}}$ converges weakly to $F$ as $i\to \infty$:
\begin{align*}
A_n r_{k_i}^{|n|}/A_0 =& \frac{1}{2\pi} \int_{(-\pi,\pi]} e^{-i n t} \diff F_{r_{k_i}}(t) \\
& \to \frac{1}{2\pi}  \int_{(-\pi,\pi]} e^{-i n t} \diff F(t) = A_n/A_0 .
\end{align*} 
Define $\diff \mu(t) = A_0 \diff F(t)$, such that $\mu$ is a finite Lebesgue-Stieltjes measure on the unit circle. Then
\begin{align*}
U(r e^{i \vartheta}) = \frac{1}{2\pi} \int_{(-\pi,\pi]} \diff \mu(t) 
\sum_{n=-\infty}^{\infty} e^{i n (\vartheta-t)} r^{|n|}.
\end{align*}
The geometric series exists for $r<1$, yielding the Poisson kernel
\begin{equation*}
P_r(\vartheta) := \frac{1-r^2}{1+ r^2- 2 r \cos \vartheta}.
\end{equation*}
Thus, a positive harmonic function $U=U(z)$ on the open unit disk, $|z|<1$, can be represented as
\begin{equation}\label{eq:representation_disk}
U(r e^{i \vartheta} ) = \frac{1}{2\pi} \int_{(-\pi,\pi]}  P_r(\vartheta-t) \diff \mu(t) ,
\end{equation}
where $\mu$ is a finite Borel measure on the unit circle.

\subsection{Positive harmonic functions on a half-plane}
The representation theorem for harmonic functions on the unit disk can be extended to the upper half-plane $\mathbb{C}_+$ via the M\"obius transform 
\begin{align*}
z \mapsto w \coloneq \frac{ {\rm i}- z}{ {\rm i} + z} = r e^{ {\rm i} \vartheta} ,
\end{align*}
which is injective on the upper half-plane $\mathbb{C}_+$ and surjective onto the open unit disk.
Points on the real line are mapped to the unit circle
\begin{align*}
\Omega \mapsto \omega := \frac{ {\rm i} -\Omega}{ {\rm i} + \Omega} = e^{ {\rm i} \tau}, \qquad \Omega \in \mathbb{R}.
\end{align*}
The Poisson kernel in complex polar coordinates is then obtained as
\begin{align*}
P_r(\vartheta-\tau) \coloneq & \sum_{n=-\infty}^\infty r^{|n|} 
e^{ {\rm i} n (\vartheta-\tau) } \\
=& 1 + \sum_{n=1}^\infty \big[ w^n (\omega^*)^n + (w^*)^n \omega^n \big] \\
=& 1 + \frac{w \omega^*}{1- w \omega^* } + \frac{w^* \omega}{1-   \omega w^* } \\
=& \frac{1 - |w|^2}{|\omega- w|^2}. 
\end{align*}
Here we used the geometric series in the third line and that $\omega^* = \omega^{-1}$ since $\omega$ is a pure phase factor. 

Transforming back to the upper half-plane $w\mapsto z= x+ i y$, the Poisson kernel becomes
\begin{align}
P_r(\vartheta-\tau) =& \frac{y}{(x-\Omega)^2 + y^2 } (1+ \Omega^2).
\label{eq:Poisson_half_plane}
\end{align}
Define a new measure $\nu$ as the transform $\nu(\Omega) = \mu(\tau)/\pi$, which implies
\begin{equation*}
\diff \mu(\tau) =  \left|\frac{\diff \tau(\Omega)}{\diff \Omega}\right| \pi \diff \nu(\Omega) = \frac{2\pi \,\diff \nu(\Omega)}{1+ \Omega^2}.
\end{equation*}
Note that the total measure of the unit circle exempting the point $\pi$ is finite, $\int_{(-\pi, \pi)} \diff \mu(\tau) < \infty$, which now entails 
\begin{equation*}
\int_{\mathbb{R}} \frac{ \diff \nu(\Omega)}{1+ \Omega^2 } < \infty.
\end{equation*}

The integrand of \eqref{eq:representation_disk} in the coordinates of the upper half-plane then reads
\begin{equation*}
P_r(\vartheta-\tau) \diff \mu(\tau) = \frac{2\pi\, y \,\diff \nu(\Omega)}{(x-\Omega)^2 + y^2}.
\end{equation*}
We have to allow for the possibility that $\mu$ possesses an atom at $\tau=\pi$, which is mapped to complex infinity. In that case, the Poisson kernel \eqref{eq:Poisson_half_plane} assumes the limit $P_r(\vartheta-\pi) = y$. 

Collecting all arguments for transforming \eqref{eq:representation_disk} from the unit disk to the upper half-plane, the positive harmonic function $V=V(x,y)$ in the upper half-plane allows for the representation 
\begin{equation*}
V(x,y) = \alpha y +  \int_{\mathbb{R}} \frac{y\,\diff \nu(\Omega)}{ (x-\Omega)^2 + y^2 }, \qquad \alpha\geq 0.
\end{equation*} 
It is now easy to identify the harmonic conjugate as
\begin{equation*}
U(x,y) = \alpha x + \beta + \int_{\mathbb{R}} \left( \frac{\Omega-x}{ (x-\Omega)^2 + y^2 } - \frac{\Omega}{1+\Omega^2} \right) \diff \nu(\Omega), 
\end{equation*}
with $\beta\in \mathbb{R}$. 
The second term in the integrand is independent of $x,y$ and is introduced to guarantee that the integral converges for large $|\Omega|$ using the integrability condition above. 

Switching back to complex functions we have proven the representation theorem for Nevanlinna functions:
\begin{propositionbox}
\textbf{Theorem (Nevanlinna representation):}
A Nevanlinna function $\hat{N}: \mathbb{C}_+ \to \mathbb{C}_+ \cup \mathbb{R}$ can be represented as
\begin{equation*}
\hat{N}(z) = \alpha z + \beta +  \int_{\mathbb{R}} \left( \frac{1}{\Omega-z} - \frac{\Omega}{1+\Omega^2} \right) \diff \nu(\Omega) ,
\end{equation*}
where $\alpha \ge 0$, $\beta\in \mathbb{R}$, and $\nu$ is a positive Lebesgue-Stieltjes measure with
\begin{equation*}
\int_{\mathbb{R}} \frac{\diff \nu(\Omega)}{1+ \Omega^2} < \infty.
\end{equation*}
\end{propositionbox}
Equivalently, one may write
\begin{equation*}
\hat{N}(z) = \alpha z + \beta +  \int_{\mathbb{R}} \frac{1 + \Omega z}{\Omega-z} \diff H(\Omega),
\end{equation*}
where $\diff H (\Omega) = \diff \nu(\Omega)/(1+\Omega^2)$, so that $H$ is a finite Lebesgue-Stieltjes measure. 

We now employ the representation theorem to prove 
\begin{propositionbox}
\textbf{Theorem (Herglotz representation):} Let $\hat{C} : \mathbb{C}_+ \to \mathbb{C}_+$ be a Nevanlinna function that satisfies either
\begin{enumerate}
\item[1.] the growth condition
\[|\hat{C}(z)| \le \frac{M}{\Imag[ z]}\quad \text{for all } z\in\mathbb{C}_+ \text{ and some } M<\infty,
\]
\end{enumerate}
or
\begin{enumerate}
\item[2.] the sectorial growth condition
$\hat{C}(z)=O(1/|z|)$ uniformly as  $|z|\to\infty$  for   $0<\delta<\arg z<\pi-\delta$.
\end{enumerate}
Then there exists a (finite) Lebesgue-Stieltjes measure $F$ on $\mathbb{R}$ such that
\[
\hat{C}(z)=\int_{\mathbb{R}} \frac{1}{\Omega-z}\,\diff F(\Omega),\qquad z\in\mathbb{C}_+.
\]
Moreover, in case \emph{(1)} one has $\int_{\mathbb{R}} \diff F(\Omega)\le M$.
\end{propositionbox}

\textbf{Proof:}
 By the Nevanlinna (Herglotz) representation theorem, there exist $\alpha\ge 0 $, $\beta\in\mathbb{R}$, and a Lebesgue-Stieltjes  measure $F$ on $\mathbb{R}$ with
\[
\int_{\mathbb{R}} \frac{1}{1+\Omega^2}\, \diff F(\Omega)<\infty,
\]
such that for all  $z\in\mathbb{C}_+$,
\[
\hat{C}(z) = \alpha z + \beta + \int_{\mathbb{R}}\!\left(\frac{1}{\Omega - z} - \frac{\Omega}{1+\Omega^2}\right)\, \diff F(\Omega).
\]

\emph{Step 1: Eliminate the linear term (and deduce finiteness of $F$).}
For $y>0$,
\[
y\,\Imag[\hat{C}(iy)] \;=\; \alpha y^2 + \int_{\mathbb{R}} \frac{y^2}{\Omega^2+y^2} \diff F(\Omega).
\]
Under (1), $|\hat{C}(iy)|\le M/y$, hence $y\,\Imag[\hat{C}(iy)] \le M$. Under (2), for any fixed $\delta\in(0,\pi/2)$ the imaginary axis (argument $\pi/2$) lies in the sector, so $\hat{C}(iy)=O(1/y)$ and thus $y\,\Imag[\hat{C}(iy)]$ is bounded. In either case, boundedness as $y\to\infty$ forces $\alpha=0$.

With $\alpha=0$, we have
\[
y\,\Imag[\hat{C}(iy)] \;=\; \int_{\mathbb{R}} \frac{y^2}{\Omega^2+y^2} \diff F(\Omega),
\]
and the integrand increases to $1$ as $y\to\infty$. By monotone convergence,
\[
\lim_{y\to\infty} y\,\Imag[\hat{C}(iy)] \;=\; F(\mathbb{R}),
\]
so $F(\mathbb{R})<\infty$ (and in case (1), $F(\mathbb{R})\le M$).

Thus, from now on $F$ is a finite measure and
\[
\hat{C}(z) = \beta + \int_{\mathbb{R}}\!\left(\frac{1}{\Omega - z} - \frac{\Omega}{1+\Omega^2}\right)\, \diff F(\Omega).
\]

\emph{Step 2: Vanishing of the constant term and pure Cauchy form.}
Letting $y\to\infty$ along $z=iy$, the assumption (1) or (2) gives $\hat{C}(iy)\to 0$. Since $F$ is finite, the function $\Omega\mapsto \Omega/(1+\Omega^2)$ is bounded and integrable with respect to $F$, and
\[
\left|\int_{\mathbb{R}} \frac{1}{\Omega-iy}\, \diff F(\Omega)\right| \le \frac{F(\mathbb{R})}{y}\xrightarrow[y\to\infty]{}0.
\]
Therefore,
\[
\lim_{y\to\infty} \hat{C}(iy) \;=\; \beta - \int_{\mathbb{R}} \frac{\Omega}{1+\Omega^2} \diff F(\Omega),
\]
and thus
\[
\beta \;=\; \int_{\mathbb{R}} \frac{\Omega}{1+\Omega^2} \diff F(\Omega).
\]
Substituting this back gives the pure Cauchy form
\[
\hat{C}(z) \;=\; \int_{\mathbb{R}} \frac{1}{\Omega - z}\, \diff F(\Omega).
\]

\emph{Step 3: Total mass bound (already obtained).}
From Step 1 we already have $F(\mathbb{R})<\infty$, and in case (1), $F(\mathbb{R})\le M$.

This yields the asserted representation
\[
\hat{C}(z)=\int_{\mathbb{R}} \frac{1}{\Omega-z}\diff F(\Omega).
\]
with $F(\mathbb{R})\le M$ in case (1) and $F(\mathbb{R})<\infty$ in case (2). \hfill $\square$

We now specialize the Nevanlinna representation theorem to positive-real (PR) functions, i.e., Nevanlinna functions $\hat{N}$ that additionally satisfy $\Real[ \hat{N}(i y) ] = 0$ for $y>0$. 

The next step for proving the Cauer representation theorem requires the following symmetry property of PR functions.
\begin{propositionbox}
\textbf{Theorem:} For all $z\in \mathbb{C}_+$, a PR function $\hat{N}: \mathbb{C}_+ \to \mathbb{C}_+\cup \mathbb{R}$ satisfies 
\[
\hat{N}(z)^* = - \hat{N}(-z^*).
\] 
\end{propositionbox}

\textbf{Proof:} Define $\tilde{N}(z) = i \hat{N}(iz)$, which is complex analytic in the right complex half-plane and fulfills $\Imag[\tilde{N}(y)]=0$ for $y>0$. This implies that all derivatives of $\tilde{N}(z)$ on the positive-real line are real-valued, $\hat{N}^{(n)}(y_0)\in \mathbb{R}, y_0>0$. From the Taylor series expansion
\[
\tilde{N}(z) = \sum_{n=0}^\infty \frac{\tilde{N}^{(n)}(y_0)}{n!} (z-y_0)^n,
\]  
it follows that $\tilde{N}(z^*) = \tilde{N}(z)^*$ in a neighborhood of each $y_0>0$, and by the identity theorem this holds for all $z$ in the right half-plane. Transforming back to $\hat{N}(z)$, this implies $\hat{N}(z)^* = - \hat{N}(-z^*)$, which is the desired property. \hfill $\square$

Using the symmetry property established above, we now complete the proof of the Cauer representation theorem for PR functions. 
The PR function $\hat{N}(z)$ can be represented as 
\begin{align*}
\hat{N}(z) &= \frac{1}{2} \left[ \hat{N}(z) - \hat{N}(-z^*)^* \right] \\
&= \alpha z + \frac{1}{2} \int_{\mathbb{R}} \left( \frac{1}{\Omega-z} - \frac{1}{\Omega + z} \right) \diff \nu(\Omega) \\
&= \alpha z + \int_{\mathbb{R}} \frac{z}{\Omega^2- z^2} \diff \nu(\Omega),
\end{align*}
where $\alpha\ge 0$ and $\nu$ is a Lebesgue-Stieltjes measure satisfying $\int_{\mathbb{R}}  \diff \nu(\Omega) / (1+ \Omega^2 ) < \infty$. Since the integral is symmetric with respect to $\Omega$, the measure $\nu$ can be symmetrized. Expressing the representation in terms of $\diff H(\Omega) = \diff \nu(\Omega) /(1+ \Omega^2)$, where $H$ is a finite Lebesgue-Stieltjes measure, the PR function can be written as
\[
\hat{N}(z) = \alpha z + \int_{\mathbb{R} } \frac{z}{\Omega^2- z^2} (1+ \Omega^2) \diff H(\Omega).
\]
This completes the proof of the Cauer representation theorem. 

\hfill $\square$


\begin{thebibliography}{27}%
\makeatletter
\providecommand \@ifxundefined [1]{%
 \@ifx{#1\undefined}
}%
\providecommand \@ifnum [1]{%
 \ifnum #1\expandafter \@firstoftwo
 \else \expandafter \@secondoftwo
 \fi
}%
\providecommand \@ifx [1]{%
 \ifx #1\expandafter \@firstoftwo
 \else \expandafter \@secondoftwo
 \fi
}%
\providecommand \natexlab [1]{#1}%
\providecommand \enquote  [1]{``#1''}%
\providecommand \bibnamefont  [1]{#1}%
\providecommand \bibfnamefont [1]{#1}%
\providecommand \citenamefont [1]{#1}%
\providecommand \href@noop [0]{\@secondoftwo}%
\providecommand \href[0]{\begingroup \@sanitize@url \@href}%
\providecommand \@href[1]{\@@startlink{#1}\@@href}%
\providecommand \@@href[1]{\endgroup#1\@@endlink}%
\providecommand \@sanitize@url [0]{\catcode `\\12\catcode `\$12\catcode
  `\&12\catcode `\#12\catcode `\^12\catcode `\_12\catcode `\%12\relax}%
\providecommand \@@startlink[1]{}%
\providecommand \@@endlink[0]{}%
\providecommand \url  [0]{\begingroup\@sanitize@url \@url }%
\providecommand \@url [1]{\endgroup\@href {#1}{\urlprefix }}%
\providecommand \urlprefix  [0]{URL }%
\providecommand \Eprint [0]{\href }%
\providecommand \doibase [0]{http://dx.doi.org/}%
\providecommand \selectlanguage [0]{\@gobble}%
\providecommand \bibinfo  [0]{\@secondoftwo}%
\providecommand \bibfield  [0]{\@secondoftwo}%
\providecommand \translation [1]{[#1]}%
\providecommand \BibitemOpen [0]{}%
\providecommand \bibitemStop [0]{}%
\providecommand \bibitemNoStop [0]{.\EOS\space}%
\providecommand \EOS [0]{\spacefactor3000\relax}%
\providecommand \BibitemShut  [1]{\csname bibitem#1\endcsname}%
\let\auto@bib@innerbib\@empty
\bibitem [{\citenamefont {Kubo}\ \emph {et~al.}(2012)\citenamefont {Kubo},
  \citenamefont {Toda},\ and\ \citenamefont
  {Hashitsume}}]{Kubo:Statistical_physics_II:2012}%
  \BibitemOpen
  \bibfield  {author} {\bibinfo {author} {\bibfnamefont {R.}~\bibnamefont
  {Kubo}}, \bibinfo {author} {\bibfnamefont {M.}~\bibnamefont {Toda}}, \ and\
  \bibinfo {author} {\bibfnamefont {N.}~\bibnamefont {Hashitsume}},\
  }\href@noop {} {\emph {\bibinfo {title} {Statistical physics II:
  nonequilibrium statistical mechanics}}},\ Vol.~\bibinfo {volume} {31}\
  (\bibinfo  {publisher} {Springer Science \& Business Media},\ \bibinfo {year}
  {2012})\BibitemShut {NoStop}%
\bibitem [{\citenamefont {Hansen}\ and\ \citenamefont
  {McDonald}(2013)}]{Hansen:Theory_of_simple_liquids:2013}%
  \BibitemOpen
  \bibfield  {author} {\bibinfo {author} {\bibfnamefont {J.-P.}\ \bibnamefont
  {Hansen}}\ and\ \bibinfo {author} {\bibfnamefont {I.~R.}\ \bibnamefont
  {McDonald}},\ }\href@noop {} {\emph {\bibinfo {title} {Theory of simple
  liquids: with applications to soft matter}}}\ (\bibinfo  {publisher}
  {Academic press},\ \bibinfo {year} {2013})\BibitemShut {NoStop}%
\bibitem [{\citenamefont {Feller}(1991)}]{Feller:Probability:1991}%
  \BibitemOpen
  \bibfield  {author} {\bibinfo {author} {\bibfnamefont {W.}~\bibnamefont
  {Feller}},\ }\href@noop {} {\emph {\bibinfo {title} {An introduction to
  probability theory and its applications, Volume 2}}},\ Vol.~\bibinfo {volume}
  {2}\ (\bibinfo  {publisher} {John Wiley \& Sons},\ \bibinfo {year}
  {1991})\BibitemShut {NoStop}%
\bibitem [{\citenamefont {Shiryaev}(2016)}]{Shiryaev:Probability:2016}%
  \BibitemOpen
  \bibfield  {author} {\bibinfo {author} {\bibfnamefont {A.~N.}\ \bibnamefont
  {Shiryaev}},\ }\href@noop {} {\emph {\bibinfo {title} {Probability}}},\
  Vol.~\bibinfo {volume} {95}\ (\bibinfo  {publisher} {Springer},\ \bibinfo
  {year} {2016})\BibitemShut {NoStop}%
\bibitem [{\citenamefont {Doi}\ \emph {et~al.}(1988)\citenamefont {Doi},
  \citenamefont {Edwards},\ and\ \citenamefont
  {Edwards}}]{Doi:Polymer_Dynamics:1988}%
  \BibitemOpen
  \bibfield  {author} {\bibinfo {author} {\bibfnamefont {M.}~\bibnamefont
  {Doi}}, \bibinfo {author} {\bibfnamefont {S.~F.}\ \bibnamefont {Edwards}}, \
  and\ \bibinfo {author} {\bibfnamefont {S.~F.}\ \bibnamefont {Edwards}},\
  }\href@noop {} {\emph {\bibinfo {title} {The theory of polymer dynamics}}},\
  Vol.~\bibinfo {volume} {73}\ (\bibinfo  {publisher} {Oxford University
  Press},\ \bibinfo {year} {1988})\BibitemShut {NoStop}%
\bibitem [{\citenamefont
  {Van~Kampen}(1992)}]{Kampen:Stochastic_Processes:1992}%
  \BibitemOpen
  \bibfield  {author} {\bibinfo {author} {\bibfnamefont {N.~G.}\ \bibnamefont
  {Van~Kampen}},\ }\href@noop {} {\emph {\bibinfo {title} {Stochastic processes
  in physics and chemistry}}},\ Vol.~\bibinfo {volume} {1}\ (\bibinfo
  {publisher} {Elsevier},\ \bibinfo {year} {1992})\BibitemShut {NoStop}%
\bibitem [{\citenamefont
  {Tuckerman}(2023)}]{Tuckerman:Statistical_Mechanics:2023}%
  \BibitemOpen
  \bibfield  {author} {\bibinfo {author} {\bibfnamefont {M.~E.}\ \bibnamefont
  {Tuckerman}},\ }\href@noop {} {\emph {\bibinfo {title} {Statistical
  {M}echanics: {T}heory and {M}olecular {S}imulation}}}\ (\bibinfo  {publisher}
  {Oxford university press},\ \bibinfo {address} {Oxford},\ \bibinfo {year}
  {2023})\BibitemShut {NoStop}%
\bibitem [{\citenamefont
  {Lovesey}(1984{\natexlab{a}})}]{Lovesey:Theory_of_neutron_scatteringI:1984}%
  \BibitemOpen
  \bibfield  {author} {\bibinfo {author} {\bibfnamefont {S.~W.}\ \bibnamefont
  {Lovesey}},\ }\href@noop {} {\emph {\bibinfo {title} {Theory of neutron
  scattering from condensed matter. Vol. 1. Nuclear scattering}}}\ (\bibinfo
  {publisher} {Clarendon Press},\ \bibinfo {year} {1984})\BibitemShut {NoStop}%
\bibitem [{\citenamefont
  {Lovesey}(1984{\natexlab{b}})}]{Lovesey:Theory_of_neutron_scatteringII:1984}%
  \BibitemOpen
  \bibfield  {author} {\bibinfo {author} {\bibfnamefont {S.~W.}\ \bibnamefont
  {Lovesey}},\ }\href@noop {} {\emph {\bibinfo {title} {Theory of neutron
  scattering from condensed matter. Vol. 2. Polarization effect and magnetic
  scattering}}}\ (\bibinfo  {publisher} {Clarendon Press},\ \bibinfo {year}
  {1984})\BibitemShut {NoStop}%
\bibitem [{\citenamefont {Berne}\ and\ \citenamefont
  {Pecora}(2000)}]{Berne:Dynamic_light_scattering:2000}%
  \BibitemOpen
  \bibfield  {author} {\bibinfo {author} {\bibfnamefont {B.~J.}\ \bibnamefont
  {Berne}}\ and\ \bibinfo {author} {\bibfnamefont {R.}~\bibnamefont {Pecora}},\
  }\href@noop {} {\emph {\bibinfo {title} {Dynamic light scattering: with
  applications to chemistry, biology, and physics}}}\ (\bibinfo  {publisher}
  {Courier Corporation},\ \bibinfo {year} {2000})\BibitemShut {NoStop}%
\bibitem [{\citenamefont {Rigler}\ and\ \citenamefont
  {Elson}(2012)}]{Rigler:Fluorescence:2012}%
  \BibitemOpen
  \bibfield  {author} {\bibinfo {author} {\bibfnamefont {R.}~\bibnamefont
  {Rigler}}\ and\ \bibinfo {author} {\bibfnamefont {E.~S.}\ \bibnamefont
  {Elson}},\ }\href@noop {} {\emph {\bibinfo {title} {Fluorescence correlation
  spectroscopy: {T}heory and applications}}},\ Vol.~\bibinfo {volume} {65}\
  (\bibinfo  {publisher} {Springer Science \& Business Media},\ \bibinfo {year}
  {2012})\BibitemShut {NoStop}%
\bibitem [{\citenamefont {H\"ofling}\ \emph {et~al.}(2011)\citenamefont
  {H\"ofling}, \citenamefont {Bamberg},\ and\ \citenamefont
  {Franosch}}]{Hofling:SM_7:2011}%
  \BibitemOpen
  \bibfield  {author} {\bibinfo {author} {\bibfnamefont {F.}~\bibnamefont
  {H\"ofling}}, \bibinfo {author} {\bibfnamefont {K.-U.}\ \bibnamefont
  {Bamberg}}, \ and\ \bibinfo {author} {\bibfnamefont {T.}~\bibnamefont
  {Franosch}},\ }\bibfield  {title} {\enquote {\bibinfo {title} {Anomalous
  transport resolved in space and time by fluorescence correlation
  spectroscopy},}\ }\href{\doibase 10.1039/C0SM00718H} {\bibfield  {journal}
  {\bibinfo  {journal} {Soft Matter}\ }\textbf {\bibinfo {volume} {7}},\
  \bibinfo {pages} {1358} (\bibinfo {year} {2011})}\BibitemShut {NoStop}%
\bibitem [{\citenamefont {H\"ofling}\ and\ \citenamefont
  {Franosch}(2013)}]{Hofling:RPP_76:2013}%
  \BibitemOpen
  \bibfield  {author} {\bibinfo {author} {\bibfnamefont {F.}~\bibnamefont
  {H\"ofling}}\ and\ \bibinfo {author} {\bibfnamefont {T.}~\bibnamefont
  {Franosch}},\ }\bibfield  {title} {\enquote {\bibinfo {title} {Anomalous
  transport in the crowded world of biological cells},}\ }\href{\doibase
  10.1088/0034-4885/76/4/046602} {\bibfield  {journal} {\bibinfo  {journal}
  {Reports on Progress in Physics}\ }\textbf {\bibinfo {volume} {76}},\
  \bibinfo {pages} {046602} (\bibinfo {year} {2013})}\BibitemShut {NoStop}%
\bibitem [{\citenamefont {Bancaud}\ \emph {et~al.}(2010)\citenamefont
  {Bancaud}, \citenamefont {Huet}, \citenamefont {Rabut},\ and\ \citenamefont
  {Ellenberg}}]{Bancaud:fluorescence:2010}%
  \BibitemOpen
  \bibfield  {author} {\bibinfo {author} {\bibfnamefont {A.}~\bibnamefont
  {Bancaud}}, \bibinfo {author} {\bibfnamefont {S.}~\bibnamefont {Huet}},
  \bibinfo {author} {\bibfnamefont {G.}~\bibnamefont {Rabut}}, \ and\ \bibinfo
  {author} {\bibfnamefont {J.}~\bibnamefont {Ellenberg}},\ }\bibfield  {title}
  {\enquote {\bibinfo {title} {Fluorescence perturbation techniques to study
  mobility and molecular dynamics of proteins in live cells: Frap,
  photoactivation, photoconversion, and flip},}\ }\href@noop {} {\bibfield
  {journal} {\bibinfo  {journal} {Cold Spring Harbor Protocols}\ }\textbf
  {\bibinfo {volume} {2010}},\ \bibinfo {pages} {pdb} (\bibinfo {year}
  {2010})}\BibitemShut {NoStop}%
\bibitem [{\citenamefont {Cerbino}\ and\ \citenamefont
  {Trappe}(2008)}]{Cerbino:PRL_100:2008}%
  \BibitemOpen
  \bibfield  {author} {\bibinfo {author} {\bibfnamefont {R.}~\bibnamefont
  {Cerbino}}\ and\ \bibinfo {author} {\bibfnamefont {V.}~\bibnamefont
  {Trappe}},\ }\bibfield  {title} {\enquote {\bibinfo {title} {Differential
  dynamic microscopy: Probing wave vector dependent dynamics with a
  microscope},}\ }\href{\doibase 10.1103/PhysRevLett.100.188102} {\bibfield
  {journal} {\bibinfo  {journal} {Phys. Rev. Lett.}\ }\textbf {\bibinfo
  {volume} {100}},\ \bibinfo {pages} {188102} (\bibinfo {year}
  {2008})}\BibitemShut {NoStop}%
\bibitem [{\citenamefont {Crocker}\ and\ \citenamefont
  {Grier}(1996)}]{Crocker:JCIS_179:1996}%
  \BibitemOpen
  \bibfield  {author} {\bibinfo {author} {\bibfnamefont {J.~C.}\ \bibnamefont
  {Crocker}}\ and\ \bibinfo {author} {\bibfnamefont {D.~G.}\ \bibnamefont
  {Grier}},\ }\bibfield  {title} {\enquote {\bibinfo {title} {Methods of
  digital video microscopy for colloidal studies},}\ }\href{\doibase
  https://doi.org/10.1006/jcis.1996.0217} {\bibfield  {journal} {\bibinfo
  {journal} {Journal of Colloid and Interface Science}\ }\textbf {\bibinfo
  {volume} {179}},\ \bibinfo {pages} {298} (\bibinfo {year}
  {1996})}\BibitemShut {NoStop}%
\bibitem [{\citenamefont {Straube}\ \emph {et~al.}(2020)\citenamefont
  {Straube}, \citenamefont {Kowalik}, \citenamefont {Netz},\ and\ \citenamefont
  {H{\"o}fling}}]{Straube:Comm_Phys_3:2020}%
  \BibitemOpen
  \bibfield  {author} {\bibinfo {author} {\bibfnamefont {A.~V.}\ \bibnamefont
  {Straube}}, \bibinfo {author} {\bibfnamefont {B.~G.}\ \bibnamefont
  {Kowalik}}, \bibinfo {author} {\bibfnamefont {R.~R.}\ \bibnamefont {Netz}}, \
  and\ \bibinfo {author} {\bibfnamefont {F.}~\bibnamefont {H{\"o}fling}},\
  }\bibfield  {title} {\enquote {\bibinfo {title} {Rapid onset of molecular
  friction in liquids bridging between the atomistic and hydrodynamic
  pictures},}\ }\href{https://www.nature.com/articles/s42005-020-0389-0#citeas}
  {\bibfield  {journal} {\bibinfo  {journal} {Communications Physics}\ }\textbf
  {\bibinfo {volume} {3}},\ \bibinfo {pages} {126} (\bibinfo {year}
  {2020})}\BibitemShut {NoStop}%
\bibitem [{\citenamefont {Gesztesy}\ and\ \citenamefont
  {Tsekanovskii}(2000)}]{Gesztesy:MathMeth_218:2000}%
  \BibitemOpen
  \bibfield  {author} {\bibinfo {author} {\bibfnamefont {F.}~\bibnamefont
  {Gesztesy}}\ and\ \bibinfo {author} {\bibfnamefont {E.}~\bibnamefont
  {Tsekanovskii}},\ }\bibfield  {title} {\enquote {\bibinfo {title} {On
  matrix--valued {H}erglotz functions},}\ }\href@noop {} {\bibfield  {journal}
  {\bibinfo  {journal} {Mathematische Nachrichten}\ }\textbf {\bibinfo {volume}
  {218}},\ \bibinfo {pages} {61} (\bibinfo {year} {2000})}\BibitemShut
  {NoStop}%
\bibitem [{\citenamefont {Zemanian}(1987)}]{Zemanian:Distribution:1987}%
  \BibitemOpen
  \bibfield  {author} {\bibinfo {author} {\bibfnamefont {A.~H.}\ \bibnamefont
  {Zemanian}},\ }\href@noop {} {\emph {\bibinfo {title} {Distribution theory
  and transform analysis: an introduction to generalized functions, with
  applications}}}\ (\bibinfo  {publisher} {Dover Publications},\ \bibinfo
  {year} {1987})\BibitemShut {NoStop}%
\bibitem [{\citenamefont {Teschl}(2014)}]{Teschl:Mathematical_methods:2014}%
  \BibitemOpen
  \bibfield  {author} {\bibinfo {author} {\bibfnamefont {G.}~\bibnamefont
  {Teschl}},\ }\href@noop {} {\emph {\bibinfo {title} {Mathematical methods in
  quantum mechanics}}},\ Vol.\ \bibinfo {volume} {157}\ (\bibinfo  {publisher}
  {American Mathematical Soc.},\ \bibinfo {year} {2014})\BibitemShut {NoStop}%
\bibitem [{\citenamefont {Krapf}\ \emph {et~al.}(2018)\citenamefont {Krapf},
  \citenamefont {Marinari}, \citenamefont {Metzler}, \citenamefont {Oshanin},
  \citenamefont {Xu},\ and\ \citenamefont {Squarcini}}]{Krapf:NJP_20:2018}%
  \BibitemOpen
  \bibfield  {author} {\bibinfo {author} {\bibfnamefont {D.}~\bibnamefont
  {Krapf}}, \bibinfo {author} {\bibfnamefont {E.}~\bibnamefont {Marinari}},
  \bibinfo {author} {\bibfnamefont {R.}~\bibnamefont {Metzler}}, \bibinfo
  {author} {\bibfnamefont {G.}~\bibnamefont {Oshanin}}, \bibinfo {author}
  {\bibfnamefont {X.}~\bibnamefont {Xu}}, \ and\ \bibinfo {author}
  {\bibfnamefont {A.}~\bibnamefont {Squarcini}},\ }\bibfield  {title} {\enquote
  {\bibinfo {title} {Power spectral density of a single {B}rownian trajectory:
  what one can and cannot learn from it},}\ }\href{\doibase
  10.1088/1367-2630/aaa67c} {\bibfield  {journal} {\bibinfo  {journal} {New
  Journal of Physics}\ }\textbf {\bibinfo {volume} {20}},\ \bibinfo {pages}
  {023029} (\bibinfo {year} {2018})}\BibitemShut {NoStop}%
\bibitem [{\citenamefont {G\"otze}(2009)}]{Goetze:Complex_Dynamics}%
  \BibitemOpen
  \bibfield  {author} {\bibinfo {author} {\bibfnamefont {W.}~\bibnamefont
  {G\"otze}},\ }\href@noop {} {\emph {\bibinfo {title} {Complex Dynamics of
  Glass-Forming Liquids -- A Mode-Coupling Theory}}}\ (\bibinfo  {publisher}
  {Oxford},\ \bibinfo {address} {Oxford},\ \bibinfo {year} {2009})\BibitemShut
  {NoStop}%
\bibitem [{\citenamefont
  {Hellwarth}(1977)}]{Hellwarth:Progress_in_Quantum_Electronic:1977}%
  \BibitemOpen
  \bibfield  {author} {\bibinfo {author} {\bibfnamefont {R.}~\bibnamefont
  {Hellwarth}},\ }\bibfield  {title} {\enquote {\bibinfo {title} {Progress in
  quantum electronics},}\ \ }(\bibinfo  {publisher} {Pergamon, New York},\
  \bibinfo {year} {1977})\BibitemShut {NoStop}%
\bibitem [{\citenamefont {Kolmogorov}(1933)}]{Kolmogorov:Grundbegriffe:1933}%
  \BibitemOpen
  \bibfield  {author} {\bibinfo {author} {\bibfnamefont {A.}~\bibnamefont
  {Kolmogorov}},\ }\href@noop {} {\emph {\bibinfo {title} {Grundbegriffe der
  Wahrscheinlichkeitsrechnung (Engl. transl. Foundations of Probability,
  Chelsea (1956))}}}\ (\bibinfo {year} {1933})\BibitemShut {NoStop}%
\bibitem [{\citenamefont
  {Kallenberg}(1997)}]{Kallenberg:Foundations_of_modern_probability:1997}%
  \BibitemOpen
  \bibfield  {author} {\bibinfo {author} {\bibfnamefont {O.}~\bibnamefont
  {Kallenberg}},\ }\href@noop {} {\emph {\bibinfo {title} {Foundations of
  modern probability}}}\ (\bibinfo  {publisher} {Springer},\ \bibinfo {year}
  {1997})\BibitemShut {NoStop}%
\bibitem [{\citenamefont {Meiners}()}]{Meiners}%
  \BibitemOpen
  \bibfield  {author} {\bibinfo {author} {\bibfnamefont {M.}~\bibnamefont
  {Meiners}},\ }\href@noop {} {}\bibinfo {note} {Private communication
  (2026)}\BibitemShut {NoStop}%
\bibitem [{\citenamefont {Koosis}(1998)}]{Koosis:Hp_spaces:1998}%
  \BibitemOpen
  \bibfield  {author} {\bibinfo {author} {\bibfnamefont {P.}~\bibnamefont
  {Koosis}},\ }\href@noop {} {\emph {\bibinfo {title} {Introduction to $H_p$
  spaces}}},\ \bibinfo {number} {115}\ (\bibinfo  {publisher} {Cambridge
  University Press},\ \bibinfo {year} {1998})\BibitemShut {NoStop}%
\end{thebibliography}
%

\end{document}